\documentclass[onecolumn,showpacs,amsfonts,amsmath,amssymb,floatfix]{revtex4}
\usepackage{url}
\usepackage{bm}		
\usepackage{graphicx}	
\usepackage{cancel}	

\begin{document}


\title[The interior structure of rotating black holes 2]{The interior structure of rotating black holes 2. Uncharged black holes}

\author{Andrew J S Hamilton}
\affiliation{JILA and
Dept.\ Astrophysical \& Planetary Sciences,
Box 440, U. Colorado, Boulder, CO 80309, USA}
\email{Andrew.Hamilton@colorado.edu}	

\newcommand{\simpropto}{\raisebox{-0.8ex}[1.5ex][0ex]{$
		\begin{array}[b]{@{}c@{\;}} \propto \\
		[-1.4ex] \sim \end{array}$}}

\newcommand{\dd}{d}
\newcommand{\ddsq}{\dd^2\mkern-1.5mu}
\newcommand{\ddd}{\dd^3\mkern-1.5mu}
\newcommand{\dddd}{\dd^4\mkern-1.5mu}
\newcommand{\DD}{D}
\newcommand{\ee}{e}
\newcommand{\im}{i}
\newcommand{\Ei}{{\rm Ei}}
\newcommand{\perpperp}{\perp\!\!\perp}
\newcommand{\ppartial}{\partial^2\mkern-1mu}
\newcommand{\nn}{\nonumber\\}

\newcommand{\diag}{{\rm diag}}
\newcommand{\Lz}{L}
\newcommand{\Msun}{{\rm M}_\odot}
\newcommand{\inn}{{\rm in}}
\newcommand{\out}{{\rm ou}}
\newcommand{\sep}{{\rm sep}}

\newcommand{\bg}{\bm{g}}
\newcommand{\bp}{\bm{p}}
\newcommand{\bv}{\bm{v}}
\newcommand{\bx}{\bm{x}}
\newcommand{\bgamma}{\bm{\gamma}}

\newcommand{\Apot}{{\cal A}}
\newcommand{\hatA}{\hat{A}}
\newcommand{\Br}{B}
\newcommand{\betar}{\lambda_r}
\newcommand{\Cx}{C_x}
\newcommand{\Cy}{C_y}
\newcommand{\Cz}{{\tilde C}}
\newcommand{\Dx}{D_x}
\newcommand{\Dy}{D_y}
\newcommand{\Deltax}{\Delta_x}
\newcommand{\Deltaxinf}{\Delta_{x , {\rm inf}}}
\newcommand{\Deltaxev}{\Delta_{x , {\rm ev}}}
\newcommand{\Deltay}{\Delta_y}
\newcommand{\Er}{E}
\newcommand{\expinf}{\xi}
\newcommand{\Fz}{{\tilde F}}
\newcommand{\starF}{\,{}^\ast\!F}
\newcommand{\Jx}{J_x}
\newcommand{\Jy}{J_y}
\newcommand{\Killing}{K}
\newcommand{\Kx}{K_x}
\newcommand{\Ky}{K_y}
\newcommand{\Lx}{L_x}
\newcommand{\Ly}{L_y}
\newcommand{\Mass}{{\cal M}}
\newcommand{\Mbh}{M_\bullet}
\newcommand{\Mdot}{\dot{M}}
\newcommand{\Mbhdot}{\dot{M}_\bullet}
\newcommand{\KCarter}{{\cal K}}
\newcommand{\NUT}{{\cal N}}
\newcommand{\NUTbh}{{\cal N}_\bullet}
\newcommand{\px}{p^x}
\newcommand{\Px}{P_x}
\newcommand{\Py}{P_y}
\newcommand{\QCarter}{{\cal Q}}
\newcommand{\Qelec}{Q}
\newcommand{\Qelecbh}{\Qelec_\bullet}
\newcommand{\Qmag}{{\cal Q}}
\newcommand{\Qmagbh}{\Qmag_\bullet}
\newcommand{\rhosep}{\rho_{\rm s}}
\newcommand{\rhosepm}{\rho_{{\rm s} , -}}
\newcommand{\rhox}{\rho_x}
\newcommand{\rhoy}{\rho_y}
\newcommand{\uel}{u}
\newcommand{\Uinf}{U}
\newcommand{\Ur}{{\cal R}}
\newcommand{\Utheta}{\Theta}
\newcommand{\rc}{{\scriptstyle R}}
\newcommand{\tc}{{\scriptstyle T}}
\newcommand{\smallrc}{{\scriptscriptstyle R}}
\newcommand{\smalltc}{{\scriptscriptstyle T}}
\newcommand{\smallzero}{{\scriptscriptstyle 0}}
\newcommand{\Ux}{U_x}
\newcommand{\Uy}{U_y}
\newcommand{\vel}{v}
\newcommand{\Wx}{W_x}
\newcommand{\Wy}{W_y}
\newcommand{\xin}{x_{\rm in}}
\newcommand{\Xx}{X_x}
\newcommand{\Xy}{X_y}
\newcommand{\cXx}{\tilde{X}_x}
\newcommand{\Yx}{Y_x}
\newcommand{\Yy}{Y_y}
\newcommand{\cYx}{\tilde{Y}_x}
\newcommand{\Zx}{Z_x}
\newcommand{\Zy}{Z_y}
\newcommand{\omegax}{\omega_x}
\newcommand{\omegay}{\omega_y}
\newcommand{\omegaxin}{\omega_{x, {\rm in}}}
\newcommand{\omegayin}{\omega_{y, {\rm in}}}
\newcommand{\omegaxout}{\omega_{x, {\rm out}}}
\newcommand{\omegayout}{\omega_{y, {\rm out}}}

\hyphenpenalty=3000

\newcommand{\ingoingconditionfig}{
    \begin{figure}[tb]
    \begin{center}
    \leavevmode
    \includegraphics[scale=.7]{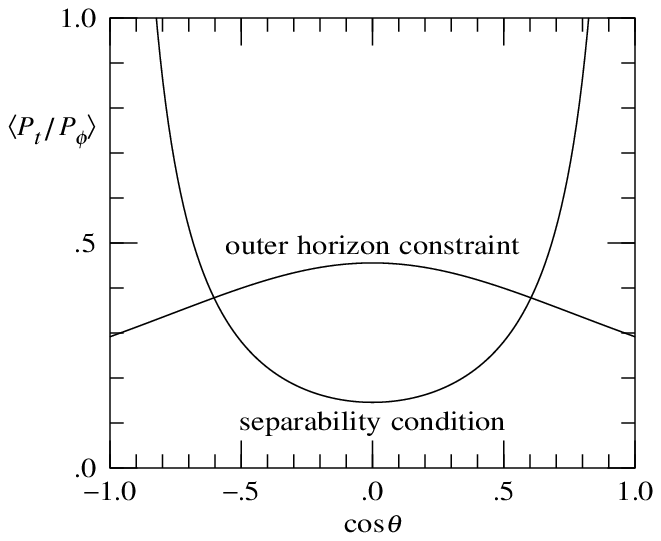}
    \caption[1]{
    \label{ingoingcondition}
Comparison of the condition on the average ratio
of $P_t$ to $P_\phi$
at the inner horizon
required by conformal separability, equation~(\protect\ref{Ptphipm}),
to the constraint on the same ratio
imposed by the condition that particles free-fall
from outside the outer horizon,
equation~(\protect\ref{Ptcondition}),
for the case of an uncharged black hole with
angular momentum parameter $a = 0.96 \Mbh$.
In the equatorial region $\cos\theta \sim 0$,
outgoing particles
(positive $P_t$ and $P_\phi$)
satisfy the constraint but ingoing particles
(negative $P_t$ and $P_\phi$)
do not.
Conversely in the polar regions $| \cos\theta | \sim 1$,
ingoing particles
satisfy the constraint but outgoing particles
do not.
    }
    \end{center}
    \end{figure}
}

\begin{abstract}
A solution is obtained for the interior structure
of an uncharged rotating black hole that accretes a collisionless fluid.
The solution is conformally stationary, axisymmetric,
and conformally separable,
possessing a conformal Killing tensor.
The solution holds approximately if the accretion rate is small but finite,
becoming more accurate
as the accretion rate tends to zero.
Hyper-relativistic counter-streaming between collisionless ingoing
and outgoing streams drives inflation at (just above) the inner horizon,
followed by collapse.
As ingoing and outgoing streams approach the inner horizon,
they focus into twin narrow beams directed along the ingoing
and outgoing principal null directions, regardless of the initial
angular motions of the streams.
The radial energy-momentum of the counter-streaming beams
gravitationally accelerates the streams even faster
along the principal directions,
leading to exponential growth in the streaming density and pressure,
and in the Weyl curvature and mass function.
At exponentially large density and curvature,
inflation stalls, and the spacetime collapses.
As the spacetime collapses, the angular motions of the
freely-falling streams grow.
When the angular motion has become comparable to the radial motion,
which happens when the conformal factor has shrunk to an exponentially
tiny scale, conformal separability breaks down,
and the solution fails.
The condition of conformal separability prescribes the form
of the ingoing and outgoing accretion flows
incident on the inner horizon.
The dominant radial part of the solution holds provided that
the densities of ingoing and outgoing streams incident
on the inner horizon are uniform, independent of latitude;
that is, the accretion flow is ``monopole.''
The sub-dominant angular part of the solution
requires a special non-radial pattern of angular motion of streams
incident on the inner horizon.
The prescribed angular pattern cannot be achieved
if the collisionless streams fall freely from outside the horizon,
so the streams must be considered as delivered ad hoc
to just above the inner horizon.
\end{abstract}


\date{\today}

\maketitle

\section{Introduction}

``No satisfactory interior solutions are known''
\cite[\S20.5]{Stephani:2003}
for rotating black holes.
The purpose of this paper is to present
nonlinear, dynamical solutions for the interior structure
of a rotating black hole
in the special case where it slowly accretes a collisionless fluid
in a ``conformally separable'' fashion.
A companion paper \cite{Hamilton:2010c},
hereafter Paper~3, extends the solutions to charged rotating black holes.
A concise derivation of the principal results
of the present paper are given in
in Paper~1 \cite{Hamilton:2010a}.
A Mathematica notebook containing many details of the calculations
is at \cite{Hamilton:notebook}.


Unlike the Schwarzschild
\cite{Schwarzschild:1916a,Schwarzschild:translation}
geometry,
the Kerr
\cite{Kerr:1963}
geometry contains an inner horizon as well as an outer horizon.
In the classic analytically extended Kerr solution,
the inner horizon is a gateway to
delightful but unrealistic pathologies,
including wormholes, white holes,
timelike singularities,
and closed timelike curves
\cite{Carter:1968a}.

Sadly, the Kerr geometry fails at the inner horizon, because
it is subject to the mass inflation instability
discovered by Poisson \& Israel (1990)
\cite{Poisson:1990eh}.
The inflationary instability is the nonlinear realization
of the infinite blueshift at the inner horizon
first pointed out by Penrose (1968)
\cite{Penrose:1968}.
Most studies of the inflationary instability
have focussed on spherical charged black holes
(see \cite{Hamilton:2008zz} for a review).
The first investigation of inflation inside rotating black holes was
that of Barrab\`es, Israel \& Poisson
\cite{Barrabes:1990},
who showed that when two light sheets pass through each other,
a mass parameter defined by the product of the expansions
along the two light bundles inflates.
Ori
\cite{Ori:1992,Ori:2001pc}
and
Brady and collaborators
\cite{Brady:1995un,Brady:1998ht}
(see also references in \cite{Hamilton:2009hu})
considered inflation driven by a Price tail of ingoing and outgoing
gravitational waves in the late time collapse of a rotating black hole.
\cite{Chan:1994rs}
found an exact mass inflation solution for a rotating black hole
in 1+2 dimensions.

Mass inflation is a generic classical mechanism
that requires one essential ingredient to precipitate it:
a source of ingoing (positive energy) and outgoing (negative energy)
streams near the inner horizon
that can stream relativistically through each other
\cite{Hamilton:2008zz}
(in this context,
positive and negative energy refer to
minus the sign of the covariant $t$-component $p_t$
of the momentum of a particle in a tetrad frame
aligned with the principal frame,
not to the conserved energy
associated with translation invariance of the coordinate time $t$).
Even the tiniest sources of ingoing and outgoing streams suffice
to trigger inflation.
As shown by
\cite{Hamilton:2004aw,Hamilton:2008zz},
in spherical charged black holes,
the smaller the streams,
the more rapidly inflation exponentiates.
This is a fierce and uncommon kind of instability,
where the smaller the trigger,
the more violent the reaction.
The sensitivity of inflation to the smallest influence
suggests that it would be difficult to avoid in a real black hole.

Inflation is not particular about the origin of
the ingoing and outgoing streams needed to trigger it: anything will do.
Most of the literature
has considered the situation where inflation is ignited by a Price
\cite{Price:1972,Dafermos:2003yw}
tail of radiation generated during the initial collapse of the black hole.
In real astronomical black holes, however,
ongoing accretion of baryons and dark matter from outside
probably soon overwhelms the initial Price tails.
To illustrate how easy it is for accretion to generate
both ingoing and outgoing streams at the inner horizon,
consider
a massive neutral dark matter particle that is slowly moving at infinity
(energy per unit mass $E = 1$)
in the equatorial plane of a Kerr black hole
of mass $\Mbh$ and angular momentum parameter $a$.
The dark matter particle
can free-fall into the black hole if its specific angular momentum $\Lz$
lies in the interval
(the following comes from solutions of the Hamilton-Jacobi equation
derived in \S\ref{HJseparation})
\begin{equation}
\label{Lrange}
  - 2 \Mbh \left( 1 + \sqrt{1 + a / \Mbh} \right)
  <
  \Lz
  <
  2 \Mbh \left( 1 + \sqrt{1 - a / \Mbh} \right)
  .
\end{equation}
The dark matter particle will become
ingoing or outgoing at the inner horizon depending on whether
its specific angular momentum $\Lz$
is lesser or greater than $L_0$ given by
\begin{equation}
  L_0
  =
  {2 a \over 1 + \sqrt{1 - a^2 / \Mbh^2}}
  \ ,
\end{equation}
which lies within the range~(\ref{Lrange})
(hitting the upper limit of (\ref{Lrange})
in the case of an extremal black hole, $a = \Mbh$).
In fact distant, slowly moving dark matter particles
can become either ingoing or outgoing
at the inner horizon provided that the inclination
of their orbit to the equatorial plane
is less than a value that varies from
$90^\circ$ (i.e.\ all inclinations are allowed) at $a = 0$,
to
$62^\circ$
at $a = 0.96 \Mbh$,
to
$\sin^{-1}\!\sqrt{1/3} \approx 35^\circ$ at the extremal limit $a = \Mbh$.
Thus cold dark matter particles falling from afar provide a natural continuing source
of both ingoing and outgoing collisionless matter near the inner horizon.

In black holes that continue to accrete,
the generic outcome following inflation in spherical charged black holes
is collapse to a spacelike singularity
\cite{Burko:2002qr,Burko:2002fv,Hamilton:2008zz}.

Even if there were no accretion from outside,
quantum mechanical pair creation would provide a source
of ingoing and outgoing radiation near the inner horizon.
Before the mechanism of mass inflation was discovered,
\cite{Novikov:1980}
showed that electromagnetic pair creation
in a spherical charged black hole would destroy the inner horizon.
They suggested that pair creation,
probably by electromagnetic rather than gravitational processes
in view of the much greater strength of electromagnetism,
would also destroy the inner horizon of a rotating black hole.
\cite{Frolov:2006is}
considered a 2-dimensional dilaton model of gravity
simple enough to allow a fully self-consistent
treatment of pair creation and its nonlinear back reaction on the
spacetime of a charged black hole.
They found that pair creation led to mass inflation near the inner horizon,
followed by collapse to a spacelike singularity.

%

The inflationary instability means that
the analytic extension of the Kerr geometry from the inner horizon inward
never occurs in real black holes.
Papers such as
\cite{Santacruz:2010ji}
that focus on the ring singularity of the Kerr geometry,
while of interest in exploring how theories beyond general
relativity might remove singularities,
do not apply to real rotating black holes.

Why does inflation take place at the inner but not outer horizon?
The essential ingredient of inflation is the simultaneous
presence of both ingoing and outgoing streams.
Ingoing and outgoing streams tend to move towards the inner horizon,
amplifying their counter-streaming.
By contrast,
outgoing streams tend to move away from the outer horizon,
de-amplifying any counter-streaming.
Classically,
all matter at the outer horizon is ingoing:
no outgoing matter can pass through the outer horizon.
Outgoing modes will however be excited quantum mechanically
near the outer horizon, leading to Hawking radiation.
There have been various speculations that
quantum effects
could cause
a quantum phase transition at the outer horizon
\cite{Chapline:2000en,Mazur:2004fk,Davidson:2010xe},
or prevent the outer horizon from forming in the first place
\cite{Barcelo:2007yk}.
The present paper assumes that the outer horizon
is not subject to a quantum instability.

To arrive at a solution, this paper builds on physical insight
gained from inflation in charged spherical black holes
\cite{Hamilton:2008zz}.
Two fundamental features of inflation point the way forward.
Firstly,
inflation is ignited by hyper-relativistic counter-streaming
between ingoing and outgoing beams just above the inner horizon.
As shown in \S\ref{focus},
regardless of their source or of their initial orbital parameters,
near the inner horizon
the counter-streaming beams become highly focussed
along the ingoing and outgoing principal null directions.
This has the consequence that the energy-momentum
tensor of the beams takes a predictable form,
\S\ref{collisionless}.
The Kerr
geometry is separable
\cite{Carter:1968c},
and the alignment of the inflationary energy-momentum
along the principal directions
suggests that the geometry could continue to be
separable during inflation.

The second key fact is that
inflation in spherical charged black holes
produces a geometry that looks like a step-function,
being close to the electrovac (Reissner-Nordstr\"om) solution
above the inner horizon,
then exponentiating to super-Planckian curvature and density over a tiny scale
of length and time.
This suggests generalizing the usual separability condition~(\ref{rhosep})
to a more general condition~(\ref{rhosepes})
that departs, at least initially,
by an amount that is tiny, but with finite derivatives.
Unless one were specifically looking for step-like solutions,
one would not think to consider such a generalization.

The strategy of this paper is to seek the simplest
inflationary solution for a rotating black hole.
I assume that the black hole is accreting at a tiny
(infinitesimal) rate,
so that the Kerr solution applies accurately down to just above the inner horizon.
This assumption is a good approximation for an astronomical black hole
during most of its lifetime,
since the accretion timescale is typically far longer than the
characteristic light crossing time of a black hole.
The approximation of small accretion rate breaks down
during the initial collapse of the black hole from a stellar core,
or during rare instances of high accretion,
such as a black hole merger.

For simplicity, I assume further that the
geometry is axisymmetric,
and possesses conformal time-translation symmetry
(self-similarity),
even though the accretion flow in real black holes is unlikely
to be axisymmetric or self-similar.
One might imagine that
the assumption of conformal time-translation invariance would,
in the limit of infinitesimal accretion rate,
be equivalent to the assumption of stationarity,
but this is false.
As explained in \S4.4 of
\cite{Hamilton:2008zz},
and elaborated further in
\S\ref{conformal}
of the present paper,
the stationary approximation is equivalent to the
assumption of symmetrically equal ingoing and outgoing streams at the inner horizon,
whereas in reality the initial conditions of the accretion flow
will generically lead to unequal streams near the inner horizon.
The stationary approximation was first introduced to inflation by
\cite{Burko:1997xa,Burko:1998az,Burko:1998jz},
who called it the homogeneous approximation
because the time direction $t$ is spacelike inside the horizon.
The present paper follow the convention of \cite[p.~203]{Carroll:2004}
in referring to time-translation symmetry as stationarity,
even when the time direction is spacelike rather than timelike.

I call the combination of conformal time-translation invariance
coupled with the limit of small accretion rates
``conformal stationarity,''
to distinguish it from strict stationarity.

Motivated by the argument of the paragraph above beginning
``To arrive at'',
I pursue the hypothesis that the spacetime is conformally separable.
By conformal separability is meant the conditions~(\ref{fnxy})
on the parameters of the line-element~(\ref{lineelement})
and electromagnetic potential~(\ref{Apot})
that emerge from requiring that the equations of motion of
massless particles be Hamilton-Jacobi separable,
\S\ref{HJseparation}.
Whereas strict stationarity requires that the conformal factor
take the separable form~(\ref{rhosep}),
conformal separability does not.
Conformal separability implies the existence of a conformal Killing tensor,
\S\ref{Killing}.

Finally,
I assume that the streams that ignite and then drive inflation
are freely-falling and collisionless.
The present paper restricts to uncharged streams,
while a companion paper \cite{Hamilton:2010c}, Paper~3,
generalizes to charged streams.
The assumption of collisionless flow
is likely to break down when centre-of-mass energies
between ingoing and outgoing particles exceed the Planck energy,
but for simplicity I neglect any collisional processes.


\section{Line element}

Choose coordinates $x^\mu \equiv \{ x , t , y , \phi \}$
in which $t$ is the conformal time with respect to which the spacetime
is conformally time-translation symmetric
(see \S\ref{conformal}),
$\phi$
is the azimuthal angle with respect to which the spacetime
is axisymmetric,
and $x$ and $y$ are radial and angular coordinates.
Appendix~\ref{reduction}
shows that under the conditions
of conformal stationarity, axisymmetry, and conformal separability
assumed in this paper,
the line-element may be taken to be
\begin{equation}
\label{lineelement}
  \dd s^2
  =
  \rho^2
  \left[
  {\dd x^2 \over \Deltax}
  -
  {\Deltax \over \sigma^4}
  \left( \dd t - \omegay \, \dd \phi \right)^2
  +
  {\dd y^2 \over \Deltay}
  +
  {\Deltay \over \sigma^4}
  \left( \dd \phi - \omegax \, \dd t \right)^2
  \right]
  \ ,
\end{equation}
where
\begin{equation}
  \sigma
  \equiv
  \sqrt{ 1 - \omegax \, \omegay }
  \ .
\end{equation}
The line-element is essentially the
Hamilton-Jacobi and Schr\"odinger separable line-element
given by
\cite[eq.~(1)]{Carter:1968c},
except that the conformal factor $\rho$ is left arbitrary,
consistent with the weaker assumptions of conformal stationarity
and conformal separability made here,
as opposed to the stronger assumptions of strict stationarity and separability
made by \cite{Carter:1968c}.

Thanks to the invariant character of the coordinates $t$ and $\phi$,
the metric coefficients
$g_{tt}$, $g_{t\phi}$, and $g_{\phi\phi}$
all have a gauge-invariant significance.
The determinant of the $2 \times 2$ submatrix
of $t$--$\phi$ coefficents
defines the radial and angular horizon functions
$\Deltax$ and $\Deltay$:
\begin{equation}
  g_{tt} g_{\phi\phi} - g_{t\phi}^2
  =
  -
  {\rho^4 \over \sigma^4}
  \Deltax \Deltay
  \ .
\end{equation}
Horizons occur when one or other of the horizon functions
$\Deltax$ and $\Deltay$
vanish.
In physically relevant cases,
horizons occur in the radial direction,
where $\Deltax$ vanishes.
The focus of this paper is inflation,
which occurs in a region just above the inner horizon,
where the radial horizon function $\Deltax$
is negative and tending to zero.
In this region the radial coordinate $x$ is timelike,
while the time coordinate $t$ is spacelike.
It is natural to choose the sign of the timelike radial coordinate $x$
so that it increases inward,
the direction of advancing proper time.

Through the identity
\begin{equation}
  \dd s^2
  =
  g_{\mu\nu}
  \, \dd x^\mu \dd x^\nu
  =
  \eta_{mn}
  e^m{}_\mu \, e^n{}_\nu
  \, \dd x^\mu \dd x^\nu
  \ ,
\end{equation}
the line-element~(\ref{lineelement}) encodes not only a metric
$g_{\mu\nu}$,
but a complete inverse vierbein $e^m{}_\mu$,
corresponding vierbein $e_m{}^\mu$,
and orthonormal tetrad
$\{ \bgamma_x , \bgamma_t , \bgamma_y , \bgamma_\phi \}$,
satisfying
$\bgamma_m \cdot \bgamma_n \equiv \eta_{mn}$
with
$\eta_{mn}$ the Minkowski metric.
The tetrad defined by the line-element~(\ref{lineelement})
aligns with the principal null tetrad,
as will become evident from the fact that the Weyl tensor
is diagonal in the tetrad frame
(it has only spin-$0$ components,
equation~(\ref{CWeylinf})).
It is convenient to choose the time axis of the tetrad
to lie in the radial $x$-direction,
since that direction is timelike near the inner horizon.
Explicitly, the
inverse vierbein
$e^m{}_\mu$
is
\begin{equation}
\label{kninversevierbein}
  e^m{}_\mu
  =
  \rho
  \left(
  \begin{array}{cccc}
  \displaystyle
  {1 \over \sqrt{- \Deltax}} &
  0 & 0 & 0\\[2ex]
  0 &
  \displaystyle
  {\sqrt{- \Deltax} \over \sigma^2} &
  0 &
  \displaystyle
  - {\omegay \sqrt{- \Deltax} \over \sigma^2}
  \\[2ex]
  0 & 0 &
  \displaystyle
  {1 \over \sqrt{\Deltay}} &
  0 \\[2ex]
  0 &
  \displaystyle
  - {\omegax \sqrt{\Deltay} \over \sigma^2} &
  0 &
  \displaystyle
  {\sqrt{\Deltay} \over \sigma^2}
  \end{array}
  \right)
  \ ,
\end{equation}
while the corresponding vierbein
$e_m{}^\mu$
is
\begin{equation}
\label{doranvierbein}
  e_m{}^\mu
  =
  {1 \over \rho}
  \left(
  \begin{array}{cccc}
  \displaystyle
  \sqrt{- \Deltax} &
  0 & 0 & 0 \\[2ex]
  0 &
  \displaystyle
  {1 \over \sqrt{- \Deltax}} & 0 &
  \displaystyle
  {\omegax \over \sqrt{- \Deltax}} \\[2ex]
  0 & 0 & \sqrt{\Deltay} & 0 \\[1ex]
  0 &
  \displaystyle
  {\omegay \over \sqrt{\Deltay}} &
  0 &
  \displaystyle
  {1 \over \sqrt{\Deltay}}
  \end{array}
  \right)
  \ .
\end{equation}

The convention in this paper
and its companions
is that dummy tetrad-frame indices are latin,
while dummy coordinate-frame indices are greek.
The tetrad-frame directed derivative is denoted $\partial_m$,
not to be confused with the coordinate-frame partial derivative
$\partial / \partial x^\mu$.
Tetrad-frame and coordinate-frame derivatives are related by
\begin{equation}
  \partial_m
  \equiv
  e_m{}^\mu {\partial \over \partial x^\mu}
  \ .
\end{equation}
Whereas coordinate-frame derivatives
$\partial / \partial x^\mu$
commute,
tetrad-frame derivatives
$\partial_m$
do not.

\section{Conformal stationarity}
\label{conformal}

A fundamental simplifying assumption made in
this paper is that the spacetime is conformally stationary,
by which is meant that the spacetime is conformally time-translation invariant
(that is, self-similar),
and that the accretion rate is asymptotically tiny.
Conformal stationarity generalizes the
stationary (or homogeneous) approximation of
\cite{Burko:1997xa,Burko:1998az,Burko:1998jz},
which required equal ingoing and outgoing streams at the inner horizon,
to the realistic case of unequal ingoing and outgoing streams.

One might imagine that
in the limit of asymptotically small accretion rate,
the spacetime would automatically become stationary,
$\partial / \partial t \equiv 0$,
but that is false.
A characteristic of inflation is that the smaller the
accretion rate, the more rapidly inflation exponentiates
\cite{Hamilton:2008zz}.
Even in the limit of infinitesimal accretion rate,
the counter-streaming energy
of ingoing and outgoing streams grows exponentially huge
during inflation.
The stationary assumption sets to zero some quantities that,
although initially infinitesimal, nevertheless grow huge.
As discussed
in \S4.4 of \cite{Hamilton:2008zz}
and \S\ref{radialbcs} of the present paper,
the stationary assumption is equivalent
to assuming equal ingoing and outgoing streams near the inner horizon.
In reality, the relative fluxes of streams near the inner horizon
depend on boundary conditions
that generically do not lead to equal streams.

In place of stationarity,
a consistent approach is to impose conformal time-translation invariance,
also known as self-similarity.
Conformal time-translation symmetry allows
the conformal factor $\rho$ in the line-element~(\ref{lineelement})
to include a time-dependent factor,
\begin{equation}
\label{rhot}
  \rho
  =
  \ee^{\vel t}
  \hat\rho ( x , y )
  \ ,
\end{equation}
where the dimensionless factor $\hat\rho ( x , y )$
is a function only of radius $x$ and angle $y$
(but not $\phi$, given axisymmetry).
Equation~(\ref{rhot}) says that
if the conformal time $t$ increases by an interval $\Delta t$,
then the spacetime expands by a factor $\ee^{\vel \Delta t}$.
The expansion is conformal,
meaning that the spacetime keeps the same shape as it expands.
The coefficient $\vel$ can be thought of as a dimensionless measure
of the rate at which the black hole is expanding.
The conformal time $t$ is dimensionless,
and there is a gauge freedom in the choice of its scaling.
A natural gauge choice is to match the change
$\dd t$
in the conformal time at some fixed conformal position
well outside the outer horizon
to the change
$\dd t_{\rm KN}$
in Kerr-Newman time measured in the natural units
$c = G = \Mbh = 1$ of the Kerr-Newman black hole:
\begin{equation}
  \dd t
  =
  {\dd t_{\rm KN} \over \Mbh}
  \ .
\end{equation}
With that gauge choice,
the accretion rate $\vel$ is just equal
to the dimensionless rate $\Mbhdot$
at which the mass $\Mbh$ of the black hole increases,
as measured by a distant observer:
\begin{equation}
  \vel
  =
  {\partial \ln \rho \over \partial t}
  =
  {\partial \ln \Mbh \over \partial t}
  =
  {\partial \Mbh \over \partial t_{\rm KN}}
  =
  \Mbhdot
  \ .
\end{equation}
The accretion rate $\vel$ is constant,
so the mass of the black hole increases linearly with time
as measured by a distant observer,
$\Mbh = \vel t_{\rm KN}$.

The limit of small accretion rate is attained when
\begin{equation}
\label{velzero}
  \vel
  \rightarrow
  0
  \ .
\end{equation}
It might seem that the small accretion rate limit~(\ref{velzero})
would be equivalent to stationarity, but that is false.
Whereas stationarity sets the accretion rate $\vel$ to zero at the outset,
conformal stationarity takes the limit~(\ref{velzero})
{\em after\/} completing all requisite calculations, not before.

In self-similar spacetimes,
all quantites are proportional to some power of the time-dependent conformal
factor $\ee^{\vel t}$,
and that power can be determined by dimensional analysis.
The metric coefficients $g_{\mu\nu}$,
inverse vierbein
$e^m{}_\mu$,
vierbein
$e_m{}^\mu$,
tetrad-frame connections
$\Gamma_{klm}$,
tetrad-frame
Riemann tensor
$R_{klmn}$,
tetrad-frame electromagnetic potential
$A_m$,
tetrad-frame electromagnetic field
$F_{mn}$,
and
tetrad-frame electromagnetic current
$j_m$,
have respective dimensions
\begin{equation}
\label{dims}
  g_{\mu\nu}
  \propto
  \rho^2
  \ , \quad
  e^m{}_\mu
  \propto \rho
  \ , \quad
  e_m{}^\mu
  \propto \rho^{-1}
  \ , \quad
  \Gamma_{klm}
  \propto
  \rho^{-1}
  \ , \quad
  R_{klmn}
  \propto
  \rho^{-2}
  \ , \quad
  A_m
  \propto
  \rho^0
  \ , \quad
  F_{mn}
  \propto
  \rho^{-1}
  \ , \quad
  j_m
  \propto
  \rho^{-2}
  \ .
\end{equation}
A quantity that is scale-free,
such as the electromagnetic potential $A_m$,
is said to be dimensionless.

The form of the time-dependent factor $\ee^{\vel t}$
in the conformal factor~(\ref{rhot})
can be regarded as following from the fact that
$\partial_m \rho$
must be dimensionless,
which in turn requires that
$\partial \ln \rho / \partial t$
must be dimensionless,
hence independent of $t$.

\section{Conformal separability}

Conformal separability is the proposition that the
Hamilton-Jacobi equations of motion are separable for massless particles,
but not necessarily for massive particles.
Operationally,
the proposition requires that the left hand side,
but not the right hand side,
of the Hamilton-Jacobi equation~(\ref{HamiltonJacobisep}) be separable.
A good part of 
\S\ref{HJseparation} and \S\ref{strictseparability} below
overlaps ground that is familiar since the work of
\cite{Carter:1968c}.
These subsections are nevertheless needed
to establish notation,
to highlight where conformal stationarity and conformal separability
differ from full stationarity and separability,
and to provide the basis for the discussion in \S\ref{focus}
and for the derivation of the energy-momentum tensor
of collisionless streams in \S\ref{collisionless}.
Subsection~\ref{conformalseparability}
covers new ground,
showing that in the inflationary regime of interest here,
even though the spacetime is only conformally separable,
the equations of motion of massive particles are nevertheless
Hamilton-Jacobi separable to an excellent approximation.
Physically,
massive particles are hyper-relativistic during inflation and collapse,
and their trajectories are approximated accurately by those
of massless particles.

\subsection{Hamilton-Jacobi separation}
\label{HJseparation}


The equation of motion of a particle
may be derived from Hamilton's equations, which express
the derivative
with respect to affine parameter $\lambda$
along the path of the particle
of its coordinates $x^\mu$
and associated generalized conjugate momenta $\pi_\mu$
in terms of its Hamiltonian
$H ( x^\mu , \pi_\mu )$:
\begin{equation}
  {\dd \pi_\mu \over \dd \lambda}
  =
  -
  {\partial H \over \partial x^\mu}
  \ , \quad
  {\dd x^\mu \over \dd \lambda}
  =
  {\partial H \over \partial \pi_\mu}
  \ .
\end{equation}
The Hamiltonian
of a test particle of rest mass $m$ and charge $q$
moving in a prescribed background
with metric $g_{\mu\nu}$
and electromagnetic potential $A_{\mu}$ is
\begin{equation}
\label{H}
  H
  =
  \frac{1}{2}
  g^{\mu\nu}
  \left(
  \pi_{\mu}
  - q A_{\mu}
  \right)
  \left(
  \pi_{\nu}
  - q A_{\nu}
  \right)
  \ .
\end{equation}
The Hamilton-Jacobi method
equates the Hamiltonian to
minus the partial derivative
of the action $S$ with respect to affine parameter,
$H = - \partial S / \partial \lambda$,
and replaces the generalized momenta with
the partial derivatives
of the action with respect to coordinates,
$\pi_\mu = \partial S / \partial x^\mu$:
\begin{equation}
\label{HamiltonJacobi}
  \frac{1}{2}
  g^{\mu\nu}
  \left(
  {\partial S \over \partial x^{\mu}}
  - q A_{\mu}
  \right)
  \left(
  {\partial S \over \partial x^{\nu}}
  - q A_{\nu}
  \right)
  =
  -
  {\partial S \over \partial \lambda}
  \ .
\end{equation}

One integral of motion,
associated with conservation of the rest mass $m$
of the particle,
follows from the fact that
the Hamiltonian does not depend explicitly on the affine parameter.
This implies that the Hamiltonian is itself a constant of motion:
\begin{equation}
\label{Hconstant}
  H
  =
  -
  {\partial S \over \partial \lambda}
  =
  -
  \frac{1}{2}
  m^2
  \ .
\end{equation}
The normalization~(\ref{Hconstant})
of the Hamiltonian in terms of the rest mass $m$
is equivalent to choosing
the affine parameter $\lambda$
to be related to the proper time $\tau$
along the path of the particle by
\begin{equation}
\label{affineparameter}
  \dd \lambda
  =
  {\dd \tau \over m}
  \ .
\end{equation}

Two more integrals of motion follow from
conformal time-translation symmetry,
and
axisymmetry.
Conformal time-translation symmetry
implies that the generalized momentum $\pi_t$
conjugate to conformal time $t$ satisfies
the equation of motion
\begin{equation}
\label{dpit}
  {\dd \pi_t \over \dd \lambda}
  =
  -
  {\partial H \over \partial t}
  =
  2 \vel H
  =
  -
  m^2 \vel
  \ ,
\end{equation}
where the factor of $- 2 \vel$
comes from the time-dependent conformal factor
$\ee^{- 2 \vel t}$ in the inverse metric $g^{\mu\nu}$
in the Hamiltonian~(\ref{H}).
Equation~(\ref{dpit}) integrates to
\begin{equation}
\label{pit}
  \pi_t
  =
  - \,
  E
  -
  m \vel \tau
  \ .
\end{equation}
In the small accretion rate limit $\vel \rightarrow 0$,
this reduces to the usual conservation of energy,
$\pi_t = - E$.
The concern expressed in \S\ref{conformal}
that some small quantities grow large during inflation
does not apply here,
because, as found in \S\ref{trajectoryparticle},
equation~(\ref{dtaudx}),
the proper time $\tau$
experienced by a particle during inflation and collapse is always tiny,
in the conformally stationary limit
(this is checked explicitly at the end of \S\ref{conformalseparability}).

The two integrals of motion
associated with conformal stationarity and axisymmetry
correspond to
conservation of
energy $E$
and azimuthal angular momentum $\Lz$,
\begin{equation}
\label{knpitphi}
  \pi_{t}
  =
  {\partial S \over \partial t}
  = - E
  \ , \quad
  \pi_{\phi}
  =
  {\partial S \over \partial \phi}
  = \Lz
  \ .
\end{equation}

Write the covariant tetrad-frame momentum $p_k$ of a particle
in terms of a set of Hamilton-Jacobi parameters $P_k$,
\begin{equation}
\label{pktetrad}
  p_k
  \equiv
  {1 \over \rho}
  \left\{
  {\Px
  \over \sqrt{- \Deltax}}
  \ ,
  {P_t
  \over \sqrt{- \Deltax}}
  \ ,
  {\Py
  \over \sqrt{\Deltay}}
  \ ,
  {P_\phi
  \over \sqrt{\Deltay}}
  \right\}
  \ ,
\end{equation}
and the covariant tetrad-frame electromagnetic potential
$A_k$
in terms of a set of Hamilton-Jacobi potentials
$\Apot_k$,
\begin{equation}
\label{Apot}
  A_k
  \equiv
  {1 \over \rho}
  \left\{
  {\Apot_x \over \sqrt{- \Deltax}}
  ,
  {\Apot_t \over \sqrt{- \Deltax}}
  ,
  {\Apot_y \over \sqrt{\Deltay}}
  ,
  {\Apot_\phi \over \sqrt{\Deltay}}
  \right\}
  \ .
\end{equation}
In Paper~3 \cite{Hamilton:2010c}
it will be found that
$\Apot_t$
is related to the enclosed electric charge within radius $x$,
while
$\Apot_\phi$
is related to the enclosed magnetic charge above latitude $y$.
If magnetic charge does not exist,
then
$\Apot_\phi$
should vanish,
but
$\Apot_\phi$
is retained here
to bring out the symmetry.
The contravariant coordinate momenta
$\dd x^\kappa / \dd \lambda = e_k{}^\kappa p^k$
are related to the Hamilton-Jacobi parameters $P_k$ by
\begin{equation}
\label{pkappa}
  {\dd x^\kappa \over \dd \lambda}
  =
  {1 \over \rho^2}
  \left\{
  -
  \Px
  \, , \ 
  {P_t \over - \Deltax}
  +
  {\omegay P_\phi \over \Deltay}
  \, , \ 
  \Py
  \, , \ 
  {\omegax P_t \over - \Deltax}
  +
  {P_\phi \over \Deltay}
  \right\}
  \ .
\end{equation}
The tetrad-frame momenta $p_k$ are related to the generalized momenta $\pi_\kappa$ by
$p_k = e_k{}^\kappa \pi_\kappa - q \Apot_k$,
which implies that the Hamilton-Jacobi parameters $P_k$ are related to the
canonical momenta $\pi_\kappa$ by
\begin{subequations}
\label{Pk}
\begin{align}
\label{Px}
  \Px
  &\equiv
  - \Deltax \pi_x - q \Apot_x
  \ ,
\\
\label{Pt}
  P_t
  &\equiv
  \pi_t + \pi_\phi \omegax - q \Apot_t
  \ ,
\\
\label{Py}
  \Py
  &\equiv
  \Delta_y \pi_y - q \Apot_y
  \ ,
\\
\label{Pphi}
  P_\phi
  &\equiv
  \pi_\phi + \pi_t \omegay - q \Apot_\phi
  \ .
\end{align}
\end{subequations}
In terms of the parameters $P_k$,
the Hamilton-Jacobi equation~(\ref{HamiltonJacobi}) is
\begin{equation}
\label{HamiltonJacobisep}
  {\Px^2 - P_t^2 
  \over \Deltax}
  +
  {\Py^2 + P_\phi^2
  \over \Deltay}
  =
  -
  m^2 \rho^2
  \ .
\end{equation}

Separation of variables of the Hamilton-Jacobi equation
proceeds by postulating that the action $S$ separates as
(equations~(\ref{Ssep}), (\ref{fnxy}), and (\ref{rhosep}) below
together constitute assumption III of \cite{Carter:1968c},
that the Hamilton-Jacobi equation
separates ``in the simplest possible way'')
\begin{equation}
\label{Ssep}
  S
  =
  \frac{1}{2} m^2 \lambda
  -
  E t
  +
  L \phi
  +
  S_x(x)
  +
  S_y(y)
  \ ,
\end{equation}
where $S_x(x)$ and $S_y(y)$
are respectively functions only of $x$ and $y$.
The left hand side of the Hamilton-Jacobi equation~(\ref{HamiltonJacobisep})
is separable
for arbitrary values of the constants $E$, $L$, and $q$ provided that
(cf.\ Appendix~\ref{reduction})
\begin{equation}
\label{fnxy}
  \begin{array}{ccccl}
  \omegax
  \ ,
  &
  \Deltax
  \ ,
  &
  \Apot_x
  \ ,
  &
  \Apot_t
  &
  \mbox{~are functions of $x$ only}
  \ ,
  \\
  \omegay
  \ ,
  &
  \Deltay
  \ ,
  &
  \Apot_y
  \ ,
  &
  \Apot_\phi
  &
  \mbox{~are functions of $y$ only}
  \ .
  \end{array}
\end{equation}
These are the conditions of conformal separability adopted in this paper.
In the remainder of this paper,
the black hole will be taken to be neutral,
so that the electromagnetic potential
$A_m$
is identically zero.
The case of a charged black hole is addressed in Paper~3.

\subsection{Strict separability}
\label{strictseparability}

Full, or strict, separability, as opposed to just conformal separability,
would require that not only the left hand side but also the right hand side
of the Hamilton-Jacobi equation~(\ref{HamiltonJacobisep}) separates.
This would require that the conformal factor $\rho$ separates, as
(this is eq.~(43) of \cite{Carter:1968c})
\begin{equation}
\label{rhosep}
  \rho^2
  =
  \rhosep^2
  =
  \rhox^2
  +
  \rhoy^2
  \ ,
\end{equation}
where
$\rhox$ is a function only of the radial coordinate $x$,
and
$\rhoy$ is a function only of the angular coordinate $y$.
Equation~(\ref{rhosep}) holds during the electrovac phase prior to inflation,
and it also holds during early inflation,
when the conformal factor $\rho$ remains at its electrovac value,
but its radial derivatives
$\partial \rho / \partial x$
and
$\ppartial \rho / \partial x^2$
are becoming large.
Equation~(\ref{rhosep}) breaks down
as the conformal factor $\rho$ begins to shrink from its electrovac value,
presaging collapse.

If $\rho$ separates as equation~(\ref{rhosep}),
then the Hamilton-Jacobi equation~(\ref{HamiltonJacobisep})
separates as
\begin{equation}
\label{HamiltonJacobisepK}
  -
  \left(
  {\Px^2 - P_t^2
  \over \Deltax}
  +
  m^2 \rhox^2
  \right)
  =
  {\Py^2 + P_\phi^2
  \over \Deltay}
  +
  m^2
  \rhoy^2
  =
  \KCarter
  \ ,
\end{equation}
with $\KCarter$ a separation constant.
The separated Hamilton-Jacobi equations~(\ref{HamiltonJacobisepK}) imply that
\begin{subequations}
\label{PUxy}
\begin{align}
\label{PUr}
  \Px
  &=
  \pm
  \sqrt{
  P_t^2
  - \left(
  \KCarter
  +
  m^2 \rhox^2
  \right)
  \Deltax
  }
  \ ,
\\
\label{PUtheta}
  \Py
  &=
  \pm
  \sqrt{
  - P_\phi^2
  +
  \left(
  \KCarter
  -
  m^2 \rhoy^2
  \right)
  \Deltay
  }
  \ .
\end{align}
\end{subequations}
The trajectory of a freely-falling particle
follows from integrating
$\dd y / \dd x = - \Py / \Px$,
equivalent to the implicit equation
\begin{equation}
\label{dxdyP}
  -
  {\dd x \over
  \Px}
  =
  {\dd y \over
  \Py}
  \ .
\end{equation}
The time and azimuthal coordinates $t$ and $\phi$
along the trajectory
are then obtained by quadratures:
\begin{equation}
\label{dtdphiP}
  \dd t
  =
  {P_t \, \dd x \over \Px \Deltax}
  +
  {\omegay P_\phi \, \dd y \over \Py \Deltay}
  \ , \quad
  \dd \phi
  =
  {\omegax P_t \, \dd x \over \Px \Deltax}
  +
  {P_\phi \, \dd y \over \Py \Deltay}
  \ .
\end{equation}

\subsection{Conformal separability}
\label{conformalseparability}

A feature of inflation, discussed in the next section, \S\ref{focus},
is that ingoing and outgoing streams of particles
move hyper-relativistically relative to each other
and to the no-going tetrad frame.
The streams remain hyper-relativistic throughout inflation
and subsequent collapse.
One should not be too surprised that the trajectories
of hyper-relativistic massive particles would be hardly distinguishable
from those of massless particles.
This subsection shows that Hamilton-Jacobi separability holds
to an excellent approximation
for massive as well as massless particles.
The arguments are confirmed formally
by showing that the difference~(\ref{dpkdx}) between the tetrad-frame momentum
predicted by the Hamilton-Jacobi solution and the true momentum,
integrated over the path of a particle during inflation and collapse,
is adequately small.

The Hamilton-Jacobi
equation~(\ref{HamiltonJacobisep}) can be separated as
\begin{subequations}
\label{PxyHJ}
\begin{align}
\label{PxHJ}
  \Px^2
  &=
  P_t^2 - \left[ \KCarter + m^2 ( \rho^2 - \rhoy^2 ) \right] \Deltax
  \ ,
\\
\label{PyHJ}
  \Py^2
  &=
  - \,
  P_\phi^2
  +
  ( \KCarter - m^2 \rhoy^2 ) \Deltay
  \ ,
\end{align}
\end{subequations}
where $\KCarter$ is the same separation constant as before.
The condition of separability is that
the right hand side of equation~(\ref{PxHJ}) is a function only of $x$,
while
the right hand side of equation~(\ref{PyHJ}) is a function only of $y$.
The latter follows from
the conformal separability conditions~(\ref{fnxy})
and the condition that $\rhoy$ is a function only of $y$,
provided that $\pi_t$ is treated as a constant.

During the electrovac and early inflationary phases,
$\rho^2$ equals its separable electrovac value $\rhosep^2$,
and equation~(\ref{PxHJ}) simplifies to its electrovac form
\begin{equation}
  \Px^2
  =
  P_t^2 - ( \KCarter + m^2 \rhox^2 ) \Deltax
  \ ,
\end{equation}
whose right hand side is a function only of $x$,
consistent with separability.
During inflation and early collapse,
the radial horizon function is tiny,
$| \Deltax | \ll 1$,
so equation~(\ref{PxHJ}) simplifies to
\begin{equation}
\label{PxHJinf}
  \Px^2
  =
  P_t^2
  \ ,
\end{equation}
whose right hand side is again a function only of $x$,
consistent with separability.
Once $| \Deltax | \gtrsim 1$,
equation~(\ref{PxHJinf}) no longer holds,
but during collapse the conformal factor $\rho$ shrinks to a tiny value,
with the net result that $\rho^2 | \Deltax | \ll 1$,
so that equation~(\ref{PxHJ}) simplifies to
\begin{equation}
\label{PxHJlate}
  \Px^2
  =
  P_t^2 - ( \KCarter - m^2 \rhoy^2 ) \Deltax
  \ .
\end{equation}
The term proportional to $\rhoy^2$ on the right hand side of
equation~(\ref{PxHJlate})
depends on $y$, apparently destroying separability.
However,
a feature of inflation and collapse,
which will be discovered in \S\ref{trajectoryparticle},
is that the coordinates $x$ and $y$ of a freely-falling particle
remain frozen at their inner horizon values
throughout inflation and collapse.
Thus $\rhoy$ remains constant along the trajectory of any particle,
and the right hand side of equation~(\ref{PxHJlate})
can be considered to be a function only of $x$,
again consistent with separability of equation~(\ref{PxHJ}).

Equation~(\ref{PxHJlate})
shows that during inflation and collapse
the trajectory of a particle of rest mass $m$
is accurately approximated by that of a massless particle
with separation constant
$\KCarter_0 = \KCarter - m^2 \rhoy^2$.

That equations~(\ref{PxyHJ})
provide accurate expressions for $\Px$ and $\Py$ for massive particles
can be confirmed by considering the total derivative
$\dd p_k / \dd x$
of the tetrad-frame momentum $p_k$,
equation~(\ref{pktetrad}),
of a neutral particle of rest mass $m$ along its trajectory.
If the parameters $\Px$ and $\Py$ are taken to be given by
equations~(\ref{PxyHJ}),
with the parameters $P_t$ and $P_\phi$ from equations~(\ref{Pk})
and $\pi_t$ constant
(as opposed to from equation~(\ref{pit})),
then the total derivative of the tetrad-frame momentum is
\begin{equation}
\label{dpkdx}
  {\dd p_k \over \dd \lambda}
  =
  {m^2 \over \rho^3}
  \left(
  {\Py \over 2} {\partial ( \rho^2 - \rhoy^2 ) \over \partial y}
  +
  {\rho^2 \vel \omegay P_\phi \over \Deltay}
  \right)
  \left\{
  {\sqrt{- \Deltax} \over \Px}
  ,
  \,
  0
  ,
  \,
  {\sqrt{\Deltay} \over \Py}
  ,
  \,
  0
  \right\}
  +
  {m^2 \vel \over \rho}
  \left\{
  {P_t \over \Px \sqrt{- \Deltax}}
  ,
  \,
  {1 \over \sqrt{- \Deltax}}
  ,
  \,
  -
  {P_\phi \omegay \over \Py \sqrt{\Deltay}}
  ,
  \,
  {\omegay \over \sqrt{\Deltay}}
  \right\}
  \ .
\end{equation}
The right hand side of equation~(\ref{dpkdx}),
which would be zero if the motion were exactly geodesic
(and is in fact zero for massless particles, $m = 0$),
is non-zero because the solution~(\ref{PxyHJ}) for $\Px$ and $\Py$ is not exact,
for massive particles.
The second of the two terms on the right hand side of equation~(\ref{dpkdx})
arises from approximating $\pi_t$ as a constant,
and would disappear if $\pi_t$ were set to the more accurate value~(\ref{pit}),
and $\tau$ were replaced by
its value as a function of respectively $x$ and $y$ in respectively
$P_t$ and $P_\phi$.

The deviation between the momentum $p_k$ predicted by equations~(\ref{PxyHJ})
and the true momentum can be obtained by integrating equation~(\ref{dpkdx})
over the path of a particle during inflation and collapse.
As shown in Appendix~\ref{estimateintegrals},
the ratio $\Delta p_k / p_k$
of the deviation
$\Delta p_k \equiv \int ( \dd p_k / \dd \lambda ) \, \dd \lambda$
to the momentum $p_k$ itself is
of order $\sim \vel^2$, equation~(\ref{Dlnpk}),
which may be considered adequately small.
This confirms the accuracy of the Hamilton-Jacobi approximation~(\ref{PxyHJ}).

\section{Focussing along the principal directions}
\label{focus}

A central feature of inflation,
demonstrated in this section,
is that as ingoing and outgoing streams
approach the inner horizon,
they see the opposite stream narrow into an increasingly
intense, blueshifted beam
focussed along the opposite principal null direction.
The fact that
near the inner horizon
the energy-momenta of the streams
becomes highly focussed along the principal null directions
regardless of the initial conditions of the streams
is what motivates the idea that the spacetime,
which is separable in the Kerr-Newman geometry,
may continue to be separable during inflation.

A particle is said to be ingoing if $P_t$,
equation~(\ref{Pt}),
is negative,
outgoing if $P_t$ is positive.
Outside the outer horizon,
$P_t$ is necessarily negative (ingoing)
while
$\Px$ can be either negative or positive.
At the outer horizon,
$P_t$ and $\Px$ are equal in magnitude,
and continuous across the horizon.
Inside the outer horizon,
$P_t$ and $\Px$ switch roles:
$\Px$ is necessarily negative
(given that the sign of the timelike radial coordinate $x$
is being chosen so that it
increases as proper time advances, equation~(\ref{pkappa})),
while
$P_t$ can be either negative (ingoing) or positive (outgoing).
A particle falling from outside the horizon
necessarily has negative $P_t$ as long as it is outside the horizon,
but its $P_t$ can change sign inside the horizon,
if its angular momentum and/or charge are sufficiently large
with the same sign as the black hole,
as exampled in the Introduction in the paragraph
containing equation~(\ref{Lrange}).

A particle at rest in the tetrad frame has by definition tetrad-frame momentum
$p^k = m \{ 1 , 0 , 0 , 0 \}$,
hence its Hamilton-Jacobi parameters $P_k$, equation~(\ref{pktetrad}), are
\begin{equation}
\label{Pktetrad}
  P_k = - m \rho \sqrt{- \Deltax} \{ 1 , 0 , 0 , 0 \}
  \ .
\end{equation}
The tetrad rest frame defines a special frame,
the no-going frame,
where $P_t = 0$,
at the boundary between ingoing and outgoing.

The tetrad-frame 4-momentum $p^k$ of a freely falling particle,
as seen in the no-going tetrad rest frame,
is given by equation~(\ref{pktetrad}).
Near the inner horizon, where $\Deltax \rightarrow - 0$,
the no-going observer sees both ingoing and outgoing streams
become hugely blueshifted
and focussed along the ingoing and outgoing principal null directions:
\begin{equation}
\label{hugepknogoing}
  p^k
  \rightarrow
  {- \Px \over \rho \sqrt{- \Deltax}}
  \{ 1 , \pm 1 , 0 , 0 \}
  \ .
\end{equation}

Instead of the no-going observer,
consider an ingoing or outgoing observer, of mass $m^\prime$,
with Hamilton-Jacobi parameters $P_k^\prime$.
Near the inner horizon,
where $\Deltax \rightarrow - 0$,
the ingoing or outgoing observer
see particles in the opposite stream with
hugely blueshifted 4-momentum $p^k$
(the following is equation~(\ref{hugepknogoing})
appropriately Lorentz-boosted in the radial direction),
\begin{equation}
\label{hugepk}
  p^k
  \rightarrow
  {2 \Px^\prime \Px \over m^\prime \rho^2 ( - \Deltax )}
  \{ 1 , \pm 1 , 0 , 0 \}
  \ ,
\end{equation}
in which the $\pm$ sign is $+$ for ingoing observers
and $-$ for outgoing observers.
An ingoing observer see a piercing beam of outgoing particles
coming from the direction towards the black hole,
focussed along the outgoing principal null direction
$\{ 1 , 1 , 0 , 0 \}$.
An outgoing observer see a similarly intense beam of ingoing particles
falling from the direction away from the black hole,
focussed along the ingoing principal null direction
$\{ 1 , - 1 , 0 , 0 \}$.

It is only particles on the opposing stream that appear highly beamed:
particles in an observer's own stream appear normal, not beamed.
The no-going observer is exceptional
in being skewered from both directions,
albeit with the square root of the energy and blueshift
that an ingoing or outgoing observer experiences.

\section{Killing tensor}
\label{Killing}

Separability is associated with the existence of a Killing tensor,
and conformal separability is associated with the existence of a conformal
Killing tensor.

%

\subsection{Electrovac Killing tensor}
\label{electrovacKilling}

The separated Hamilton-Jacobi equation~(\ref{HamiltonJacobisepK})
can be written
\begin{equation}
  \Killing^{mn}
  p_m p_n
  =
  \KCarter
  \ ,
\end{equation}
where
$p_m$ is the covariant tetrad-frame momentum~(\ref{pktetrad}),
and $\Killing^{mn}$ is the tetrad-frame Killing tensor
\begin{equation}
\label{K}
  \Killing^{mn}
  =
  \diag ( \rhoy^2 \, , \  - \rhoy^2 \, , \  \rhox^2 \, , \  \rhox^2 )
  \ .
\end{equation}
The Killing tensor
$\Killing^{mn}$
satisfies Killing's equation
\begin{equation}
\label{DK}
  D_{(k} \Killing_{mn)}
  =
  0
  \ ,
\end{equation}
where $D_k$
denotes covariant differentiation,
and parentheses denote symmetrization over enclosed indices.

\subsection{Early inflationary Killing tensor}
\label{inflationKilling}

The separation of the conformal factor $\rho$ as equation~(\ref{rhosep})
leads to the usual electrovac solutions
\cite{Carter:1968a,Carter:1968c},
but not to inflation.
To admit inflation,
it is necessary to go beyond equation~(\ref{rhosep}).
This section proposes a modification of the conformal factor,
and shows that there is an associated Killing tensor during early inflation.

In spherically symmetric models,
and in the limit of slow accretion,
inflation has a step-function character at the inner horizon.
This suggests generalizing
the separation of the conformal factor $\rho$
by allowing it to depart infinitesimally
from equation~(\ref{rhosep}),
but with finite derivatives in the radial $x$ direction.
Specifically, the modified conformal factor is
\begin{equation}
\label{rhosepes}
  \rho
  =
  \rhosep
  \ee^{\vel t - \expinf}
  \ ,
\end{equation}
where $\rhosep$
is the usual separable conformal factor, equation~(\ref{rhosep}),
and $\expinf$,
a function of $x$ and $y$ (not $t$ or $\phi$),
is negligibly small,
but with finite radial derivatives
satisfying the hierarchy of inequalities
\begin{equation}
\label{econditionsx}
  0
  \approx
  \expinf
  \ll
  {\partial \expinf \over \partial x}
  \ll
  {\ppartial \expinf \over \partial x^2}
  \quad
  ( \mbox{early inflation} )
  \ ,
\end{equation}
and with negligible angular derivatives,
\begin{equation}
\label{econditionsy}
  0
  \approx
  {\partial \expinf \over \partial y}
  \approx
  {\ppartial \expinf \over \partial y^2}
  \ .
\end{equation}
The inequalities~(\ref{econditionsx}) and (\ref{econditionsy})
mean that
$\expinf$
is mainly a function of the radial coordinate $x$,
its derivatives with respect to the angular coordinate $y$
being small.
The kind of function $\rho$ that equation~(\ref{rhosepes})
describes is one that takes a sharp turn,
like a step function, in the radial direction.
In \S\ref{ignition},
it will be found that the initial conditions
in the electrovac phase lead to a function
$\expinf$ that 
satisfies the conditions~(\ref{econditionsx})
and (\ref{econditionsy})
during early inflation.

The separation~(\ref{rhosepes}) of the conformal factor $\rho$
indeed proves to admit a Killing tensor,
satisfying Killing's equation~(\ref{DK}),
subject to the conditions~(\ref{econditionsx}) and (\ref{econditionsy}).
The tetrad-frame Killing tensor $\Killing^{mn}$
associated with the separation~(\ref{rhosepes}) is
\begin{equation}
\label{Ke}
  \Killing^{mn}
  =
  \diag ( \rhoy^2 \, , \  - \rhoy^2 \, , \  \rho^2 - \rhoy^2 \, , \  \rho^2 - \rhoy^2 )
  \ ,
\end{equation}
in which, to linear order in the small parameters
$\vel$ and $\expinf$,
\begin{equation}
\label{Krho}
  \rho^2 - \rhoy^2
  =
  \rhox^2
  +
  2 \rhosep^2 ( \vel t - \expinf )
  \ .
\end{equation}
Since
$\vel$ and
$\expinf$
are negligibly small,
it might appear that the modified Killing tensor
given by equation~(\ref{Ke})
is identical to the original tensor, equation~(\ref{K}).
Indeed,
$\vel$ can be set to zero in equation~(\ref{Krho})
without further delay.
However,
Killing's equation~(\ref{DK})
involves derivatives of the Killing tensor,
which bring in non-negligible derivatives of
$\expinf$.
Thus the modified Killing tensor~(\ref{Ke})
differs non-trivially from the original~(\ref{K}).
The Killing tensor~(\ref{Ke}) satisfies Killing's equation~(\ref{DK})
provided that
not only $\vel$ and $\expinf$ are negligible,
but also $\partial \expinf / \partial y$ is negligible,
but $\partial \expinf / \partial x$ may be large,
consistent with conditions~(\ref{econditionsx}) and (\ref{econditionsy}).

The Killing tensor~(\ref{Ke})
applied to the tetrad-frame momentum $p_k$ given by
equations~(\ref{pktetrad}) and (\ref{PxyHJ}) gives
\begin{equation}
  \Killing^{mn}
  p_m p_n
  =
  \KCarter
  \ ,
\end{equation}
confirming that
equations~(\ref{PxyHJ})
constitute a valid separation of the Hamilton-Jacobi equations
for massive particles during early inflation.

\subsection{Conformal Killing tensor}
\label{conformalKilling}

The previous two subsections,
\S\ref{electrovacKilling} and \S\ref{inflationKilling},
showed that
the spacetime possesses a Killing tensor
in the electrovac and early inflationary stages,
but not later.
However, the spacetime possesses a conformal Killing tensor at all times,
from electrovac through inflation and collapse.
During early inflation,
the inflationary exponent $\expinf$ remains negligibly small 
while its radial derivatives grow large,
conditions~(\ref{econditionsx}),
but during later inflation and collapse
the inflationary exponent $\expinf$ grows huge.

The traceless part $\hat{K}^{mn}$
of the Killing tensor~(\ref{Ke}),
\begin{equation}
\label{tracelessKe}
  \hat{K}^{mn}
  =
  \Killing^{mn}
  -
  {\textstyle \frac{1}{4}}
  \eta^{mn} \Killing^k_k
  =
  {\textstyle \frac{1}{2}}
  \rho^2
  \diag ( 1 , -1 , 1 , 1 )
  \ ,
\end{equation}
is a conformal Killing tensor
\cite[\S{35.3}]{Stephani:2003},
satisfying
\begin{equation}
\label{DtracelessKe}
  D_{(k} \hat{K}_{mn)}
  -
  {\textstyle \frac{1}{3}}
  \eta_{(km} D^l \hat{K}_{n)l}
  =
  0
  \ .
\end{equation}
The tensor
$\hat{K}^{mn}$
satisfies the condition~(\ref{DtracelessKe})
to be a conformal Killing tensor
provided that the conformal separability conditions~(\ref{fnxy}) hold,
but without any restriction on the conformal factor $\rho$,
and in particular without any restriction on the
inflationary exponent $\expinf$.

In \S\ref{nonlinearinflationeqs} it will be found that
during inflation and collapse
the horizon function $\Deltax$ remains a function of radius $x$,
as required by the conformal separability conditions~(\ref{fnxy}),
only provided that the inflationary exponent $\expinf$ is purely radial:
\begin{equation}
\label{econdition}
  \expinf
  ~
  \mbox{is a function of $x$ only}
  \ .
\end{equation}

\section{Collisionless freely-falling streams}
\label{collisionless}

The essential ingredient that triggers mass inflation
is the presence near the inner horizon
of ingoing and outgoing streams
that can stream relativistically through each other.
This paper adopts a general collisionless fluid
as the source of energy-momentum that ignites and then drives inflation.
In astronomical black holes,
streams near the inner horizon will typically originate
from accretion of baryons and cold dark matter.
A combination of collisions and magnetohydrodynamic processes
\cite{Balbus:1998,Balbus:2003xh}
are likely to keep baryons, electrons, and photons
tightly coupled above the inner horizon,
forcing them into a common ingoing or outgoing stream
before inflation ignites.
Dark matter, and also gravitational waves,
which should behave like a collisionless fluid of gravitons
in the high-frequency limit,
can occupy the opposing stream,
and stream relativistically through the baryonic stream without collisions,
driving inflation.
In the limit of small accretion rate considered here,
the geometry above the inner horizon is accurately approximated
by the electrovac ($\Lambda$-Kerr-Newman) solution,
so the precise behaviour of the gas there is irrelevant.
The strategy adopted in this paper and its companions
is to seek solutions that hold from just above the inner horizon inward.

The approximation of a collisionless fluid will probably break down
when center-of-mass collision energies between ingoing and outgoing particles
exceed the Planck energy.
Such super-Planckian collisional processes, though probably important,
are neglected in the present paper.

It should be commented that general relativistic numerical treatments
that model the energy-momentum as a single fluid with
wave speed less than the speed of light are not satisfactory
near the inner horizon,
since such a fluid cannot support relativistic counter-streaming,
and therefore artificially suppresses the inflation that would occur
if even the tiniest admixture of an oppositely going fluid were admitted.

\subsection{Occupation number}

The distribution of particles in a collisionless fluid
is described by a scalar occupation number
$f( x^\mu , \bp )$
that specifies
the number $\dd N$ of particles
at position $x^\mu$
with tetrad-frame momentum $p^m \equiv \{ \px , \bp \}$
in a Lorentz-invariant 6-dimensional volume of phase space,
\begin{equation}
\label{dN}
  \dd N
  =
  f ( x^\mu , \bp ) \ 
  {\ddd x \, \ddd p \over (2\pi\hbar)^3}
  \ .
\end{equation}
Here $\ddd x$ denotes the proper tetrad-frame 3-volume element
measured by an observer at rest in the tetrad frame,
not a coordinate-frame 3-volume element.
The collisionless Boltzmann equation is
\begin{equation}
\label{Boltzmannp}
  {\dd
  f ( x^\mu , \bp )
  \over \dd \lambda}
  =
  p^m \partial_m f
  + {\dd p^m \over \dd \lambda} {\partial f \over \partial p_m}
  =
  0
  \ ,
\end{equation}
where the affine parameter is $\dd \lambda \equiv \dd \tau / m$,
with $\tau$ the proper time of
an observer at rest in the tetrad frame.
The Boltzmann equation~(\ref{Boltzmannp})
asserts that the occupation number
$f ( x^\mu , \bp )$
is constant along phase-space trajectories.

The tetrad-frame momentum 3-volume element
$\ddd p$
is related to the scalar 4-volume element
$\dddd p$
by
\begin{equation}
\label{momentumvolumeelement}
  ( 2 \pi \hbar )
  \delta_D ( p^k p_k + m^2 ) \,
  {\dddd p \over (2\pi\hbar)^4}
  =
  {\ddd p \over 2 \px (2\pi\hbar)^3}
  \ ,
\end{equation}
where the Dirac delta-function enforces conservation of rest mass $m$.
The Lorentz-invariant tetrad-frame momentum volume element
from equation~(\ref{momentumvolumeelement})
translates into a 4-volume element of parameters $P_m$
with the Jacobian from
the relation~(\ref{pktetrad})
between $p_m$ and $P_m$:
\begin{equation}
\label{Pvolumeelement}
  {\ddd p \over 2 \px (2\pi\hbar)^3}
  =
  ( 2 \pi \hbar ) \delta_D \!
  \left(
  {\Px^2 - P_t^2
  \over \rho^2 \Deltax}
  +
  {\Py^2 + P_\phi^2
  \over \rho^2 \Deltay}
  +
  m^2
  \right) \,
  {\dd P^4 \over (2\pi\hbar)^4 \rho^4 \Deltax \Deltay}
  \ .
\end{equation}
This in turn translates into an element of orbital constants of motion,
the energy
$E \equiv - \pi_t$,
angular momentum
$L \equiv \pi_\phi$,
and separation constant $\KCarter$,
with the Jacobian calculated from equations~(\ref{Pt}), (\ref{Pphi}),
amd (\ref{PxyHJ}):
\begin{equation}
\label{Evolumeelent}
  {\ddd p \over 2 \px (2\pi\hbar)^3}
  =
  {\sigma^2 \over \rho^2 \Px \Py}
  {\dd E \, \dd L \, \dd \KCarter \over 4 (2\pi\hbar)^3}
  \ .
\end{equation}

\subsection{Number current}

The number density and flux of particles
at any position
form a tetrad-frame 4-vector
$n^k$,
\begin{equation}
\label{nstream}
  n^k
  =
  \int
  p^k \,
  f ( x^\mu , \bp ) \,
  {\ddd p \over 2 \px (2\pi\hbar)^3}
  =
  \int
  p^k \,
  f ( x^\mu , \bp ) \,
  {\sigma^2 \over \rho^2 \Px \Py}
  {\dd E \, \dd L \, \dd \KCarter \over 4 (2\pi\hbar)^3}
  \ .
\end{equation}
Covariant number conservation follows from
the collisionless Boltzmann equation~(\ref{Boltzmannp}),
\begin{equation}
  D_k
  n^k
  =
  \int
  p^k \,
  D_k
  f
  \,
  {\ddd p \over 2 \px (2\pi\hbar)^3}
  =
  \int
  {\dd f \over \dd \lambda}
  \,
  {\ddd p \over 2 \px (2\pi\hbar)^3}
  =
  0
  \ .
\end{equation}

For a single stream with fixed constants of motion
$E$, $L$, and $\KCarter$,
it follows from
equation~(\ref{nstream})
and
the constancy of the occupation number $f$ along phase-space trajectories,
equation~(\ref{Boltzmannp}),
that
the number current $n^k$ along the stream varies as
\begin{equation}
\label{nsingle}
  n^k
  =
  N
  p^k
  \ , \quad
  N
  \propto
  {\sigma^2 \over \rho^2 \Px \Py}
  \ .
\end{equation}
The denominators $\Px$ and $\Py$
would vanish where orbits turned around in radius $x$ or angle $y$,
and there would be cusps in the number current at such points.
However, this never happens in the inflationary zone
of interest in the present paper,
because the radius $x$ is a timelike coordinate,
so $\Px$ never changes sign,
while $\Py$, given to an excellent approximation by equation~(\ref{PyHJ}),
remains essentially constant along any stream throughout inflation and collapse.

For massive particles,
$\Px$ and $\Py$ along a stream
are approximated accurately by equations~(\ref{PxyHJ}).
The accuracy of the approximations~(\ref{PxyHJ})
can be checked by seeing how closely
the covariant divergence $D_k n^k$ that they predict vanishes.
With $\Px$ and $\Py$ given by equations~(\ref{PxyHJ}),
along with $P_t$ and $P_\phi$ from equations~(\ref{Pk})
and $\pi_t$ constant (as opposed satisfying equation~(\ref{pit})),
the covariant divergence of the single-stream number current $n^k$ given by
equation~(\ref{nsingle}) is
\begin{equation}
\label{Dknk}
  D_k n^k
  =
  {m^2 N \over \rho^2 \Px^2}
  \left[
  \Deltax
  \left(
  {\Py \over 2} {\partial ( \rho^2 - \rhoy^2 ) \over \partial y}
  +
  {\rho^2 \vel \omegay P_\phi \over \Deltay}
  \right)
  -
  \rho^2 \vel P_t
  \right]
  \ .
\end{equation}
As expected, the divergence vanishes identically for massless particles,
$m = 0$,
but not for massive particles,
because Hamilton-Jacobi separation is exact for massless particles,
but not quite exact for massive particles.
Since
\begin{equation}
  {\dd \ln N \over \dd \lambda}
  =
  {1 \over N}
  D_k n^k
  -
  D_k p^k
  \ ,
\end{equation}
and the momentum $p^k$ has already been checked,
from equation~(\ref{dpkdx}),
to be given accurately by expressions~(\ref{PxyHJ}) for $\Px$ and $\Py$,
the deviation between the predicted and true
logarithmic density $\ln N$ can be estimated
by integrating equation~(\ref{Dknk}) over the path of a particle
during inflation and collapse.
As shown in Appendix~\ref{estimateintegrals},
the resulting deviation
$\Delta \ln N \equiv \int ( \dd \ln N / \dd \lambda ) \, \dd \lambda$
is of order $\vel^2$,
equation~(\ref{DlnN}),
which may be considered adequately small.
This again confirms the accuracy of the Hamilton-Jacobi approximation~(\ref{PxyHJ}),
and the consequent expressions~(\ref{nsingle})
for the number density $N$ and number current $n^k$.

\subsection{Energy-momentum}
\label{energymomentumcollisionless}

The tetrad-frame energy-momentum density
$T_{kl}$
of a system of freely-falling particles is
\begin{equation}
  T_{kl}
  =
  \int
  p_k p_l \,
  f ( x^\mu , \bp ) \,
  {\ddd p \over 2 \px (2\pi\hbar)^3}
  \ .
\end{equation}
Covariant energy-momentum conservation follows from
the collisionless Boltzmann equation~(\ref{Boltzmannp}),
\begin{equation}
  D^k
  T_{kl}
  =
  \int
  p_k p_l \,
  D^k
  f
  \,
  {\ddd p \over 2 \px (2\pi\hbar)^3}
  =
  \int
  p_l \,
  {\dd f \over \dd \lambda}
  \,
  {\ddd p \over 2 \px (2\pi\hbar)^3}
  =
  0
  \ .
\end{equation}

For a single stream with fixed constants of motion
$E$, $L$, and $\KCarter$,
the energy-momentum is
\begin{equation}
  T_{kl}
  =
  n_k p_l
  =
  N p_k p_l
  \ ,
\end{equation}
where the number current $n^k$ and number density $N$ are given
by equations~(\ref{nsingle}).
Covariant energy-momentum conservation follows from
number conservation and the geodesic equation,
\begin{equation}
  D^k T_{kl}
  =
  D^k n_k p_l
  =
  p_l D^k n_k
  +
  N {\dd p_l \over \dd \lambda}
  =
  0
  \ .
\end{equation}
In accordance with
the expression~(\ref{pktetrad}) for the
tetrad-frame momentum $p_k$
and
the proportionality~(\ref{nsingle}) for $N$,
the energy-momentum of a single stream is
\begin{equation}
\label{Tklsingle}
  T_{kl}
  =
  {N P_k P_l \over \rho^2 \sqrt{| \Delta_k \Delta_l |}}
  \propto
  {\sigma^2 P_k P_l \over \rho^4 \Px \Py \sqrt{| \Delta_k \Delta_l |}}
  \ ,
\end{equation}
where
\begin{equation}
  \Delta_k
  \equiv
  \left\{
  \begin{array}{ll}
  \Deltax & \mbox{for } m = x \, , \, t \, , \\
  \Deltay & \mbox{for } m = y \, , \, \phi \, .
  \end{array}
  \right.
\end{equation}

The behaviour of the collisionless energy-momentum
will be explored in detail in \S\ref{inflation},
but qualitative features of the behaviour are already apparent from
equation~(\ref{Tklsingle}).
As a stream approaches the inner horizon,
$\Deltax \rightarrow - 0$,
the radial components of its energy-momentum grow large,
because of the inverse factors of the radial horizon function in
equation~(\ref{Tklsingle}).
During inflation,
the behaviour of the energy-momentum continues to be
dominated by the radial horizon function $\Deltax$,
which is driven to an exponentially tiny value,
causing the radial components of the energy-momentum
to grow exponentially huge.
During collapse,
the conformal factor $\rho$ shrinks,
amplifying all components of the energy-momentum.

The trace of the energy-momentum tensor of a stream of particles
of rest mass $m$,
\begin{equation}
\label{Tkksingle}
  T^k_k
  =
  N p^k p_k
  =
  - N m^2
  \ ,
\end{equation}
can always be treated as negligibly small.
During inflation,
the trace is negligible because the incident accretion flow is negligibly small,
in the conformally stationary limit.
As inflation develops and collapse begins,
the density $N \propto \rho^{-2}$
increases as the conformal factor $\rho$ shrinks,
but the individual components of the energy-momentum
increase more rapidly, as $T_{kl} \propto \rho^{-4}$,
equation~(\ref{Tklsingle}),
so the trace is never significant.

\section{Inflationary solutions}
\label{inflation}

This section presents the inflationary solutions
that emerge from Einstein's equations with a collisionless source
under the conditions of conformal stationarity, axisymmetry,
and conformal separability assumed in this paper.

Above the inner horizon,
the $\Lambda$-Kerr-Newman electrovac geometry,
\S\ref{electrovac},
provides the background
in which inflation ignites, \S\ref{ignition}.
As inflation develops,
it back-reacts on the geometry,
driving the conformal factor $\rho$ and the radial horizon function $\Deltax$
from their electrovac forms.
The equations governing the evolution of the conformal factor and
radial and angular horizon functions
are obtained in \S\ref{nonlinearinflationeqs}
by separation of variables in two components of the Einstein tensor
that have negligible collisionless source.
The equations are solved
to obtain the evolution of the conformal factor and radial horizon function
in \S\ref{nonlinearinflation}.
The implications
for the trajectories and densities of freely-falling streams
are presented in \S\ref{trajectoryparticle}.
The solutions for the conformal factor and horizon functions
are inserted into the remaining 8 Einstein components in
\S\ref{inflationeinstein},
and in \S\ref{streaming}
it is shown that the 8 Einstein components fit the form
of the energy-momentum tensor of two collisionless streams,
one ingoing and one outgoing.
The solutions indicate that during collapse
the angular ($y$ and $\phi$) motion of the collisionless streams grows,
and would begin to dominate once the radial horizon function
is no longer small,
$| \Deltax | \gtrsim 1$.
The last two subsections,
\S\ref{angularenergymomenta} and \S\ref{inflationnextorder}
address the effect of the angular motions to higher order,
showing that, while the earlier results remain robust as long
as angular motions are small, $| \Deltax | \ll 1$,
the solutions fail when the angular motions become important,
$| \Deltax | \gtrsim 1$.

\subsection{Charged black hole}
\label{charge}

The case of a charged black hole is deferred to Paper~3,
because the inclusion of electromagnetic currents, fields,
and energy-momenta adds a whole extra layer of complexity
to the solutions.
Of course, a charged black hole is physically less interesting
than the simpler case of an uncharged black hole,
which is the focus of the present paper.

Nevertheless, much of the results of this section and the next,
\S\ref{inflation} and \S\ref{masscurvature},
carry over essentially unchanged to the case of a charged black hole.
In particular, the expressions~(\ref{GUv}) and (\ref{GUvcsimp})
for the Einstein tensor are unchanged in the presence of
an electromagnetic field produced by a collisionless source,
and the solution, \S\ref{nonlinearinflation},
for the evolution of the horizon function and conformal factor
is unchanged,
the only difference being that
the derivative
$\Deltax^\prime$
of the electrovac radial horizon function at the inner horizon is altered by the
presence of charge
(because the electrovac horizon function~(\ref{DeltaxLKN}) depends on charge),
effectively changing the boundary conditions
of the solution.
The various differences between the charged and uncharged cases
are detailed in Paper~3.

To avoid repetition of the same results in Paper~3,
the spacetime that provides the boundary conditions for the inflationary
solution is referred to hereafter as ``electrovac'' rather than just ``vacuum''.

\subsection{Electrovac initial conditions}
\label{electrovac}

Electrovac solutions
\cite{Carter:1968c,Stephani:2003},
of which the physically relevant solutions
are Kerr-Newman with a cosmological constant $\Lambda$,
provide the boundary conditions for the inflationary solutions.

The electrovac solutions are strictly stationary, satisfying $\vel = 0$,
and strictly separable, so the conformal factor $\rho$ is separable,
equation~(\ref{rhosep}).
Solution of the Einstein equations,
Appendix~\ref{electrovacappendix},
leads to the standard results
\begin{subequations}
\label{rhoseps}
\begin{gather}
  \rhosep^2
  =
  \rhox^2 + \rhoy^2
  =
  {
  \sigma^2
  \over ( f_0 + f_1 \omegax ) ( f_1 + f_0 \omegay )}
  \ ,
\\
\label{rhoxy}
  \rhox
  =
  \sqrt{
  {g_0 - g_1 \omegax
  \over
  ( f_0 g_1 + f_1 g_0 )
  ( f_0 + f_1 \omegax ) }
  }
  \ , \quad
  \rhoy
  =
  \sqrt{
  {g_1 - g_0 \omegay
  \over
  ( f_0 g_1 + f_1 g_0 )
  ( f_1 + f_0 \omegay )}
  }
  \ ,
\end{gather}
\end{subequations}
and
\begin{subequations}
\label{domegadvarpi}
\begin{align}
\label{domega}
  {\dd \omegax \over \dd x}
  &=
  2
  \sqrt{
  \left( f_0 + f_1 \omegax \right)
  \left( g_0 - g_1 \omegax \right)
  }
  \ ,
\\
\label{dvarpi}
  {\dd \omegay \over \dd y}
  &=
  2
  \sqrt{
  \left( f_1 + f_0 \omegay \right)
  \left( g_1 - g_0 \omegay \right)
  }
  \ ,
\end{align}
\end{subequations}
where
$f_0$, $f_1$, $g_0$, and $g_1$ are constants
set by boundary conditions.
The sign of the square root for $\dd \omegax / \dd x$
is the same as that for $\rhox$,
while
the sign of the square root for $\dd \omegay / \dd y$
is the same as that for $\rhoy$.
Inflation leaves the separable factor $\rhosep$
in the conformal factor $\rho$, equation~(\ref{rhot}),
and the vierbein coefficients $\omegax$ and $\omegay$
unchanged from their electrovac values.

For $\Lambda$-Kerr-Newman,
the constants
$f_0$, $f_1$, $g_0$, and $g_1$ are
\begin{equation}
  f_0 = 0
  \ , \quad
  f_1 = a^{-1/2}
  \ , \quad
  g_0 = a^{3/2}
  \ , \quad
  g_1 = a^{5/2}
  \ ,
\end{equation}
where $a$ is the usual angular momentum parameter.
The conformal factor and vierbein coefficients are given by
\begin{equation}
\label{rhoLKN}
  \rho_x
  =
  r
  =
  a \cot ( a x )
  \ , \quad
  \rho_y
  =
  a \cos\theta
  =
  - a y
  \ ,
\end{equation}
\begin{equation}
\label{omegaLKN}
  \omegax
  =
  {a \over R^2}
  \ ,
  \quad
  \omegay
  =
  a \sin^2\!\theta
  \ ,
  \quad
  \sigma
  \equiv
  1 - \omegax \omegay
  =
  {\rhosep \over R}
  \ , \quad
  R
  \equiv
  \sqrt{r^2 + a^2}
  \ .
\end{equation}
The radial and angular horizon functions $\Deltax$ and $\Deltay$ are
\begin{subequations}
\label{DeltaLKN}
\begin{align}
\label{DeltaxLKN}
  \Deltax
  &=
  {1 \over R^2}
  \left(
  1 - {2 \Mbh r \over R^2} + {\Qelecbh^2 + \Qmagbh^2 \over R^2} - {\Lambda r^2 \over 3}
  \right)
  \ ,
\\
\label{DeltayLKN}
  \Deltay
  &=
  \sin^2\!\theta
  \left(
  1
  +
  {\Lambda a^2 \cos^2\!\theta \over 3}
  \right)
  \ ,
\end{align}
\end{subequations}
where $\Mbh$ is the black hole's mass,
$\Qelecbh$ and $\Qmagbh$ are its electric and magnetic charge,
and $\Lambda$ is the cosmological constant.
Inflation modifies the radial horizon function $\Deltax$,
but leaves the angular horizon function $\Deltay$,
along with $\omegax$ and $\omegay$, unchanged.

The full conformal factor $\rho$,
equation~(\ref{rhosepes}),
involves an additional time-dependent factor of
$\ee^{\vel t}$
(as well as an inflationary factor $\ee^{- \expinf}$
that equals unity away from the inner horizon).
The parameters $\Mbh$, $\Qelecbh$, $\Qmagbh$, $a$ of the black hole
coincide with the actual mass, charge, and specific angular momentum
of the black hole at conformal time $t = 0$,
and increase linearly with proper external time $t_{\rm KN}$.
Physically,
the cosmological constant $\Lambda$ should not increase with time,
but since it becomes completely overwhelmed during inflation
by other exponentially growing energy-momenta,
it can be included consistently in the description of the
spacetime outside the inflationary regime
(see the remarks in the paragraph following eq.~(\ref{XYxhomog})).

\subsection{Ignition}
\label{ignition}

This subsection outlines the behaviour of the inflationary exponent $\expinf$
during the earliest phase of inflation,
when the geometry is still electrovac,
and collisionless streams are approaching the inner horizon.
A more precise treatment that is valid throughout electrovac,
inflation, and collapse starts in the next subsection,
\S\ref{nonlinearinflationeqs}.

As discussed in \S\ref{focus},
collisionless streams become highly focussed
along the ingoing and outgoing principal null directions
as they approach the inner horizon,
causing the radial components
$T_{xx}$, $T_{xt}$, and $T_{tt}$
of their energy-momentum to grow large.
The Einstein combination
$G_{xx} + G_{tt}$,
which has zero electrovac source,
is
\begin{equation}
\label{Gxxttsep}
  \rho^2
  \left(
  G_{xx} + G_{tt}
  \right)
  =
  - \,
  2 \Deltax
  \left[
  {\ppartial \expinf \over \partial x^2}
  +
  \left(
  {\partial \expinf \over \partial x}
  \right)^2
  -
  2
  {\partial
  \ln ( \rhosep / \sigma )
  \over \partial x}
  {\partial \expinf \over \partial x}
  \right]
  -
  {2 \vel^2 \over \Deltax}
  \ .
\end{equation}
The dominant term in this expression is the one proportional
to the second derivative
$\ppartial \expinf / \partial x^2$
of the inflationary exponent.
Equating the
Einstein component
$G_{xx} + G_{tt}$
to the energy-momentum
$8\pi ( T_{xx} + T_{tt} )$
of collisionless streams,
equation~(\ref{Tklsingle}),
yields to leading order in $1/\Deltax$
\begin{equation}
\label{Pxxtte}
  {\ppartial \expinf \over \partial x^2}
  =
  {8 \pi \sum N \Px^2
  \over \Deltax^2}
  \ ,
\end{equation}
where the sum is over collisionless streams.
Because collisionless streams are hyper-relativistic near the inner horizon,
the Hamilton-Jacobi parameters of every stream satisfy
$P_t^2 = \Px^2$
to an excellent approximation.
The parameters $N$ and $\Px$ of each component of the collisionless
streams are sensibly constant as a function of radius $x$
near the inner horizon preceeding inflation and during early inflation
(but may be a function of angle $y$),
and thus $\sum N \Px^2$ is sensibly constant.
The sum over streams
$\sum N \Px^2$
is of the order of the accretion rate $\vel$.

In the situation of small accretion rate considered in this paper,
the electrovac geometry provides an excellent approximation
down to just above the inner horizon.
Near the inner horizon,
where $\Deltax \rightarrow - 0$,
the horizon function $\Deltax$ may be approximated by
\begin{equation}
\label{Drinflate}
  \Deltax
  =
  ( x - \xin )
  \Deltax^\prime
  \ ,
\end{equation}
where
$\Deltax^\prime \equiv \left. \dd \Deltax / \dd x \right|_{\xin}$
is the (positive) derivative of the electrovac horizon function at the
electrovac inner horizon at
$x = \xin$.
The derivative
$\Deltax^\prime$
is non-zero provided that the black hole is non-extremal,
as should be true for any astronomically realistic black hole.
Introduce the quantity $\uel$,
a small positive parameter of order the accretion rate, $\uel \sim \vel$,
defined by
\begin{equation}
\label{uel}
  \uel
  \equiv
  \left.
  {8 \pi \sum N \Px^2
  \over \Deltax^\prime}
  \right|_{\xin}
  \ ,
\end{equation}
evaluated at the inner horizon $\xin$.
Later, \S\ref{radialbcs},
the overall accretion rates of ingoing and outgoing streams
on to the inner horizon will be found to be proportional
respectively to the combinations
$\uel \mp \vel$.
Potentially $\uel$ could be a function of angle $y$,
but in \S\ref{nonlinearinflationeqs} it will be found that
separability continues to hold as inflation develops
only if $\uel$ is independent of angle $y$.
In terms of $\uel$,
the second derivative~(\ref{Pxxtte}) of $\expinf$ is
\begin{equation}
\label{epp}
  {\ppartial \expinf \over \partial x^2}
  =
  {\uel / \Deltax^\prime \over ( \xin - x )^2}
  \ .
\end{equation}
Integrating equation~(\ref{epp}) gives
\begin{equation}
\label{ep}
  {\partial \expinf \over \partial x}
  =
  {\uel / \Deltax^\prime \over \xin - x}
  =
  -
  {\uel \over \Deltax}
  \ ,
\end{equation}
where the constant of integration,
established well outside the inner horizon,
has been dropped because it is tiny compared to the retained term,
which is diverging at the inner horizon $x \rightarrow \xin$.
Integrating equation~(\ref{ep}) in turn yields
\begin{equation}
\label{e}
  \expinf
  =
  -
  {\uel \over \Deltax^\prime}
  \ln \left( {\xin - x \over \xin} \right)
  \ ,
\end{equation}
where the constant of integration follows from requiring that
$\expinf$ starts at zero well outside the inner horizon,
where $x \rightarrow 0$.
Equation~(\ref{e}) shows that the inflationary exponent
$\expinf$
is the product of a small factor $\uel \sim \vel$
and a term that diverges logarithmically at the inner horizon.
Thus $\expinf$ remains small even while its derivatives are becoming large.
The inflationary exponent $\expinf$ fulfills the
conditions~(\ref{econditionsx}) and (\ref{econditionsy})
anticipated in \S\ref{inflationKilling}.

\subsection{Equations governing evolution of the horizon function and conformal factor}
\label{nonlinearinflationeqs}

Inflation alters the spacetime geometry by changing the radial horizon function
$\Deltax$
and conformal factor
$\rho$
from their initial electrovac forms.
Equations governing the evolution of the horizon function
and conformal factor
are obtained from the Einstein equations for the
two diagonal components
$G_{xx} - G_{tt}$
and
$G_{yy} + G_{\phi\phi}$.
At least initially, these two components have negligible collisionless source
in the conformally stationary limit.
The angular component
$G_{yy} + G_{\phi\phi}$
has negligible collisionless source because the flow incident on
the inner horizon has negligible energy-momentum,
and inflation amplifies only radial components, not angular components.
The component
$G_{xx} - G_{tt}$
has negligible collisionless source because the trace of the
energy-momentum of a collisionless source is always negligible,
equation~(\ref{Tkksingle}).

In \S\ref{inflationnextorder}
a non-vanishing collisionless source for
$G_{xx} - G_{tt}$
and
$G_{yy} + G_{\phi\phi}$
will be taken into account,
and it will be found that the results of the present subsection are robust.

Since the collisionless source is negligible
for these two Einstein components,
it is natural to seek homogeneous solutions of equations~(\ref{GUv})
by separation.
To achieve the desired separation, introduce
$\Ux$ and $\Uy$
defined by
\begin{subequations}
\label{Uxy}
\begin{align}
\label{Ux}
  \Ux
  &\equiv
  -
  {\partial \expinf \over \partial x} \Deltax
  \ ,
\\
\label{Uy}
  \Uy
  &\equiv
  {\partial \expinf \over \partial y} \Deltay
  \ .
\end{align}
\end{subequations}
Initially,
in the electrovac phase just above the inner horizon,
$\Ux$ equals the small parameter $\uel$ defined by
equation~(\ref{uel}),
\begin{equation}
  \Ux
  =
  \uel
  \ ,
\end{equation}
and $\Uy$ is similarly small.
As will be seen in \S\ref{nonlinearinflation},
$\Ux$ is driven by inflation to large values,
but $\Uy$ remains always small.
Further,
define $\Xx$, $\Xy$, $\Yx$, and $\Yy$ by
\begin{subequations}
\label{XYxy}
\begin{align}
\label{Xx}
  \Xx
  &\equiv
  {\partial \Ux \over \partial x}
  +
  2 {\Ux^2 - \vel^2 \over \Deltax}
  \ ,
\\
\label{Xy}
  \Xy
  &\equiv
  {\partial \Uy \over \partial y}
  -
  2 {\Uy^2 + \vel^2 \omegay^2 \over \Deltay}
  \ ,
\\
\label{Yx}
  \Yx
  &\equiv
  {\dd \Deltax \over \dd x}
  +
  3 \Ux
  -
  \Deltax
  {\dd \over \dd x}
  \ln
  \left[ ( f_0 {+} f_1 \omegax ) {\dd \omegax \over \dd x} \right]
  \ ,
\\
\label{Yy}
  \Yy
  &\equiv
  {\dd \Deltay \over \dd y}
  -
  3 \Uy
  -
  \Deltay
  {\dd \over \dd y}
  \ln
  \left[ ( f_1 {+} f_0 \omegay ) {\dd \omegay \over \dd y} \right]
  \ .
\end{align}
\end{subequations}
In terms of
$\Ux$, $\Uy$, $\Xx$, $\Xy$, $\Yx$, and $\Yy$,
the Einstein components
$G_{xx} - G_{tt}$ and $G_{yy} + G_{\phi\phi}$ are
\begin{subequations}
\label{GUv}
\begin{align}
\label{GxxmttUv}
  \rho^2 \left( G_{xx} - G_{tt} \right)
  &=
  {1 \over \sigma^2}
  \left(
  \Yx {\dd \ln \omegax \over \dd x}
  -
  \Yy {\dd \ln \omegay \over \dd y}
  \right)
  -
  2 \Xx
  +
  \Yx
  {\dd
  \over \dd x}
  \ln
  \left(
  {f_0 {+} f_1 \omegax \over \omegax}
  \right)
  +
  \Xy
  -
  {\partial \Yy \over \partial y}
  +
  \Yy
  {\dd
  \over \dd y}
  \ln
  \left[
  {\omegay ( f_1 {+} f_0 \omegay ) \over \dd \omegay / \dd y}
  \right]
\nonumber
  \\
  & \quad
  + \,
  \Ux
  {\partial
  \over \partial x}
  \ln
  \left[ \sigma^2 ( f_0 {+} f_1 \omegax ) \right]
  -
  \Uy
  {\partial \over \partial y}
  \ln \left[
  {( g_1 {-} g_0 \omegay ) \over \sigma^2} {\dd \omegay \over \dd y}
  \right]
  \ ,
\\
\label{GyypphiphiUv}
  \rho^2 \left( G_{yy} + G_{\phi\phi} \right)
  &=
  {1 \over \sigma^2}
  \left(
  \Yx {\dd \ln \omegax \over \dd x}
  -
  \Yy {\dd \ln \omegay \over \dd y}
  \right)
  -
  2 \Xy
  -
  \Yy
  {\dd \over \dd y}
  \ln \left( {f_1 {+} f_0 \omegay \over \omegay} \right)
  +
  \Xx
  +
  {\partial \Yx \over \partial x}
  -
  \Yx
  {\dd \over \dd x}
  \ln
  \left[
  {\omegax ( f_0 {+} f_1 \omegax ) \over \dd \omegax / \dd x}
  \right]
\nonumber
  \\
  & \quad
  + \,
  \Uy
  {\partial \over \partial y}
  \ln \left[ \sigma^2 ( f_1 {+} f_0 \omegay ) \right]
  -
  \Ux
  {\partial \over \partial x}
  \ln \left[
  {( g_0 {-} g_1 \omegax ) \over \sigma^2} {\dd \omegax \over \dd x}
  \right]
  \ .
\end{align}
\end{subequations}
Homogeneous solutions of these equations can be found by supposing that
$\Ux$, $\Xx$, and $\Yx$ are all functions of radius $x$,
while
$\Uy$, $\Xy$, and $\Yy$ are all functions of angle $y$,
and by separating each of the equations as
\begin{equation}
\label{sepG}
  {1 \over \sigma^2}
  \left(
  {f_0 h_0 {+} h_2 \omegax {+} f_1 h_1 \omegax^2 \over \omegax}
  -
  {f_1 h_1 {+} h_2 \omegay {+} f_0 h_0 \omegay^2 \over \omegay}
  \right)
  -
  {f_0 h_0 {+} h_3 \omegax \over \omegax}
  +
  {f_1 h_1 {+} h_3 \omegay \over \omegay}
  =
  0
  \ ,
\end{equation}
for some constants $h_0$, $h_1$, $h_2$, and $h_3$.
If one attempts to separate equations~(\ref{GUv}) exactly,
then the attempt fails unless $\Ux$ and $\Uy$ are identically zero,
which is the usual electrovac case.
But if $\Ux$ is taken to be small but finite,
then separation succeeds,
and inflation emerges.
If $\Ux$ and $\Uy$ on the second lines of equations~(\ref{GUv})
are treated as negligibly small,
then separating the first lines of each of equations~(\ref{GUv})
according to the pattern of equation~(\ref{sepG})
leads to the homogeneous solutions
\begin{subequations}
\label{XYxhomog}
\begin{align}
\label{Xxhomog}
  \Xx
  &=
  0
  \ ,
\\
\label{Xyhomog}
  \Xy
  &=
  0
  \ ,
\\
\label{Yxhomog}
  \Yx
  &=
  {( f_0 + f_1 \omegax ) ( h_0 + h_1 \omegax ) \over \dd \omegax / \dd x}
  \ ,
\\
\label{Yyhomog}
  \Yy
  &=
  {( f_1 + f_0 \omegay ) ( h_1 + h_0 \omegay ) \over \dd \omegay / \dd y}
  \ .
\end{align}
\end{subequations}

If $\Ux = \Uy = 0$,
then solution of the differential equations~(\ref{Yx}) and (\ref{Yy})
with the homogenous solutions~(\ref{Yxhomog}) and (\ref{Yyhomog})
for $\Yx$ and $\Yy$,
yields,
subject to appropriate boundary conditions,
the radial and angular horizon functions $\Deltax$ and $\Deltay$
of the Kerr line-element.
The separable solutions generalize to other electrovac spacetimes
by admitting appropriate sources for $\Yx$ and $\Yy$,
Appendix~\ref{electrovacappendix}.
The solutions with a static radial electromagnetic field have
a contribution
$G^e_{mn} = \left[ ( \Qelecbh^2 + \Qmagbh^2 ) / ( \rho^2 \rhosep^2 ) \right] \diag ( 1 , -1 , 1 , 1 )$,
and those with a cosmological constant have
$G^\Lambda_{mn} = - ( \rhosep^2 / \rho^2 ) \Lambda \eta_{mn}$.
These electrovac contributions cease to describe
a static radial electromagnetic field or cosmological constant when
$\rho \neq \rhosep$,
but this happens only from the onset of collapse,
by which time any electrovac contribution is
overwhelmed by the collisionless energy-momentum,
so the failure is unimportant
(see \S{IV\,F} of Paper~3 for a more precise treatment).

Solution of the angular behaviour during inflation is immediate.
The vanishing,
equation~(\ref{Xyhomog}),
of $\Xy$ defined by equation~(\ref{Xy})
implies that $\Uy$,
which is initially neglible in the conformally stationary limit
$\vel \rightarrow 0$,
remains negligible throughout,
and may be set to zero
\begin{equation}
  \Uy
  =
  0
  \ .
\end{equation}
The expression~(\ref{Yy}) for $\Yy$
governing the angular horizon funtion $\Deltay$
is then unchanged from its electrovac form,
and the angular horizon function $\Deltay$
thus retains its electrovac form
during inflation and collapse.

Down to just above the inner horizon,
solution of the radial
equations~(\ref{Xx}), (\ref{Yx}), (\ref{Xxhomog}), and (\ref{Yxhomog})
for $\Xx$ and $\Yx$ leads to the usual electrovac form
of the radial horizon function $\Deltax$.
But near the inner horizon, $\Deltax \rightarrow -0$,
the term proportional to $1/\Deltax$ on the right hand side
of equation~(\ref{Xx}) for $\Xx$ starts to diverge,
presaging inflation.
In the vicinity of the inner horizon, where $\Deltax$ is near zero,
expression~(\ref{Yx}) for $\Yx$ simplifies to
\begin{equation}
  \Yx
  =
  {\dd \Deltax \over \dd x}
  +
  3 \Ux
  \ .
\end{equation}
It will be found in the next subsection, \S\ref{nonlinearinflation},
that the radius $x$ remains frozen at its inner horizon value $\xin$
throughout inflation and subsequent collapse.
Consequently $\omegax$, hence $\Yx$, are also frozen at their inner horizon
values during inflation and collapse.
Thus $\Yx$ is given by its electrovac value incident on the iner horizon,
$\Yx = \Deltax^\prime$
where
$\Deltax^\prime \equiv \left. \dd \Deltax / \dd x \right|_{\xin}$
is the derivative of the electrovac horizon function at the inner horizon
$x = \xin$.
It follows that
in the vicinity of the inner horizon
the equations~(\ref{Xx}), (\ref{Yx}), (\ref{Xxhomog}), and (\ref{Yxhomog})
governing the evolution of $\Ux$ and $\Deltax$ reduce to
\begin{subequations}
\label{DDUx}
\begin{align}
\label{DUx}
  {\partial \Ux \over \partial x}
  +
  2 
  {\Ux^2 - \vel^2 \over \Deltax}
  &=
  0
  \ ,
\\
\label{DDx}
  {\dd \Deltax \over \dd x}
  +
  3 \Ux
  &=
  \Deltax^\prime
  \ .
\end{align}
\end{subequations}
During the electrovac and earliest phase of inflation,
when not only $\Ux$ but also $\partial \Ux / \partial x$
are negligible,
which is true when $| \Deltax | \gtrsim \uel$,
the Einstein components~(\ref{GUv}) separate without requiring that $\Ux$,
which at this early stage equals $\uel$,
be a function only of $x$.
As inflation progresses however,
continued separability,
which requires that $\Deltax$ be a function only of $x$,
requires also that $\Ux$ be a function only of $x$.
It follows that the inflationary exponent $\expinf$,
given initially by equation~(\ref{e}),
must also be a function only of $x$.

Below, equations~(\ref{GUvcsimp}),
it will be found that the energy-momenta
of ingoing and outgoing collisionless streams
are proportional respectively to $\Ux \mp \vel$.
If one or other stream vanished,
then $\Ux^2 - \vel^2$ would vanish,
so equation~(\ref{DUx}) would have no diverging term,
and there would be no inflation.
This is consistent with the physical argument
that inflation requires the simultaneous presence
of both ingoing and outgoing streams at the inner horizon.

Equation~(\ref{DUx}) indicates an instability only at the inner horizon,
not the outer horizon.
As streams approach the inner horizon,
$\Ux$ is driven away from zero
because the horizon function is negative and tending to zero,
$\Deltax \rightarrow -0$.
By contrast, 
as infalling streams approach the outer horizon,
the horizon function is positive and tending to zero,
$\Deltax \rightarrow 0$,
causing $\Ux$ to decay rather than grow.

\subsection{Evolution of the horizon function and conformal factor during inflation and collapse}
\label{nonlinearinflation}

This subsection integrates equations~(\ref{DDUx}),
along with equation~(\ref{Ux}),
to determine the evolution of the radial horizon function $\Deltax$
and conformal factor $\rho$ during inflation and collapse.

The separation of the Einstein components~(\ref{GUv})
in the previous subsection, \S\ref{nonlinearinflationeqs},
was premised on $\Ux$ (and $\Uy$) being negligibly small.
However,
the separation continues to remain valid during inflation
and collapse when $\Ux$ grows huge.
The reason for this is that the dominant terms
in the Einstein components~(\ref{GUv})
during inflation and collapse
are of order $\Ux^2 / \Deltax$,
coming from the expression~(\ref{Xx}) for $\Xx$.
Thus,
once $\Ux$ ceases to be negligible,
the condition for the validity of the separation becomes
$\Ux \ll \Ux^2 / | \Deltax |$,
or equivalently
$| \Deltax | \ll \Ux$.
Thus the condition for the validity of the separation
of the Einstein components~(\ref{GUv}) is
\begin{equation}
\label{sepcondition}
  \mbox{either}
  \quad
  \Ux \ll 1
  \quad
  \mbox{or}
  \quad
  | \Deltax | \ll \Ux
  \ .
\end{equation}
It will be found in \S\ref{angularenergymomenta}
that the conformally separable Einstein equations cease to be satisfied with
a collisionless source once the angular motion of the collisionless
streams becomes large, which happens when $| \Deltax | \gtrsim 1$.
At this point $\Ux$ is exponentially huge, equation~(\ref{UD1}).
Thus condition~(\ref{sepcondition})
remains well satisfied throughout inflation and collapse.

The fact that the separation of the Einstein components~(\ref{GUv})
remains valid even when $\Ux$ grows large,
subject only to the condition~(\ref{sepcondition}),
is confirmed in \S\ref{inflationnextorder},
where the equations are solved to the next higher order in $\Deltax / \Ux$.

Integrating equation~(\ref{Ux}) for $\Ux$
and using equation~(\ref{DUx}),
gives the inflationary exponent $\expinf$
as a function of $\Ux$:
\begin{equation}
\label{eUinf}
  \expinf
  =
  -
  \int {\Ux \, \dd x \over \Deltax}
  =
  \int
  {\Ux \, \dd \Ux \over 2 ( \Ux^2 - \vel^2 )}
  =
  \frac{1}{4}
  \ln \left( {\Ux^2 - \vel^2 \over \uel^2 - \vel^2} \right)
  \ ,
\end{equation}
the constant of integration coming from $\expinf = 0$ at $\Ux = \uel$.
Equivalently, the inflationary part $\ee^{- \expinf}$
of the conformal factor $\rho$ is
\begin{equation}
\label{rhoinf}
  \ee^{- \expinf}
  =
  \left(
  {\uel^2 - \vel^2
  \over
  \Ux^2 - \vel^2}
  \right)^{1/4}
  \ .
\end{equation}
Inverting equation~(\ref{rhoinf}) gives $\Ux$ in terms of $\expinf$:
\begin{equation}
\label{Ueinf}
  \Ux
  =
  \sqrt{
  \vel^2
  +
  ( \uel^2 - \vel^2 ) 
  \ee^{4 \expinf}
  }
  \ .
\end{equation}

Equation~(\ref{DDx}) divided by equation~(\ref{DUx})
yields an equation for the horizon function $\Deltax$:
\begin{equation}
\label{DU}
  {\dd \ln \Deltax \over \dd \Ux}
  =
  {3 \Ux - \Deltax^\prime
  \over 2 ( \Ux^2 - \vel^2 )}
  \ .
\end{equation}
Equation~(\ref{DU}) integrates to
\begin{equation}
\label{Dinf}
  \Deltax
  =
  -
  \left( {\Ux^2 - \vel^2 \over \uel^2 - \vel^2} \right)^{3/4}
  \left[
  {
  ( \Ux + \vel )
  ( \uel - \vel )
  \over
  ( \Ux - \vel )
  ( \uel + \vel )
  }
  \right]^{\Deltax^\prime / ( 4 \vel )}
  \ ,
\end{equation}
where the constant of integration is established by
$\Deltax \sim -1$ at $\Ux = \uel$.
The precise constant is not important
since near $\Ux \approx \uel$,
the second factor on the right hand side of equation~(\ref{Dinf})
is a number near unity taken to a large power
$\Deltax^\prime / ( 4 \vel )$.
The conformal factor $\rho$
starts to depart significantly from its electrovac value $\rhosep$
when $\expinf$ departs appreciably from zero,
which occurs when $\Ux$ is a factor
somewhat greater than unity times $\uel$.
At this point the horizon function is of order
\begin{equation}
  \Deltax
  \sim
  \ee^{- 1 / \vel}
  \ ,
\end{equation}
which is exponentially tiny.

The horizon function goes through an extremum,
$\dd \Deltax / \dd \Ux = 0$,
where,
according to equation~(\ref{DU}),
\begin{equation}
\label{Dmax}
  \Ux
  =
  {\Deltax^\prime \over 3}
  \ ,
\end{equation}
which is of order unity.
At horizon extremum, the inflationary exponent $\expinf$ is
\begin{equation}
  \expinf
  =
  \frac{1}{2}
  \ln
  \left(
  {\Deltax^\prime
  \over 3 \sqrt{\uel^2 - \vel^2}}
  \right)
  \ ,
\end{equation}
and consequently the inflationary part of the conformal factor $\rho$ is
\begin{equation}
  \ee^{-\expinf}
  =
  \left(
  {3 \sqrt{\uel^2 - \vel^2}
  \over
  \Deltax^\prime}
  \right)^{1/2}
  \sim
  \vel^{1/2}
  \ ,
\end{equation}
which is becoming small.
The horizon function at its extremum is
\begin{equation}
  \Deltax
  =
  -
  \left(
  {\ee \Deltax^\prime \over 3 \sqrt{\uel^2 - \vel^2}}
  \right)^{3/2}
  \left(
  {\uel - \vel \over \uel + \vel}
  \right)^{\Deltax^\prime / ( 4 \vel )}
  \sim
  \ee^{-1/\vel}
  \ ,
\end{equation}
which is exponentially tiny.

The value of the radial coordinate $x$
can be found by integrating equation~(\ref{DUx}),
\begin{equation}
\label{xUinf}
  x - \xin
  =
  -
  \int
  {\Deltax \, \dd \Ux \over 2 ( \Ux^2 - \vel^2 )}
  \ .
\end{equation}
The integral can be expressed analytically as an incomplete beta-function,
but the expression is not useful.
Physically, equation~(\ref{xUinf}) says
that the radius $x$ is frozen at its inner horizon value $\xin$
during inflation and collapse,
where $\Ux$ is growing, while $\Deltax$ remains small.
The radial coordinate $x$ remains frozen
even while the conformal factor $\rho \propto \ee^{- \expinf}$
is shrinking.

After the horizon function $\Deltax$ goes through its extremum,
it starts increasing in absolute value
as $\Ux^{3/2}$ according to equation~(\ref{Dinf}),
or equivalently as $\ee^{3 \expinf}$ according to equation~(\ref{Ueinf}):
\begin{equation}
\label{DUlate}
  \Deltax
  \approx
  \Ux^{3/2}
  \Delta_0
  \approx
  \ee^{3 \expinf}
  \Delta_0
  \ , \quad
  \Delta_0
  \equiv
  -
  \left(
  {\uel - \vel \over \uel + \vel}
  \right)^{\Deltax^\prime / 4 \vel}
  \ ,
\end{equation}
where factors of order unity have been dropped compared to the
exponentially huge factor $\Delta_0$.
In \S\ref{angularenergymomenta}
it will be found that the conformally separable Einstein equations
can no longer be solved with a collisionless source
once $| \Deltax | \gtrsim 1$,
the point at which the motion of freely-falling ingoing and outgoing streams
become predominantly angular rather than radial.
Equation~(\ref{DUlate})
implies that at this point $\Ux$ is
\begin{equation}
\label{UD1}
  \Ux \approx | \Delta_0^{-2/3} |
  \quad\mbox{at~}
  | \Deltax | \approx 1
  \ ,
\end{equation}
which is exponentially huge.
In particular,
the condition~(\ref{sepcondition})
for the validity of the separation of Einstein components
is well satisfied throughout inflation and collapse.

During collapse,
$\vel$ and $\Deltax^\prime$ in equations~(\ref{DDUx})
may be neglected, and the equations
yield a simplified equation for
the evolution of the radius $x$,
\begin{equation}
  {\dd \ln ( \Deltax / \Ux ) \over \dd x}
  =
  -
  {\Ux \over \Deltax}
  \ ,
\end{equation}
which integrates to
\begin{equation}
\label{xDU}
  x - \xin
  =
  -
  {\Deltax \over \Ux}
  \ ,
\end{equation}
the constant of integration coming from $x = \xin$ at $\Deltax = 0$.
Since $| \Deltax | / \Ux \ll 1$
throughout collapse,
the radial coordinate $x$ remains frozen at its inner horizon value $\xin$
to high precision
through inflation and collapse.


\subsection{Trajectory and density of a freely-falling stream}
\label{trajectoryparticle}

In the previous subsection, \S\ref{nonlinearinflation},
it was found  that inflation and collapse takes place over an extremely
narrow zone in (conformal) radial coordinate $x$
about the inner horizon value $\xin$.
Consequently the radial coordinate $x$ attached to a freely-falling stream
remains essentially frozen at its inner horizon value
throughout inflation and collapse.
The radial coordinate remains frozen even while the conformal factor $\rho$
is collapsing to an exponentially tiny scale.

The proper time $\tau$ that elapses on the stream satisfies,
equation~(\ref{pkappa}),
\begin{equation}
\label{dtaudx}
  {\dd \tau \over \dd x}
  =
  m {\dd \lambda \over \dd x}
  =
  -
  {m \rho^2 \over \Px}
  \ .
\end{equation}
The right hand side of equation~(\ref{dtaudx})
is approximately constant during early inflation,
while the conformal factor $\rho$ is still
close to its separable value $\rhosep$,
but then decreases as the conformal factor shrinks.
Thus very little proper time elapses on a stream during
the entire of inflation and collapse.
During collapse, the proper time $\tau$ is even more frozen
than the radial coordinate $x$, which itself is frozen.

The angular coordinate $y$ along the trajectory of the stream satisfies,
equation~(\ref{dxdyP}),
\begin{equation}
\label{dydx}
  {\dd y \over \dd x}
  =
  - {\Py \over \Px}
  \ .
\end{equation}
The right hand side of equation~(\ref{dydx}) is a number of order unity
as long as $| \Deltax | \ll 1$.
and then becomes less than unity
when $| \Deltax | \gtrsim 1$.
Thus the angular coordinate $y$ is also frozen
along the trajectory of a stream.
It follows that quantities such as $\omegax$ and $\omegay$, hence $\sigma$,
are likewise frozen
along the trajectory of a stream.

The conformal time $t$ and azimuthal angle $\phi$
coordinates along the trajectory are given by equations~(\ref{dtdphiP}),
which can be written
\begin{equation}
  {\dd t \over \dd x}
  =
  {1 \over \Px}
  \left(
  {P_t \over \Deltax}
  -
  {\omegay P_\phi \over \Deltay}
  \right)
  \ , \quad
  {\dd \phi \over \dd x}
  =
  {1 \over \Px}
  \left(
  {\omegay P_t \over \Deltax}
  -
  {P_\phi \over \Deltay}
  \right)
  \ .
\end{equation}
As long as
$| \Deltax | \ll 1$,
the conformal time coordinate $t$
along an ingoing ($+$) or outgoing ($-$) stream
is given by
\begin{equation}
\label{tparticle}
  t
  =
  \pm
  \int {\dd x \over \Deltax}
  =
  \mp
  \int {\dd \Ux \over 2 ( \Ux^2 - \vel^2 )}
  =
  \pm
  {1 \over 4 \vel}
  \ln \left[
  {( \Ux + \vel ) ( \uel - \vel ) \over ( \Ux - \vel ) ( \uel + \vel )}
  \right]
  \ ,
\end{equation}
the constant of integration being established by $t = 0$ at
$\Ux = \uel$.
Consequently
the time part $\ee^{\vel t}$ of the conformal factor $\rho$ is
\begin{equation}
\label{evtparticle}
  \ee^{\vel t}
  =
  \left[
  {( \Ux + \vel ) ( \uel - \vel ) \over ( \Ux - \vel ) ( \uel + \vel )}
  \right]^{\pm 1/4}
  \ .
\end{equation}
Unlike $x$ and $y$, the coordinate $t$
is not frozen along the trajectory
of the particle.
Rather $\ee^{\vel t}$, equation~(\ref{evtparticle}),
varies by a factor of order unity as $\Ux$ increases from $\uel$
to some large value.
Once $\Ux$ increases to some value much larger than $\uel$
(such as $\Ux \sim 1$, since $\uel \ll 1$),
the time coordinate $t$ freezes out.
Once $| \Deltax | \gtrsim 1$,
the relation $t = \pm \int \dd x / \Deltax$ fails,
but by that time $t$ is already frozen out,
so in practice equation~(\ref{evtparticle})
is valid accurately throughout inflation and collapse.

The no-going tetrad frame has the special property that $P_t = 0$.
In the no-going frame
the potentially large term proportional to $P_t/\Deltax$
in equation~(\ref{tparticle}) vanishes,
and $t$, like $x$, remains frozen throughout inflation and collapse.
Physically,
the factor $\ee^{\vel t}$ in equation~(\ref{evtparticle})
is the expansion factor of the black hole
at the time that particles were accreted
relative to the time that no-going particles were accreted.
Equation~(\ref{evtparticle})
says that
outgoing particles were accreted to the past
of when no-going particles were accreted,
when the black hole was smaller,
while ingoing particles were accreted to the future
of when no-going particles were accreted,
when the black hole was larger.

The furthest into the past that ingoing particles see
(or into the future that outgoing particles see)
is at the end of inflation when the geometry is collapsing.
At this point the ratio of the sizes of the black hole
at the times the ingoing ($+$) and outgoing ($-$) particles were accreted is,
from equation~(\ref{evtparticle}) with large $\Ux$,
\begin{equation}
\label{rhopmparticle}
  {\rho^+_{\rm accrete} \over \rho^-_{\rm accrete}}
  =
  \left(
  {\uel + \vel \over \uel - \vel}
  \right)^{1/2}
  \ ,
\end{equation}
which is a number of order unity or a few.
Equation~(\ref{rhopmparticle})
shows that what happens deep inside the black hole
depends only on the finite past and future of the black hole,
not on what happens at the initial moments
of collapse, nor on the indefinite future.

The density $N$ along a stream
varies according to proportionality~(\ref{nsingle}),
which given that $\sigma$ is frozen is
\begin{equation}
  N
  \propto
  {1 \over \rho^2 \Px \Py}
  \ .
\end{equation}
The parameters $\Px$ and $\Py$
are given by equations~(\ref{PxyHJ}),
which are exact for massless particles,
and whose accuracy for massive particles was established
in \S\ref{conformalseparability}.

As long as
$| \Deltax | \ll 1$,
the Hamilton-Jacobi parameters
$\Px$ and $\Py$ are constant
along the trajectory of a stream,
and the density $N^{\pm}$
of an ingoing ($+$) or outgoing ($-$) stream simplifies to
\begin{equation}
\label{Npmearly}
  N^\pm
  \propto
  {1 \over \rho^2}
  \propto
  \ee^{2 ( \xi - \vel t )}
  \propto
  \Ux \mp \vel
  \ ,
\end{equation}
where the inflationary exponent $\xi$ and conformal time $t$
have been eliminated in favour of $\Ux$
using equations~(\ref{rhoinf}) and (\ref{evtparticle}).
In this regime,
the tetrad-frame momentum $p^\pm_k$ is hyper-relativistic,
and focussed along the radial direction,
\begin{equation}
\label{ppmearly}
  p^\pm_k
  \propto
  {1 \over \rho}
  \left\{
  -
  {1 \over \sqrt{- \Deltax}}
  \, , \ 
  \mp
  {1 \over \sqrt{- \Deltax}}
  \, , \ 
  \mu_y
  \, , \ 
  \mu_\phi
  \right\}
  \ , 
  \quad
  \mu_k
  \equiv
  {P_k \over | P_t | \sqrt{\Deltay}}
  \ .
\end{equation}
Note that $\mu_k$,
which are constant along the trajectory of the stream,
are generically of order unity.

It will be found in \S\ref{inflationnextorder}
that the conformally separable Einstein equations cease
to be satisfied by the energy-momentum of collisionless streams
once the angular motions of the streams begin to exceed their radial motions,
which happens when
$| \Deltax | \gtrsim 1$.
To see how this happens, it is necessary to (attempt to) follow the behaviour
of collisionless streams into this regime.
For $| \Deltax | \gtrsim 1$,
as long as the spacetime remains conformally separable,
the Hamilton-Jacobi parameters
$P_t$, $\Py$, and $P_\phi$ all remain constant,
but $\Px$ is no longer constant,
varying in accordance with equation~(\ref{PxHJlate}),
which can be written
(the argument of the square root in the following equation
is positive since $\Deltax$ is negative)
\begin{equation}
\label{PxPt}
  {\Px \over P_t}
  =
  \pm
  \sqrt{1 - ( \mu_y^2 + \mu_\phi^2 ) \Deltax}
  \ ,
\end{equation}
where
$\mu_y^2 + \mu_\phi^2$,
a constant along the trajectory of the stream,
is
\begin{equation}
  \mu_y^2 + \mu_\phi^2
  =
  {\KCarter - m^2 \rhoy^2 \over P_t^2}
  \ .
\end{equation}
Putting together the dependence on $\rho$, equation~(\ref{Npmearly}),
and $\Px$, equation~(\ref{PxPt}),
yields the density $N^\pm$ along an ingoing ($+$) or outgoing ($-$) stream,
\begin{equation}
\label{Npmlate}
  N^\pm
  \propto
  {\Ux \mp \vel \over \sqrt{1 - ( \mu_y^2 + \mu_\phi^2 ) \Deltax}}
  \ .
\end{equation}
The corresponding tetrad-frame momentum $p^\pm_k$ is
\begin{equation}
\label{ppmlate}
  p^\pm_k
  \propto
  {1 \over \rho}
  \left\{
  -
  {\sqrt{1 - ( \mu_y^2 + \mu_\phi^2 ) \Deltax} \over \sqrt{- \Deltax}}
  \, , \ 
  \mp
  {1 \over \sqrt{- \Deltax}}
  \, , \ 
  \mu_y
  \, , \ 
  \mu_\phi
  \right\}
  \ ,
\end{equation}
whose angular components $p_y$ and $p_\phi$
exceed the radial component
$p_t$ once $| \Deltax | \gtrsim 1$.

\subsection{Einstein tensor}
\label{inflationeinstein}

The two Einstein components~(\ref{GUv})
led to equations for the evolution of the conformal factor
and radial horizon function during inflation and collapse.
The remaining 8 components of the tetrad-frame Einstein tensor $G_{kl}$,
which have zero electrovac source,
may be written in terms
of the expressions~(\ref{Uxy}) for $\Ux$ and $\Uy$,
and (\ref{XYxy}) for $\Xx$, $\Xy$, $\Yx$ and $\Yy$:
\begin{subequations}
\label{GUvc}
\begin{align}
\label{GxtxUv}
  \rho^2
  \left(
  {G_{xx} + G_{tt} \over 2}
  \,\pm\,
  G_{xt}
  \right)
  &=
  ( \Ux \mp \vel )
  \left[
  {\Yx \pm \vel \over - \Deltax}
  -
  {\dd \over \dd x}
  \ln \left( \dd \omegax \over \dd x \right)
  \right]
  +
  \Xx
  \ ,
\\
\label{GxtyUv}
  \rho^2
  \left(
  G_{xy}
  \,\pm\,
  G_{ty}
  \right)
  &=
  -
  {1 \over \sqrt{- \Deltax \Deltay}}
  ( \Ux \mp \vel )
  \left(
  \Deltay
  {\partial \ln \rhosep^2 \over \partial y}
  -
  2 \Uy
  \right)
  -
  \sqrt{- \Deltax \over \Deltay}
  \left(
  \Uy
  {\partial \ln \rhosep^2 \over \partial x}
  \, \pm \,
  {\vel \omegay \over \sigma^2} {\dd \omegay \over \dd y}
  \right)
  \ ,
\\
\label{GxtphiUv}
  \rho^2
  \left(
  G_{x\phi}
  \,\pm\,
  G_{t\phi}
  \right)
  &=
  \pm
  {1 \over \sqrt{- \Deltax \Deltay}}
  ( \Ux \mp \vel )
  \left(
  {\Deltay \over \sigma^2}
  {\dd \omegax \over \dd x}
  \mp
  2 \vel \omegay
  \right)
  \,\mp\,
  \sqrt{- \Deltax \over \Deltay}
  \left(
  {\Uy \over \sigma^2} {\dd \omegay \over \dd y}
  \mp
  \vel \omegay {\partial \ln \rhosep^2 \over \partial x}
  \right)
  \ ,
\\
\label{GyphiyUv}
  \rho^2
  \left(
  {G_{yy} - G_{\phi\phi} \over 2}
  \,\pm\,
  \im
  G_{y\phi}
  \right)
  &=
  ( \Uy \mp \im \vel \omegay )
  \left[
  {- \,
  \Yy
  \pm
  \im \vel \omegay \over \Deltay}
  -
  {\dd \over \dd y}
  \ln \left( \dd \omegay \over \dd y \right)
  \right]
  +
  \Xy
  \,\mp\,
  \im \vel {\dd \omegay \over \dd y}
  \ .
\end{align}
\end{subequations}
Most of the terms
in equations~(\ref{GUvc})
are negligible in the conformally stationary limit
$\vel \rightarrow 0$.
Terms are negligible because:
(a) they are proportional to one of $\Xx$ or $\Xy$, which vanish;
(b) they are proportional to $\Uy$,
which remains negligibly small in the conformally stationary limit;
or
(c) they are proportional to $\vel$,
and not proportional to inverse factors of $\Deltax$,
so remain negligible in the conformally stationary limit $\vel \rightarrow 0$.
In addition,
the term proportional to
$\dd \ln ( \dd \omegax / \dd x ) / \dd x$
inside square brackets in equation~(\ref{GxtxUv}) may be neglected.
This term might potentially become important
when $| \Deltax | \gtrsim 1$,
but it will be found below, equation~(\ref{GxtxUvlate}),
that this term in any case disappears
when the Einstein equations are solved to next order in $\Deltax / \Ux$.

With all negligible and sub-dominant terms discarded,
equations~(\ref{GUvc}) simplify
in the conformally stationary limit to
\begin{subequations}
\label{GUvcsimp}
\begin{align}
\label{GxtxUvsimp}
  \rho^2
  \left(
  {G_{xx} + G_{tt} \over 2}
  \,\pm\,
  G_{xt}
  \right)
  &=
  {1 \over - \Deltax}
  ( \Ux \mp \vel )
  \left(
  \Deltax^\prime \pm \vel
  \right)
  \ ,
\\
\label{GxtyUvsimp}
  \rho^2
  \left(
  G_{xy}
  \,\pm\,
  G_{ty}
  \right)
  &=
  -
  {1 \over \sqrt{- \Deltax \Deltay}}
  ( \Ux \mp \vel )
  \Deltay
  {\partial \ln \rhosep^2 \over \partial y}
  \ ,
\\
\label{GxtphiUvsimp}
  \rho^2
  \left(
  G_{x\phi}
  \,\pm\,
  G_{t\phi}
  \right)
  &=
  \pm
  {1 \over \sqrt{- \Deltax \Deltay}}
  ( \Ux \mp \vel )
  \left(
  {\Deltay \omegax^\prime \over \sigma^2}
  \mp
  2 \vel \omegay
  \right)
  \ ,
\\
\label{GyphiyUvsimp}
  \rho^2
  \left(
  {G_{yy} - G_{\phi\phi} \over 2}
  \,\pm\,
  \im
  G_{y\phi}
  \right)
  &=
  0
  \ ,
\end{align}
\end{subequations}
in which
$\omegax^\prime \equiv \left. \dd \omegax / \dd x \right|_{\xin}$
and $\Deltax^\prime \equiv \left. \dd \Deltax / \dd x \right|_{\xin}$
are the derivatives of $\omegax$
and the electrovac horizon function $\Deltax$ at the inner horizon
$x = \xin$.
The $\Deltax^\prime$
in equation~(\ref{GxtxUvsimp})
comes from
$\Yx = \Deltax^\prime$,
equation~(\ref{DDUx}).
The $\omegax^\prime$
in equation~(\ref{GxtphiUvsimp})
comes from replacing $\dd \omegax / \dd x$,
which remains frozen through inflation and collapse,
by its inner horizon value $\omegax^\prime$,
as determined by equation~(\ref{domega}).

Should not the factor
$\Deltax^\prime \pm \vel$
on the right hand side of equation~(\ref{GxtxUvsimp})
be replaced by
$\Deltax^\prime$
in the conformally stationary limit $\vel \rightarrow 0$,
and likewise the factor
$\Deltay \omegax^\prime / \sigma^2 \mp 2 \vel \omegay$
on the right hand side of equation~(\ref{GxtphiUvsimp})
be replaced by
$\Deltay \omegax^\prime / \sigma^2$?
No.
Each equation describes not one but two Einstein components,
and the full expressions given are needed to capture both accurately.
In the first case,
the difference of equations~(\ref{GxtxUvsimp}) gives
$\rho^2 G_{xt} = \vel ( \Ux - \Deltax^\prime ) / ( - \Deltax)$,
and in the second case
the sum of equations~(\ref{GxtphiUvsimp}) gives
$\rho^2 G_{x\phi} = - \vel ( \Deltay \omegax^\prime / \sigma^2 + 2 \omegay \Ux ) / \sqrt{- \Deltax \Deltay}$.

\subsection{Streaming energy-momenta}
\label{streaming}

The form~(\ref{GUvcsimp}) of the Einstein components,
derived under the conditions of conformal stationarity
and conformal separability,
fits to the form
of the energy-momentum tensor $T_{kl}$ of a
collisionless fluid
consisting of two streams, one ingoing ($+)$
and one outgoing ($-$):
\begin{equation}
  T_{kl}
  =
  T^+_{kl}
  +
  T^-_{kl}
  \ .
\end{equation}
The energy-momenta of the ingoing
(positive energy, $p_t < 0$)
and outgoing
(negative energy, $p_t > 0$)
streams are
\begin{equation}
\label{Tpmkl}
  T^\pm_{kl}
  =
  N^\pm p^\pm_k p^\pm_l
  \ ,
\end{equation}
with densities
\begin{equation}
\label{Npm}
  N^\pm
  =
  {1 \over 16\pi}
  ( \Ux \mp \vel )
  ( \Deltax^\prime \pm \vel )
  \ ,
\end{equation}
and tetrad-frame momenta $p^\pm_k$
\begin{equation}
\label{ppm}
  p^\pm_k
  =
  {1 \over \rho}
  \left\{
  -
  {1 \over \sqrt{- \Deltax}}
  \, , \ 
  \mp
  {1 \over \sqrt{- \Deltax}}
  \, , \ 
  {1 \over \sqrt{\Deltay}}
  \left(
  {\Deltay
  \partial \ln \rhosep^2 / \partial y
  \over
  \Deltax^\prime \pm \vel}
  \right)
  \, , \ 
  \mp
  {1 \over \sqrt{\Deltay}}
  \left(
  {\Deltay \omegax^\prime / \sigma^2 \mp 2 \vel \omegay
  \over
  \Deltax^\prime \pm \vel}
  \right)
  \right\}
  \ .
\end{equation}
Equations~(\ref{Npm}) and (\ref{ppm})
are defined up to an arbitrary normalization that leaves $T_{kl}$ constant:
the tetrad-frame momentum $p_k$ and density $N$ can be multiplied
by some constant $\alpha$ and $\alpha^{-2}$ respectively.
The corresponding Hamilton-Jacobi parameters $P^\pm_k$,
equation~(\ref{pktetrad}),
of the collisionless streams are
\begin{equation}
\label{Ppm}
  P^\pm_k
  =
  \left\{
  - 1
  , \,
  \mp 1
  , \,
  {\Deltay
  \partial \ln \rhosep^2 / \partial y
  \over
  \Deltax^\prime \pm \vel}
  , \,
  \mp
  {\Deltay \omegax^\prime / \sigma^2 \mp 2 \vel \omegay
  \over
  \Deltax^\prime \pm \vel}
  \right\}
  \ ,
\end{equation}
again up to an arbitrary normalization factor $\alpha$.
Equations~(\ref{Npm}) and (\ref{ppm})
agree with the behaviour~(\ref{Npmearly})
and (\ref{ppmearly})
of collisionless streams in the regime where
the radial horizon function is small,
\begin{equation}
  | \Deltax | \ll 1
  \ .
\end{equation}

Equations~(\ref{Npm}) and (\ref{ppm})
do not describe correctly the behaviour~(\ref{Npmlate}) and (\ref{ppmlate})
of collisionless streams when the horizon function approaches and exceeds unity,
$| \Deltax | \gtrsim 1$.
Moreover
the tetrad-frame momentum~(\ref{ppm})
predicts that the purely angular components of the collisionless energy-momentum
grow, becoming dominant when $| \Deltax | \gtrsim 1$,
whereas it has been assumed from \S\ref{nonlinearinflationeqs}
up to the present point that the angular components are negligible.
The next two subsections,
\S\ref{angularenergymomenta} and \S\ref{inflationnextorder},
address these issues.

\subsection{Angular energy-momenta imposed by conformal separability}
\label{angularenergymomenta}

As long as conformal separability holds,
the angular components
$G_{yy} - G_{\phi\phi}$ and $G_{y\phi}$
of the Einstein tensor must necessarily vanish,
so that the $2 \times 2$ angular submatrix of the Einstein tensor
must be isotropic
(proportional to the $2 \times 2$ unit matrix).
The reason for this is that the expressions~(\ref{GyphiyUv})
for those Einstein components are functions only of angle $y$,
so the only kind of source of energy-momentum that they admit
is one that depends only on $y$.
By contrast,
the densities $N^\pm$, equation~(\ref{Npm}),
of the collisionless streams arising from inflation
depend strongly on radius through $\Ux$.
The only way that the collisionless streams can
source the angular components of the Einstein tensor
is that their angular energy-momentum tensor must be isotropic.

Of course,
as long as the angular components are sub-dominant,
which is true when $| \Deltax | \ll 1$,
there is no need for the angular components of the Einstein equations
to be satisfied accurately,
because the sub-dominant components
have no effect on the remaining Einstein equations.
This expectation is confirmed explicitly in the next subsection,
\S\ref{inflationnextorder}.

Nevertheless,
as long as $| \Deltax | \ll 1$,
an isotropic collisionless angular energy-momentum tensor
can be accomplished,
by allowing not just one ingoing and one outgoing stream,
but several streams $a$, with densities
\begin{equation}
  N_a^\pm
  \propto
  N^\pm
  \ ,
\end{equation}
that sum to the totals prescribed by equation~(\ref{Npm})
\begin{equation}
  \sum_a N_a^+
  =
  N^+
  \ , \quad
  \sum_a N_a^-
  =
  N^-
  \ .
\end{equation}
Denote the means and mean squares of the angular components
$\mu_y$ and $\mu_\phi$
of the tetrad-frame momenta of the streams,
equation~(\ref{ppmearly}),
by
\begin{subequations}
\begin{gather}
  \langle \mu^\pm_y \rangle
  \equiv
  {\sum_a N_a^\pm
  \mu_{a,y}
  \over N^\pm}
  \ , \quad
  \langle \mu^\pm_\phi \rangle
  \equiv
  {\sum_a N_a^\pm
  \mu_{a,\phi}
  \over N^\pm}
  \ ,
\\
  \langle \mu_y^2 \rangle
  \equiv
  {\sum_a N_a
  \mu_{a,y}^2
  \over
  N^+ + N^-}
  \ , \quad
  \langle \mu_\phi^2 \rangle
  \equiv
  {\sum_a N_a
  \mu_{a,\phi}^2
  \over
  N^+ + N^-}
  \ , \quad
  \langle \mu_y \mu_\phi \rangle
  \equiv
  {\sum_a N_a
  \mu_{a,y}
  \mu_{a,\phi}
  \over
  N^+ + N^-}
  \ ,
\end{gather}
\end{subequations}
in which for the means the sum is over either the ingoing or outgoing stream,
while for the mean squares the sum is over
both ingoing and outgoing streams combined.
The angular components $\mu_y$ and $\mu_\phi$
must average to the values
prescribed by equation~(\ref{ppm}),
\begin{subequations}
\label{mumean}
\begin{align}
  \langle \mu^\pm_y \rangle
  &=
  {1 \over \sqrt{\Deltay}}
  \left(
  {\Deltay
  \partial \ln \rhosep^2 / \partial y
  \over
  \Deltax^\prime \pm \vel}
  \right)
  \ ,
\\
  \langle \mu^\pm_\phi \rangle
  &=
  \mp
  {1 \over \sqrt{\Deltay}}
  \left(
  {\Deltay \omegax^\prime / \sigma^2 \mp 2 \vel \omegay
  \over
  \Deltax^\prime \pm \vel}
  \right)
  \ .
\end{align}
\end{subequations}
The condition of angular isotropy requires that
the mean squares of the angular components are equal,
and the mean of their product is zero,
\begin{equation}
\label{muisotropic}
  \langle \mu_y^2 \rangle
  =
  \langle \mu_\phi^2 \rangle
  =
  \mu^2
  \ , \quad
  \langle \mu_y \mu_\phi \rangle
  =
  0
  \ ,
\end{equation}
for some $\mu^2$.
The angular components of the collisionless energy-momentum
then form the isotropic matrix
\begin{equation}
\label{Tangularisotropic}
  \rho^2 T_{yy}
  =
  \rho^2 T_{\phi\phi}
  =
  ( N^+ + N^- ) \mu^2
  =
  {\Ux \Deltax^\prime - \vel^2 \over 8\pi}
  \mu^2
  \ , \quad
  \rho^2 T_{y\phi}
  =
  0
  \ .
\end{equation}
The Schwarz inequality,
which states that the mean square of a distribution
must exceed the squared mean,
requires that
\begin{equation}
\label{mu2condition}
  \mu^2
  \geq
  \max \left[
  \left(
  {N^+
  \langle \mu_y^+ \rangle
  +
  N^-
  \langle \mu_y^- \rangle
  \over
  N^+ + N^-}
  \right)^2
  \, , \ 
  \left(
  {N^+
  \langle \mu_\phi^+ \rangle
  +
  N^-
  \langle \mu_\phi^- \rangle
  \over
  N^+ + N^-}
  \right)^2
  \right]
  \ .
\end{equation}
While contrived,
there is no difficulty to construct a distribution of collisionless streams
with the required mean momenta~(\ref{mumean})
and the conditions~(\ref{muisotropic}) and (\ref{mu2condition})
on the mean squared momenta.

On the other hand,
the collisionless energy-momentum cannot be contrived
to have isotropic angular components
once $| \Deltax | \gtrsim 1$.
In this regime, as long as conformal separability holds,
the densities $N^\pm$
and tetrad-frame momenta of streams
are given by equations~(\ref{Npmlate}) and (\ref{ppmlate}).
The condition that the angular components be isotropic requires that,
generalizing~(\ref{muisotropic}),
\begin{equation}
\label{muisotropiclate}
  \sum_a N_a
  {\mu_{a,y}^2 \over \sqrt{1 - ( \mu_{a,y}^2 + \mu_{a,\phi}^2 ) \Deltax}}
  =
  \sum_a N_a
  {\mu_{a,\phi}^2 \over \sqrt{1 - ( \mu_{a,y}^2 + \mu_{a,\phi}^2 ) \Deltax}}
  \ .
\end{equation}
If it happened that the means~(\ref{mumean})
of $\mu_y$ and $\mu_\phi$ were the same,
then the isotropy condition~(\ref{muisotropiclate}) could be satisfied
for all values of the horizon function $\Deltax$.
But generically (e.g.\ for Kerr) the mean momenta differ,
and the isotropy condition~(\ref{muisotropiclate})
on the squared momenta cannot hold for all $\Deltax$.

\subsection{Inflation and collapse to next order}
\label{inflationnextorder}

In \S\ref{nonlinearinflationeqs},
equations governing the evolution of the horizon function and conformal factor
were derived from the assumption that
the Einstein components
$G_{xx} - G_{tt}$
and
$G_{yy} + G_{\phi\phi}$
had negligible collisionless source.
It is true that the trace of the collisionless energy-momentum
always remains negligible, equation~(\ref{Tkksingle}).
However,
in \S\ref{streaming} it was found that the angular components
of the collisionless energy-momentum,
though initially negligible, grow, becoming dominant
when $| \Deltax | \gtrsim 1$.
In this subsection,
the angular components are taken into account,
which involves taking the Einstein equations to next order in $\Deltax / \Ux$.
It is found that the earlier results are robust as long as $| \Deltax | \ll 1$,
but fail when angular motions become important,
$| \Deltax | \gtrsim 1$.

A collisionless source for the Einstein components
$G_{xx} - G_{tt}$
and
$G_{yy} + G_{\phi\phi}$
can be accommodated
by admitting source terms for the quantities $\Xx$ and $\Yx$
defined by equations~(\ref{Xx}) and (\ref{Yx}),
in much the same way that electrovac sources
(a radial electromagnetic field and a cosmological constant)
could be accommodated by admitting source terms for
$\Yx$ and $\Yy$.
The sources for $\Xx$ and $\Yx$ are conveniently written
in terms of two arbitrary functions $F_X$ and $F_Y$:
\begin{subequations}
\label{XYxinfcollapse}
\begin{align}
\label{Xxinfcollapse}
  \Xx
  &=
  -
  \Ux
  \left(
  {\partial \ln ( \rhosep / \sigma^2 ) \over \partial x}
  +
  F_X
  \right)
  \ ,
\\
\label{Yxinfcollapse}
  \Yx
  &=
  \Deltax
  \left(
  {\partial \ln \rhosep \over \partial x}
  +
  {2 ( f_0 g_1 {+} f_1 g_0 ) \over \dd \omegax / \dd x}
  -
  F_Y
  \right)
  \ .
\end{align}
\end{subequations}
During early inflation, when both $\Ux$ and $\Deltax$
are negligibly small,
both sets of source terms are negligible.
The source terms are of order $\Deltax / \Ux$ compared
to the principal terms in $\Xx$ and $\Yx$,
and remain sub-dominant through inflation and collapse.
The contributions
${\partial \ln ( \rhosep / \sigma^2 ) / \partial x}$
and
${\partial \ln \rhosep / \partial x}$
to the source terms depend on angle $y$ as well as radius $x$,
breaking separability,
but the breakdown is unimportant as long as
condition~(\ref{sepcondition}) holds,
which is well satisfied through inflation and collapse.

It is convenient to define new quantities $\cXx$ and $\cYx$
that concatenate $\Xx$ and $\Yx$ with their source terms:
\begin{subequations}
\label{cXYxinfcollapse}
\begin{align}
\label{cXxinfcollapse}
  \cXx
  &\equiv
  {\partial \Ux \over \partial x}
  +
  2 {\Ux^2 - \vel^2 \over \Deltax}
  +
  \Ux
  \left(
  {\partial \ln ( \rhosep / \sigma^2 ) \over \partial x}
  +
  F_X
  \right)
  \ ,
\\
  \cYx
  &\equiv
  {\partial \Deltax \over \partial x}
  +
  3 \Ux
  +
  \Deltax
  \left(
  {\partial \ln ( \rhosep^3 / \sigma^4 ) \over \partial x}
  +
  F_Y
  \right)
  \ .
\end{align}
\end{subequations}
The equations governing the evolution of $\Ux$ and $\Deltax$ are then
\begin{subequations}
\label{DDUxc}
\begin{align}
\label{DUxc}
  \cXx
  &=
  0
  \ ,
\\
\label{DDxc}
  \cYx
  &=
  \Deltax^\prime
  \ ,
\end{align}
\end{subequations}
generalizing equations~(\ref{DDUx}).
Equations~(\ref{DDUxc}) constitute the evolution equations for
$\Ux$ and $\Deltax$
taken to next order in $\Deltax / \Ux$.
As long the condition~(\ref{sepcondition})
for the validity of the earlier evolution equations holds,
the evolution of $\Ux$ and $\Deltax$
is essentially unchanged,
and all the results of \S\ref{nonlinearinflation} carry through unchanged.
This is as expected physically:
the evolution of the conformal factor and horizon function
should be essentially unaffected by sub-dominant contributions
to the energy-momentum.

Expressions for the Einstein components
$G_{xx} - G_{tt}$
and
$G_{yy} + G_{\phi\phi}$
in terms of the higher order $\cXx$ and $\cYx$
defined by equations~(\ref{Xxinfcollapse}) and (\ref{Yxinfcollapse})
are
\begin{subequations}
\label{GUvcollapse}
\begin{align}
\label{Gxxmttcollapse}
  \rho^2
  \left( G_{xx} - G_{tt} \right)
  &=
  - \,
  2 \cXx
  -
  2 \cYx
  {\partial \ln \rhosep \over \partial x}
  +
  \Xy
  -
  {\partial \Yy \over \partial y}
  -
  \Yy
  {\partial \over \partial y}
  \ln \left( {\rhosep^2 \over \sigma^4} {\dd \omegay \over \dd y} \right)
  +
  2
  \Ux
  F_X
\nonumber
\\
  & \quad
  + \,
  2
  \Deltax
  \left[
  {1 \over 4}
  \left(
  {1 \over \sigma^2} {\dd \omegay \over \dd y}
  \right)^2
  +
  F_Y
  {\partial \ln \rhosep \over \partial x}
  \right]
  +
  \Uy
  {\partial \over \partial y}
  \ln \left[ {\rhosep^{10} \over \sigma^{20}}
  \left( {\dd \omegay \over \dd y} \right)^3
  \right]
  \ ,
\\
\label{Gyypphiphicollapse}
  \rho^2
  \left(
  G_{yy}
  +
  G_{\phi\phi}
  \right)
  &=
  - \,
  2 \Xy
  +
  2 \Yy
  {\partial \ln \rhosep \over \partial y}
  +
  \cXx
  +
  {\partial \cYx \over \partial x}
  +
  \cYx
  \left[
  {\partial \ln ( \rhosep / \sigma^2 ) \over \partial x}
  -
  F_Y
  \right]
  +
  \Ux
  \left(
  - \,
  F_X
  +
  3 F_Y
  \right)
\nonumber
\\
  & \quad
  + \,
  \Deltax
  \left[
  {3 \over 4}
  \left(
  {1 \over \sigma^2} {\dd \omegay \over \dd y}
  \right)^2
  -
  {\partial F_Y \over \partial x}
  +
  F_Y^2
  +
  F_Y
  {\partial \ln ( \rho^2 / \sigma^2 ) \over \partial x}
  \right]
  -
  2 \Uy
  {\partial \ln ( \rhosep / \sigma^2 ) \over \partial y}
  \ ,
\end{align}
\end{subequations}
while
the Einstein component~(\ref{GxtxUv}) becomes
\begin{equation}
\label{GxtxUvlate}
  \rho^2
  \left(
  {G_{xx} + G_{tt} \over 2}
  \,\pm\,
  G_{xt}
  \right)
  =
  ( \Ux \mp \vel )
  \left(
  {\cYx \pm \vel \over - \Deltax}
  -
  F_X
  +
  F_Y
  \right)
  +
  \cXx
  \,\mp\,
  \vel
  \left(
  {\partial \ln ( \rhosep / \sigma^2 ) \over \partial x}
  +
  F_X
  \right)
  \ ,
\end{equation}
with expressions~(\ref{GxtyUv}), (\ref{GxtphiUv}), and (\ref{GyphiyUv})
remaining unchanged.

During collapse,
when $\Ux$ is large,
the dominant terms contributing to the right
hand sides of equations~(\ref{GUvcollapse})
are those proportional to $\Ux$.
Expressions for the functions $F_X$ and $F_Y$
follow from requiring that the $\Ux$-dependent terms
on the right hand sides of equations~(\ref{GUvcollapse}) are,
in accordance with Einstein's equations,
equal to $8\pi$ times the corresponding collisionless energy-momenta.
For $| \Deltax | \ll 1$,
the relevant collisionless energy-momenta are
the isotropic angular components~(\ref{Tangularisotropic}),
and a negligible collisionless trace, $T^k_k = 0$.
The resulting functions $F_X$ and $F_Y$ are
\begin{equation}
\label{FXY}
  F_X
  =
  F_Y
  =
  \left( \Deltax^\prime - {\vel^2 \over \Ux} \right)
  \mu^2
  \ .
\end{equation}
The functions are of order unity,
and thus, as anticipated, their inclusion has negligible effect on the
evolution equations~(\ref{DDUxc}) as long as condition~(\ref{sepcondition})
is true,
which as shown in \S\ref{nonlinearinflation}
is well satisfied through inflation and collapse.

If the functions $F_X$ and $F_Y$ from equations~(\ref{FXY})
are inserted into the right hand side of
equation~(\ref{GxtxUvlate}),
then the result agrees with the earlier expression~(\ref{GxtxUv})
except that the sub-dominant term proportional to
$\dd \ln ( \dd \omegax / \dd x ) / \dd x$
in the earlier expression disappears.
The net result is that, when the conformally stationary limit is taken,
the same set of expressions~(\ref{GUvcsimp})
is obtained for the Einstein components as found previously.
This confirms that
taking into account the purely angular components of the collisionless
energy-momentum has essentially no effect on the solution,
as long $| \Deltax | \ll 1$.

On the other hand,
as already discussed in \S\ref{angularenergymomenta},
once $| \Deltax | \gtrsim 1$,
generically (e.g.\ for Kerr)
the angular components of the collisionless energy-momentum
can no longer be arranged to be isotropic,
equation~(\ref{muisotropiclate}),
so the conformally separable Einstein equations cannot
be satisfied with a collisionless source.
Even if it could be arranged that the angular components were isotropic,
once $| \Deltax | \gtrsim 1$
the situation is further complicated by the fact that
$\dd \omegax / \dd x$,
which up to this point has been frozen at its electrovac value,
must also change in order to satisfy the Einstein equation for $G_{t\phi}$.
Further discussion is unwarranted in this paper.

In conclusion,
the solution found earlier in this section,
up to \S\ref{streaming},
holds as long as $| \Deltax | \ll 1$,
but breaks down when
the angular motions of the collisionless
streams exceed their radial motions,
which happens when $| \Deltax | \gtrsim 1$.
What happens after angular motions become important is undetermined.

\section{Mass and curvature}
\label{masscurvature}

\subsection{Mass inflation}
\label{massinflation}

The term mass inflation
comes from the fact that in charged spherically symmetric models
of black holes,
the interior mass, or Misner-Sharp \cite{Misner:1964} mass,
inflates exponentially.
%
In spherical black holes,
the interior mass $M$ can be defined by
\begin{equation}
  {2 M \over r} - 1
  \equiv
  - 
  \partial^m r
  \,
  \partial_m r
  \ ,
\end{equation}
where $r$ is the circumferential radius,
which plays the role of a conformal factor,
and $\partial_m$ is the directed derivative
in any tetrad frame.
An analogous scalar quantity in the rotating black holes
considered in the present paper is
\begin{equation}
\label{Mass}
  {2 M \over \rho} - 1
  \equiv
  \Mass
  =
  \Mass_x
  +
  \Mass_y
  =
  -
  \partial^m \rho
  \,
  \partial_m \rho
  \ ,
\end{equation}
where $\Mass_x$ and $\Mass_y$
are dimensionless radial and angular mass parameters defined by
\begin{subequations}
\begin{align}
\label{Massx}
  \Mass_x
  \equiv
  ( \partial_x \rho )^2
  -
  ( \partial_t \rho )^2
  &=
  -
  {1 \over \Deltax}
  \left[
  \left( {\partial \ln \rho \over \partial x} \Deltax \right)^2
  -
  \vel^2
  \right]
  \ ,
\\
\label{Massy}
  \Mass_y
  \equiv
  - \,
  ( \partial_y \rho )^2
  -
  ( \partial_\phi \rho )^2
  &=
  -
  {1 \over \Deltay}
  \left[
  \left( {\partial \ln \rho \over \partial y} \Deltay \right)^2
  +
  \vel^2 \omegay^2
  \right]
  \ .
\end{align}
\end{subequations}
The mass parameter proposed by
\cite{Barrabes:1990}
for rotating black holes
is the radial mass parameter $\Mass_x$.
In the Kerr-Newman geometry ($\Lambda = 0$),
the interior mass $M$ defined by equation~(\ref{Mass})
goes over to the black hole mass $\Mbh$ far from the black hole,
\begin{equation}
  M
  \rightarrow
  \Mbh
  \quad
  \mbox{as $r \rightarrow \infty$}
  \ .
\end{equation}
In terms of the quantities
$\Ux$ and $\Uy$ defined by equations~(\ref{Uxy}),
the dimensionless mass parameters are
\begin{subequations}
\label{MassU}
\begin{align}
\label{MassUx}
  \Mass_x
  &=
  -
  {1 \over \Deltax}
  \left[
  \left( {\partial \ln \rhosep \over \partial x} \Deltax + \Ux \right)^2
  -
  \vel^2
  \right]
  \ ,
\\
\label{MassUy}
  \Mass_y
  &=
  -
  {1 \over \Deltay}
  \left[
  \left( {\partial \ln \rhosep \over \partial y} \Deltay - \Uy \right)^2
  +
  \vel^2 \omegay^2
  \right]
  \ .
\end{align}
\end{subequations}


During the electrovac phase prior to inflation,
the radial mass parameter $\Mass_x$
is proportional to the horizon function $\Deltax$,
\begin{equation}
  \Mass_x
  =
  -
  \left( {\partial \ln \rhosep \over \partial x} \right)^2
  \Deltax
  \ ,
\end{equation}
which approaches some small value near the inner horizon
$\Deltax \rightarrow -0$.
The radial mass parameter $\Mass_x$
given by equation~(\ref{MassUx}) reaches a minimum,
signalling the start of inflation,
when
(here $\Ux$ equals its initial value $\uel$)
\begin{equation}
  {\partial \ln \rhosep \over \partial x} \Deltax
  =
  \sqrt{ \uel^2 - \vel^2}
  \ .
\end{equation}
At this point the streaming energy and pressure in the no-going tetrad frame
are comparable to unity in natural black hole units,
$c = G = \Mbh = 1$,
while the mass parameter is small, of order $\Mass_x \sim \vel$.
Once inflation gets going, the mass parameter is
\begin{equation}
  \Mass_x
  \approx
  {\Ux^2 - \vel^2 \over - \Deltax}
  \ ,
\end{equation}
which can be recognized as the principal term driving the evolution of $\Ux$,
equation~(\ref{DUx}).
During inflation the radial mass parameter increases exponentially.

During inflation and collapse,
the ratio of the mass parameter $\Mass_x$ to the
streaming energy $T_{xx}$ in the no-going frame is
\begin{equation}
\label{MassT}
  {\Mass_x \over 8\pi T_{xx}}
  =
  \rho^2
  {\Ux^2 - \vel^2 \over \Ux \Deltax^\prime}
  =
  \rhosep^2
  {\sqrt{\uel^2 - \vel^2} \over \Deltax^\prime}
  \sqrt{1 - {\vel^2 \over \Ux^2}}
  \ ,
\end{equation}
which remains always of order $\vel$,
increasing mildly as $\Ux$ increases from $\uel$
to some large value.

After the horizon function has gone through its extremum,
it increases in absolute value as $\Deltax \propto \Ux^{3/2}$,
so the dimensionless mass parameter varies as $\Mass_x \propto \Ux^{1/2}$,
or equivalently as $\Mass_x \propto 1 / \rho$.
The interior mass $M$ defined by equation~(\ref{Mass})
thus goes to a (huge) constant during collapse,
consistent with behaviour of the interior mass during collapse
in charged spherical black holes,
\S4.3 of \cite{Hamilton:2008zz}.

\subsection{Weyl curvature}
\label{weylcurvature}

The only non-vanishing component of the Weyl tensor is the
complex spin-$0$ component.
The fact that only the spin-$0$ component is non-zero
defines the spacetime as Petrov type D.
Subject to the conditions of
conformal time invariance
(not necessarily conformal stationarity)
and conformal separability
assumed in this paper,
the polar (real) part of spin-$0$ Weyl component,
equations~(\ref{CWeyl})--(\ref{CWeylpolarxy}),
may be written in terms of the quantities
$\Uy$, $\Xy$, and $\Yy$
defined by equations~(\ref{Uy}), (\ref{Xy}), and (\ref{Yy}),
as well as the mass parameter $\Mass$ defined by equation~(\ref{Mass}),
as
\begin{align}
  \rho^2 C^{(p)}
  &=
  \rho^2 \left(
  \frac{1}{6}
  G_{xx}
  -
  \frac{1}{6}
  G_{tt}
  +
  \frac{1}{12}
  G_{yy}
  +
  \frac{1}{12}
  G_{\phi\phi}
  \right)
  -
  \frac{1}{2}
  \Mass
  -
  \frac{3}{4}
  \left( {1 \over \sigma^2} {\dd \omegay \over \dd y} \right)^2
  \Deltax  
  -
  \frac{3}{4}
  \left( {1 \over \sigma^2} {\dd \omegax \over \dd x} \right)^2
  \Deltay
\nonumber
\\
  &\quad
  + \,
  \frac{1}{2}
  \Xy
  +
  \frac{1}{2}
  {\partial \Yy \over \partial y}
  +
  \frac{1}{2}
  \Yy
  {\partial \over \partial y}
  \ln
  \left(
  {1 \over \sigma^4}
  {\dd \omegay \over \dd y}
  \right)
  +
  \frac{1}{2}
  \Uy
  {\partial \over \partial y}
  \ln
  \left[
  {\rhosep^4 \over \sigma^8}
  \left( {\dd \omegay \over \dd y} \right)^3
  \right]
  \ ,
\end{align}
which is valid throughout the electrovac,
inflationary, and collapse regimes.
Since $\Uy$ and $\Xy$ are negligible
in the conformally stationary limit,
the polar spin-$0$ Weyl component reduces to
\begin{align}
\label{CWeylinf}
  \rho^2 C^{(p)}
  &=
  \rho^2 \left(
  \frac{1}{6}
  G_{xx}
  -
  \frac{1}{6}
  G_{tt}
  +
  \frac{1}{12}
  G_{yy}
  +
  \frac{1}{12}
  G_{\phi\phi}
  \right)
  -
  \frac{1}{2}
  \Mass
  -
  \frac{3}{4}
  \left( {1 \over \sigma^2} {\dd \omegay \over \dd y} \right)^2
  \Deltax  
  -
  \frac{3}{4}
  \left( {1 \over \sigma^2} {\dd \omegax \over \dd x} \right)^2
  \Deltay
\nonumber
\\
  &\quad
  + \,
  \frac{1}{2}
  {\partial \Yy \over \partial y}
  +
  \frac{1}{2}
  \Yy
  {\partial \over \partial y}
  \ln
  \left(
  {1 \over \sigma^4}
  {\dd \omegay \over \dd y}
  \right)
  \ .
\end{align}
I have not been able to find any enlightening expression
for the axial (imaginary) spin-$0$ component of Weyl tensor,
beyond that already given as equation~(\ref{CWeylaxial}).
%
%

During inflation and collapse,
the Weyl curvature is dominated by the mass term,
\begin{equation}
  \rho^2 C
  \approx
  - \frac{1}{2} \Mass
  \ .
\end{equation}

\section{Boundary conditions}

In \S\ref{inflation}
it was found that a conformally stationary, conformally separable
solution exists for the interior structure of a rotating black hole,
from electrovac through inflation to collapse,
in which the Einstein equations are sourced by the energy-momentum
of ingoing and outgoing collisionless streams.
Because the accretion rate is asymptotically tiny,
the streams have negligible energy-momentum above the inner horizon,
and therefore have no effect on the geometry above the inner horizon.
From the point of view of boundary conditions,
what is important is the form of the collisionless streams
incident on the inner horizon.

\subsection{Density of collisionless streams incident on the inner horizon}
\label{radialbcs}

The most important boundary conditions are those on the radial ($x$-$t$)
components of the collisionless energy-momentum incident on the inner horizon.
The radial components of the energy-momentum
of ingoing ($+$) and outgoing ($-$) collisionless
during inflation and collapse are,
equations~(\ref{Npm}) and (\ref{ppm}),
\begin{equation}
  T^\pm_{xx} = \pm T^\pm_{xt} = T^\pm_{tt}
  =
  {N^\pm
  \over \rho^2 | \Deltax |}
  \ .
\end{equation}
The initial values of these components
are set by the densities $N^\pm$, equation~(\ref{Npm}),
of the ingoing and outgoing collisionless streams
incident on the inner horizon,
which are, since $\Ux = \uel$ initially,
\begin{equation}
\label{Npminit}
  N^\pm
  =
  {1 \over 16\pi}
  ( \uel \mp \vel )
  ( \Deltax^\prime \pm \vel )
  \ .
\end{equation}
The densities $N^\pm$ of incident ingoing and outgoing streams
are proportional to $\uel \mp \vel$ respectively.
Both ingoing and outgoing streams must be present for inflation to occur,
so both densities must be strictly positive.
Moreover, at least classically, the black hole must expand as it accretes,
so $\vel$ must be positive.
Thus the accretion rates $\uel$ and $\vel$,
both of which are to be considered small,
must satisfy
\begin{equation}
\label{ugtv}
  \uel > \vel > 0
  \ .
\end{equation}
Positive $\vel$ implies that in the no-going tetrad frame
the density of the outgoing stream exceeds that of the ingoing stream
during inflation and collapse.
Equivalently,
the center-of-mass frame is outgoing.
An outgoing density that exceeds the ingoing density
ensures that the black hole's angular momentum,
which in self-similar solutions is determined by the
angular momentum of the accreted streams,
is positive.
The case $\vel = 0$ corresponds to equal ingoing and outoing streams,
which is the stationary (or homogeneous) approximation first applied
to inflation by
\cite{Burko:1997xa,Burko:1998az,Burko:1998jz}.

Equation~(\ref{Npminit}) prescribes that the densities $N^\pm$
of incident ingoing and outgoing streams
must be uniform, independent of latitude $y$.
In other words, the accretion flow must be ``monopole'' in density.
It makes physical sense that the condition of conformal separability
would require a high degree of symmetry of the incident accretion flow.

The small accretion rate parameters $\uel$ and $\vel$
completely characterize the inflationary solution.
The evolution of the inflationary exponent $\xi$ and horizon function $\Deltax$
found in \S\ref{nonlinearinflation} is determined entirely by these
parameters, along with the derivative $\Deltax^\prime$ of the
electrovac horizon function at the inner horizon.

The positivity of the ingoing and outgoing densities $N^\pm$,
equation~(\ref{Npminit}),
requires also that $\Deltax^\prime \pm \vel$ be positive.
This means that the inflationary solutions do not apply to
extremal black holes, whose inner horizons coincide with their outer horizons,
and for which $\Deltax^\prime = 0$.

\subsection{Angular motion of collisionless streams incident on the inner horizon}

The angular components of the momenta of the collisionless streams
are sub-dominant during inflation and collapse.
As seen in \S\ref{inflationnextorder},
the evolution of the inflationary exponent $\xi$ and horizon function $\Deltax$,
and of the radial components of the collisionless energy-momentum,
are unaffected by angular motions until angular motions become important,
at which point the solution fails,
which happens when the geometry has collapsed to exponentially tiny scale.
As far as the radial solution is concerned,
a sufficient condition for conformal separability to hold
is the condition~(\ref{Npminit})
on the density of the accretion flow.
If however the Einstein equations are required to hold also
for the sub-dominant radial-angular components of the collisionless
energy-momentum,
which is a more stringent constraint on conformal separability,
then the angular components of the tetrad-frame momenta
of the ingoing and outgoing collisionless streams must be as given
by equation~(\ref{ppm}).
The corresponding Hamilton-Jacobi parameters $P^\pm_k$ are
given by equation~(\ref{Ppm}).
The condition~(\ref{Ppm})
prescribes that the angular components
of the Hamilton-Jacobi parameters
$P^\pm_k$
do not vanish,
but rather vary in the given fashion with latitude $y$.

Equation~(\ref{Ppm}) requires that the collisionless streams
have some net motion in the angular $y$ direction.
One might perhaps have anticipated that conformally stationary flow
might require that each particle in the collisionless stream
fall along a surface of constant latitude $y$,
which would happen if $\Py$ for each particle were identically zero.
Trajectories at constant $y$ occur when $\Py$ is not only zero
but also an extremum with respect to variation of $x$ or $y$
at fixed constants of motion $E$, $L$, $\KCarter$.
In fact equation~(\ref{Ppm}) shows that the required $\Py$ is not zero.

\ingoingconditionfig

\subsection{The angular boundary conditions cannot be achieved with collisionless streams accreted from outside the outer horizon}
\label{fail}

Can the relations~(\ref{Ppm})
governing the angular motions of streams incident on the inner horizon
be accomplished by collisionless streams that accrete from outside the horizon?
As will now be shown, the answer is no.

Equation~(\ref{Ppm})
shows that the ratio $P^\pm_t / P^\pm_\phi$
incident on the inner horizon
for each of the ingoing ($+$) and outgoing ($-$) streams
is
\begin{equation}
\label{Ptphipm}
  {P^\pm_t \over P^\pm_\phi}
  =
  {\Deltax^\prime \pm \vel \over \Deltay \omegax^\prime / \sigma^{2} \mp 2 \vel \omegay}
  \ .
\end{equation}
If the ingoing and outgoing streams contain multiple components,
then the ratio~(\ref{Ptphipm})
is a density-weighted average.
If one insists that every particle originate from outside the outer horizon,
then $P_t$ must be negative at the outer horizon for every particle.
The relations~(\ref{Pk})
with vanishing electromagnetic potential
(the case of a charged black hole is deferred to Paper~3)
then imply that
$P_t$ and $P_\phi$ of the particle at the inner horizon
must satisfy the inequality
\begin{equation}
\label{Ptcondition}
  P_t
  \,\leq\,
  P_\phi
  {\omegaxin - \omegaxout \over 1 - \omegaxout \omegayin}
  \ ,
\end{equation}
where the subscripts ${\rm in}$ and ${\rm out}$ denote values
respectively at the inner and outer horizon.
The condition~(\ref{Ptcondition}) excludes
half of the $P_t$--$P_\phi$ plane.
As illustrated in Figure~\ref{ingoingcondition},
the right hand sides of equations~(\ref{Ptphipm}) and (\ref{Ptcondition})
have different dependences on latitude.
While the inequality~(\ref{Ptcondition}) may be satisfied
at any latitude
by either the ingoing or the outgoing stream,
the inequality cannot be satisfied simultaneously by
both ingoing and outgoing streams
at that latitude.

\subsection{Dispersion of angular motions incident on the inner horizon}

The purely angular components of the collisionless energy-momentum
are sub-sub-dominant.
During inflation, the angular components are negligible
because the incident accretion flow is by assumption tiny,
and inflation does not amplify the angular components of the
momenta of the streams.
During collapse, the angular components of the momentum grow
faster than the radial components, eventually becoming comparable
to the radial components, at which point the solution fails.
As argued in \S\ref{angularenergymomenta},
if the Einstein equations are required to hold also
for the sub-sub-dominant angular components of the collisionless
energy-momentum,
then conformal separability requires that the angular energy-momentum
be isotropic (proportional to the unit $2 \times 2$ matrix).
If the angular energy-momentum is initially isotropic,
then the behaviour of freely-falling streams ensures that
it will remain so during inflation and collapse.

An isotropic angular energy-momentum incident on the inner horizon
can be contrived by allowing multiple ingoing and outgoing streams
whose angular components of momentum
satisfy condition~(\ref{Ppm}) in the mean,
but are isotropic in their mean squares,
which can be arranged.

It is apparent that the required boundary conditions become more special
and more contrived as the condition of conformal separability is tightened.
To satisfy the radial components of the Einstein equations,
only condition~(\ref{Npminit}) on the incident densities $N^\pm$ is required.
To satisfy the sub-dominant radial-angular components of the Einstein equations,
condition~(\ref{Ppm}) on the angular components of the momenta is required.
To satisfy the sub-sub-dominant angular components of the Einstein equations,
the angular energy-momentum tensor must be isotropic.

\section{Conclusions}

This paper has presented details of a new set of
conformally stationary, axisymmetric, conformally separable solutions
for the interior structure of a rotating uncharged black hole
that accretes a collisionless fluid.
The solutions are generalized to charged black holes in Paper~3
\cite{Hamilton:2010c}.
A concise derivation appears in Paper~1
\cite{Hamilton:2010a}.

The solutions confirm explicitly,
for the special case of conformal separability,
that inflation develops in rotating black holes
as anticipated by
\cite{Barrabes:1990}.

Conformal stationarity means that the black hole accretes
steadily at an asymptotically tiny rate,
in such a way that the spacetime expands conformally
(self-similarly), without change of shape.
The solutions presented should be a good approximation,
in the sense of perturbation theory,
when the accretion rate is small,
and should become increasingly accurate
as the accretion rate tends to zero.


The most important equations in this paper are equations~(\ref{GUv})
for two of the diagonal Einstein components,
which have negligible collisionless source of energy-momentum.
These equations hold over the entire regime of interest,
from electrovac through inflation and collapse.
The equations can be separated, and their solution seamlessly yields
both electrovac and inflationary solutions.
The equations recover several known physical features of inflation:
inflation occurs at the inner horizon but not at the outer horizon;
inflation requires the simultaneous presence of both ingoing and
outgoing streams;
and the smaller the accretion rate, the more violently inflation exponentiates.

A central feature of separable electrovac ($\Lambda$-Kerr-Newman) solutions
is that as ingoing ($p_t < 0$) and outgoing ($p_t > 0$)
collisionless streams approach the inner horizon,
they concentrate into narrow, intense beams
focussed along the ingoing and outgoing principal null directions,
regardless of the initial angular motion of the streams.
If there were no back-reaction on the electrovac geometry,
then the streams would exceed the speed of light relative to each other
and fall through distinct ingoing and outgoing inner horizons
into two causally separated regions of spacetime.
In reality,
the energy and pressure of the counter-streaming ingoing and outgoing beams
builds to the point that it becomes a significant source of gravity,
however tiny the accretion rate.
The gravity of the streaming energy-momentum acts so as to accelerate
the ingoing and outgoing beams even faster through each other
along the principal null directions.
The result is the inflationary instability,
in which the energy-momentum of the hyper-relativistically
counter-streaming beams grows,
along with the Weyl curvature and mass parameter,
to exponentially huge values,
while the radial horizon function decreases exponentially.

At an exponentially tiny value,
the horizon function goes through a minimum (in absolute value),
whereupon the spacetime collapses,
the conformal factor shrinking to an exponentially tiny scale.
This is consistent with the argument of \cite{Hamilton:2008zz}
that inflation leads to collapse, not a null singularity,
if the black hole continues to accrete,
as is ensured in the present solutions
by the presumption of conformal time-translation invariance
(self-similarity).

During collapse, the angular motion of the infalling streams grows
(the tetrad is chosen to align with the principal frame,
so the angular directions are orthogonal to the principal null directions).
The solutions presented here break down
when the angular motion of the streams
becomes comparable to the radial motion,
but this occurs only when the spacetime has collapsed to an
exponentially tiny scale.
That the solutions fail when angular motions become large
is consistent with the physical idea that conformal separability
can persist only so long as the energy-momentum is focussed
along the radial direction.
The result is also consistent with the slowly rotating black hole
solutions of \cite{Hamilton:2009hu}.

Until angular motions become important,
the radial components of the collisionless energy-momentum
dominate the angular components, satisfying the hierarchy of inequalities
\begin{equation}
  | T_{xx} | \gg | T_{xy} | \gg | T_{yy} |
  \ ,
\end{equation}
where $x$ signifies radial ($x$-$t$) directions,
and $y$ angular ($y$-$\phi$) directions.
The radial Einstein equations are unaffected
by the sub-dominant radial-angular Einstein equations,
which in turn are unaffected by the sub-sub-dominant angular Einstein equations.

Conformal separability imposes a corresponding hierarchy of boundary conditions
on the collisionless ingoing and outgoing streams incident on the inner horizon.
The indispensible boundary condition is set by requiring
that the radial Einstein equations be satisfied.
Conformal separability
requires that the accretion flow be ``monopole''
in the sense that the densities of ingoing and outgoing streams must be
independent of latitude, equation~(\ref{Npminit}).
It makes physical sense that conformal separability would require
this high degree of symmetry on the incident accretion flow.
If the accretion rate were different at different latitudes,
then the spacetime would collapse faster where the accretion rate is higher,
destroying the symmetry.

The evolution of the conformal factor and radial horizon function
during inflation and collapse is determined
by two small constant parameters $\uel$ and $\vel$
set by the incident densities of ingoing and outgoing streams,
which are proportional to $\uel \mp \vel$.
Positivity of both densities, coupled with the requirement that
the black hole expand as it accretes, imposes $\uel > \vel > 0$.
The case $\vel = 0$ corresponds to the stationary (homogeneous) approximation of
\cite{Burko:1997xa,Burko:1998az,Burko:1998jz}.

If the sub-dominant radial-angular Einstein equations
are required to be satisfied, then conformal separability
requires that the angular components of the momenta of the
incident ingoing and outgoing streams have Hamilton-Jacobi parameters
satisfying equation~(\ref{Ppm}).

A limitation of the required angular motions is that they cannot be accomplished
by collisionless streams that fall freely from outside the outer horizon.
Particles that fall from outside are necessarily ingoing at the
outer horizon.
This condition excludes half the phase space available
to freely falling particles,
and makes it impossible to fulfill the required angular conditions
at the inner horizon.
Thus, if the angular conditions are imposed,
then the ingoing and outgoing streams must be regarded as being delivered
ad hoc to just above the inner horizon.

If the sub-sub-dominant purely angular Einstein equations
are required to be satisfied,
then conformal separability requires that the mean squares of the
angular components of the collisionless momenta be isotropic,
which can be contrived.

\begin{acknowledgements}
I thank Gavin Polhemus for numerous conversations
that contributed materially to the development of the ideas herein,
and Carlos Herdeiro
for bringing attention to the possibility of a conformal Killing tensor.
This work was supported by NSF award
AST-0708607.
\end{acknowledgements}

\section*{References}

\bibliographystyle{unsrt}
\bibliography{bh}

\appendix

\section{Reduction of the line-element}
\label{reduction}

This Appendix shows how the assumptions of
conformally stationarity, axisymmetry, and conformal separability
lead to the line-element~(\ref{lineelement})
and associated vierbein~(\ref{doranvierbein})
adopted in the text.

By conformal separability is meant conditions on the vierbein
$e_m{}^\mu$
and electromagnetic potential
$A_m$
that follow from assuming that
the action $S$ governing the free motion
of neutral or charged particles separates
as a sum of terms each depending only on a single coordinate,
equation~(\ref{Ssep}),
and that
the left hand side of the resulting Hamilton-Jacobi equation,
after multiplication by an arbitrary overall factor,
separates
``in the simplest possible way,''
assumption III of \cite{Carter:1968c},
as a sum of terms
each depending only on either radius $x$ or angle $y$.

Conformal separability differs from separability
in that the resulting Hamilton-Jacobi equation separates exactly
only for massless particles, $m = 0$.
Consequently the spacetime has a conformal Killing tensor.


The assumptions of conformal stationarity and axisymmetry
imply that the canonical momenta $\pi_t$ and $\pi_\phi$ are constants,
equations~(\ref{knpitphi}).
The assumption of conformal separability implies that the canonical momenta
$\pi_x = \partial S / \partial x$
and
$\pi_y = \partial S / \partial y$
are functions only of $x$ and $y$ respectively.

Let
$\hat{e}_m{}^\mu$
and
$\hat{A}_m$
denote the vierbein coefficients
and tetrad-frame electromagnetic potential
with an overall conformal factor $\rho$ factored out:
\begin{equation}
  \hat{e}_m{}^\mu
  \equiv
  \rho 
  e_m{}^\mu
  \ , \quad
  \hat{A}_m
  \equiv
  \rho 
  A_m
  \ .
\end{equation}
The conformal factor $\rho$
here could be any arbitrary function of all the coordinates.
In terms of the scaled vierbein
$\hat{e}_m{}^\mu$
and
electromagnetic potential
$\hat{A}_m$,
the Hamilton-Jacobi equation~(\ref{HamiltonJacobi})
is
\begin{equation}
\label{HamiltonJacobihat}
  \eta^{mn}
  \left( \hat{e}_m{}^\mu \pi_\mu - q \hat{A}_m \right)
  \left( \hat{e}_n{}^\nu \pi_\nu - q \hat{A}_n \right)
  =
  -
  m^2
  \rho^2
  \ .
\end{equation}
Following assumption III of \cite{Carter:1968c},
assume that the left hand side of equation~(\ref{HamiltonJacobihat})
separates
in the simplest possible way
as a sum of terms each of which depends only on $x$ or only on $y$.
Since $\pi_t$ and $\pi_\phi$ are constants,
while
$\pi_x$ and $\pi_y$ are respectively functions of $x$ and $y$,
the assumption implies that
\begin{equation}
\label{conditioneA}
  \mbox{for each $m$,~}
  \left\{
  \begin{array}{cl}
  \mbox{either}
  &
  \mbox{
  $\hat{e}_m{}^\mu$
  for all $\mu$,
  and
  $\hat{A}_m$,
  are functions of $x$ only,
  and
  $\hat{e}_m{}^y = 0$
  ,
  }
  \\
  \mbox{or}
  &
  \mbox{
  $\hat{e}_m{}^\mu$
  for all $\mu$,
  and
  $\hat{A}_m$,
  are functions of $y$ only,
  and
  $\hat{e}_m{}^x = 0$
  .
  }
  \end{array}
  \right.
\end{equation}
The case that matches the $\Lambda$-Kerr-Newman geometry,
which provides the boundary conditions
for the inflationary solutions considered in this paper
and its companions, is
\begin{equation}
\label{conditioneAkn}
  \mbox{the~}
  \left\{
  \begin{array}{l}
  \mbox{top}
  \\
  \mbox{bottom}
  \end{array}
  \right\}
  \mbox{~condition of (\ref{conditioneA}) holds for~}
  \left\{
  \begin{array}{l}
  m = x \mbox{~and~} t
  \\
  m = y \mbox{~and~} \phi
  \end{array}
  \right\}
  \ .
\end{equation}
Thus conformal separability
``in the simplest possible way''
consistent with
$\Lambda$-Kerr-Newman
requires that
\begin{equation}
\label{exyzero}
  \hat{e}_x{}^y
  =
  \hat{e}_t{}^y
  =
  \hat{e}_y{}^x
  =
  \hat{e}_\phi{}^x
  =
  0
  \ .
\end{equation}

Given the conformal separability conditions~(\ref{exyzero}),
the vierbein coefficients
$\hat{e}_t{}^x$
and
$\hat{e}_\phi{}^y$
can be transformed to zero
by a tetrad gauge transformation consisting of
a Lorentz boost by velocity $\hat{e}_t{}^x / \hat{e}_x{}^x$
between tetrad axes $\bgamma_x$ and $\bgamma_t$,
and a (commuting) spatial rotation by angle
$\tan^{-1} ( \hat{e}_\phi{}^y / \hat{e}_y{}^y )$
between tetrad axes $\bgamma_y$ and $\bgamma_\phi$.
Thus without loss of generality,
\begin{equation}
\label{etxzero}
  \hat{e}_t{}^x
  =
  \hat{e}_\phi{}^y
  =
  0
  \ .
\end{equation}
The gauge conditions~(\ref{etxzero})
having been effected,
the vierbein coefficients
$\hat{e}_x{}^t$,
$\hat{e}_y{}^t$,
$\hat{e}_x{}^\phi$,
and
$\hat{e}_y{}^\phi$
can be eliminated
by coordinate gauge transformations
$t \rightarrow t^\prime$
and
$\phi \rightarrow \phi^\prime$
defined by
\begin{equation}
\label{tphitransform}
  \dd t
  =
  \dd t^\prime
  +
  {\hat{e}_x{}^t \over \hat{e}_x{}^x}
  \, \dd x
  +
  {\hat{e}_y{}^t \over \hat{e}_y{}^y}
  \, \dd y
  \ , \quad
  \dd \phi
  =
  \dd \phi^\prime
  +
  {\hat{e}_x{}^\phi \over \hat{e}_x{}^x}
  \, \dd x
  +
  {\hat{e}_y{}^\phi \over \hat{e}_y{}^y}
  \, \dd y
  \ .
\end{equation}
Equations~(\ref{tphitransform})
are integrable because
$\hat{e}_x{}^\mu$
and
$\hat{e}_y{}^\mu$
are respectively functions of $x$ and $y$ only.
The transformations~(\ref{tphitransform}) of $t$ and $\phi$
are admissible because they preserve the Killing vectors
$\partial / \partial t$
and
$\partial / \partial \phi$,
\begin{equation}
  \left.
  {\partial \over \partial t}
  \right|_{x, y, \phi}
  =
  \left.
  {\partial \over \partial t^\prime}
  \right|_{x, y, \phi}
  \ , \quad
  \left.
  {\partial \over \partial \phi}
  \right|_{x, t, y}
  =
  \left.
  {\partial \over \partial \phi^\prime}
  \right|_{x, t, y}
  \ .
\end{equation}
Thus without loss of generality
\begin{equation}
\label{extzero}
  \hat{e}_x{}^t
  =
  \hat{e}_y{}^t
  =
  \hat{e}_x{}^\phi
  =
  \hat{e}_y{}^\phi
  =
  0
  \ .
\end{equation}
Finally,
coordinate transformations of the $x$ and $y$ coordinates
\begin{equation}
  x
  \rightarrow
  x^\prime
  \ , \quad
  y
  \rightarrow
  y^\prime
  \ ,
\end{equation}
can be chosen such that
$\hat{e}_x{}^x$
and
$\hat{e}_y{}^y$
satisfy
\begin{equation}
\label{exxone}
  \hat{e}_x{}^x
  \hat{e}_t{}^t
  =
  \hat{e}_y{}^y
  \hat{e}_\phi{}^\phi
  =
  1
  \ .
\end{equation}
The conformal separability conditions~(\ref{conditioneA})
with~(\ref{conditioneAkn}),
which imply conditions~(\ref{exyzero}),
coupled with the gauge conditions~(\ref{etxzero}), (\ref{extzero}),
and (\ref{exxone}),
lead to the
line-element~(\ref{lineelement})
and
vierbein~(\ref{doranvierbein})
adopted in this paper.

\section{Tetrad-frame connections, Einstein and Weyl tensors}
\label{einsteingeneral}

This Appendix gives expressions for the tetrad-frame connections,
and Einstein and Weyl tensors,
in the case that the conformal separability conditions~(\ref{fnxy}) hold,
and the conformal factor $\rho$ is any arbitrary function
not only of $x$ and $y$, but also of $t$ and $\phi$.
There are
18 non-vanishing tetrad-frame connections
$\Gamma_{klm}$
(counting 
$\Gamma_{klm} = - \Gamma_{lkm}$
as one),
comprising 8 distinct connections,
\begin{subequations}
\begin{align}
  \Gamma_{yxy}
  =
  \Gamma_{\phi x\phi}
  &=
  \partial_x \ln \rho
  \ ,
\\
  \Gamma_{txx}
  =
  \Gamma_{yty}
  =
  \Gamma_{\phi t\phi}
  &=
  \partial_t \ln \rho
  \ ,
\\
  \Gamma_{yxx}
  =
  \Gamma_{tyt}
  &=
  \partial_y \ln \rho
  \ ,
\\
  \Gamma_{\phi xx}
  =
  \Gamma_{t\phi t}
  =
  \Gamma_{y\phi y}
  &=
  \partial_\phi \ln \rho
  \ ,
\\
  \Gamma_{xt\phi}
  =
  \Gamma_{x\phi t}
  =
  \Gamma_{\phi tx}
  &=
  {\sqrt{\Deltay} \over 2 \rho \sigma^2}
  {\dd \omegax \over \dd x}
  \ ,
\\
  \Gamma_{yt\phi}
  =
  \Gamma_{y\phi t}
  =
  \Gamma_{t\phi y}
  &=
  {\sqrt{- \Deltax} \over 2 \rho \sigma^2}
  {\dd \omegay \over \dd y}
  \ ,
\\
  \Gamma_{txt}
  &=
  \partial_x \ln \rho
  +
  {\sigma^2 \over \rho}
  {\partial \sqrt{- \Deltax} / \sigma^2 \over \partial x}
  \ ,
\\
  \Gamma_{\phi y\phi}
  &=
  \partial_y \ln \rho
  +
  {\sigma^2 \over \rho}
  {\partial \sqrt{\Deltay} / \sigma^2 \over \partial y}
  \ ,
\\
  \Gamma_{xty}
  =
  \Gamma_{xyt}
  =
  \Gamma_{tyx}
  =
  \Gamma_{xy\phi}
  =
  \Gamma_{x\phi y}
  =
  \Gamma_{y\phi x}
  &=
  0
  \ .
\end{align}
\end{subequations}
Since the tetrad-frame connections
$\Gamma_{klm}$
constitute a set of 4 bivectors,
it is possible to combine the connections into complex combinations,
but no additional insightful result emerges from those combinations.

The 10 tetrad-frame components $G_{kl}$ of the Einstein tensor satisfy
\begin{equation}
\label{einsteingeneralrho}
  {2 \over \rho}
  D_k \partial_l \rho
  +
  G_{kl}
  +
  \eta_{kl}
  \left(
  {\textstyle \frac{1}{3}}
  R
  -
  {1 \over \rho^2}
  \partial_m \rho
  \,
  \partial^m \rho
  \right)
  =
  \left\{
  \begin{array}{l@{\qquad}l}
  4 \Cx - 2 \Cy
  +
  \Gamma_{xt\phi}^2
  +
  \Gamma_{yt\phi}^2
  &
  xx
  \\[.5ex]
  - \, 4 \Cx + 2 \Cy
  -
  \Gamma_{xt\phi}^2
  +
  \Gamma_{yt\phi}^2
  &
  tt
  \\[.5ex]
  - \, 4 \Cy + 2 \Cx
  +
  \Gamma_{xt\phi}^2
  +
  \Gamma_{yt\phi}^2
  &
  yy
  \\[.5ex]
  - \, 4 \Cy + 2 \Cx
  -
  \Gamma_{xt\phi}^2
  +
  \Gamma_{yt\phi}^2
  &
  \phi\phi
  \\[.5ex]
  -
  6 
  \Gamma_{xt\phi}
  \Gamma_{yt\phi}
  &
  xy
  \\[.5ex]
  -
  2
  (
  \Wx
  -
  \Wy
  )
  &
  t\phi
  \\[.5ex]
  0
  &
  xt , x\phi , yt , y\phi
  \end{array}
  \right.
\end{equation}
where
$R$ is the Ricci scalar,
and $\Cx$, $\Cy$, $\Wx$, and $\Wy$
are related to the Weyl tensor as described immediately below.
The Weyl tensor can be thought of as a matrix with bivector indices,
and as such has a natural complex structure
\cite{Doran:2003},
embodied in the complexified, self-dual Weyl tensor
$\Cz_{klmn}$
defined in terms of the usual Weyl tensor
$C_{klmn}$
by
\begin{equation}
\label{complexifiedWeyl}
  \Cz_{klmn}
  \equiv
  \frac{1}{4}
  \left(
  \delta_k^p \delta_l^q
  +
  \frac{\im}{2} \,
  \varepsilon_{kl}{}^{pq}
  \right)
  \left(
  \delta_m^r \delta_n^s
  +
  \frac{\im}{2} \,
  \varepsilon_{mn}{}^{rs}
  \right)
  C_{pqrs}
  \ ,
\end{equation}
where
$\varepsilon_{klmn}$
is the totally antisymmetric tensor,
normalized here to
$\varepsilon^{klmn} = [klmn]$
in an orthonormal tetrad frame.
The distinct components of the complexified Weyl tensor
form a $3 \times 3$ traceless, symmetric, complex matrix,
whose 5 distinct complex components form objects of spin
$0$, $\pm 1$, and $\pm 2$.
The spin-$0$ part of the Weyl tensor is
\begin{equation}
  C
  =
  \Cz_{xtxt}
  =
  -
  \frac{1}{2}
  \Cz_{xyxy}
  =
  -
  \frac{1}{2}
  \Cz_{xtx\phi}
  \ .
\end{equation}
The real and imaginary parts of the spin-$0$ Weyl tensor
are commonly called its polar $(p)$ and axial $(a)$ parts,
the polar part remaining unchanged,
while the axial part changes sign,
under a flip
$\bgamma_\phi \rightarrow - \gamma_\phi$ of the azimuthal tetrad axis:
\begin{equation}
\label{CWeyl}
  C
  =
  C^{(p)} + \im C^{(a)}
  \ .
\end{equation}
In the present case, the polar spin-$0$ component of the Weyl tensor is
\begin{equation}
  C^{(p)}
  =
  \Cx + \Cy
  \ ,
\end{equation}
where
\begin{subequations}
\label{CWeylpolarxy}
\begin{align}
  \rho^2
  \Cx
  &=
  \frac{\sigma^2}{12}
  {\partial \over \partial x}
  \left[
  \sigma^2
  {\partial ( \Deltax / \sigma^4 ) \over \partial x}
  \right]
  -
  {1 \over 6}
  \left(
  {1 \over \sigma^2}
  {\dd \omegax \over \dd x}
  \right)^2
  \Deltay
  \ ,
\\
  \rho^2
  \Cy
  &=
  \frac{\sigma^2}{12}
  {\partial \over \partial y}
  \left[
  \sigma^2
  {\partial ( \Deltay / \sigma^4 ) \over \partial y}
  \right]
  -
  {1 \over 6}
  \left(
  {1 \over \sigma^2}
  {\dd \omegay \over \dd y}
  \right)^2
  \Deltax
  \ ,
\end{align}
\end{subequations}
while the axial spin-$0$ component is given by
\begin{equation}
\label{CWeylaxial}
  \rho^2
  C^{(a)}
  =
  \frac{\sigma^2}{4}
  \left[
  {\dd \omegay \over \dd y}
  {\partial ( \Deltax / \sigma^4 ) \over \partial x}
  -
  {\dd \omegax \over \dd x}
  {\partial ( \Deltay / \sigma^4 ) \over \partial y}
  \right]
  \ .
\end{equation}
The only other non-vanishing component of the Weyl tensor is
the spin-$1$ component
\begin{equation}
\label{weylspin1}
  \Cz_{xtx\phi}
  =
  \Wx + \Wy
  \ ,
\end{equation}
where
\begin{subequations}
\begin{align}
  \rho^2
  \Wx
  &=
  {\sqrt{- \Deltax \Deltay} \over 4}
  {\partial \over \partial x}
  \left(
  {1 \over \sigma^2}
  {\dd \omegax \over \dd x}
  \right)
  \ ,
\\
  \rho^2
  \Wy
  &=
  {\sqrt{- \Deltax \Deltay} \over 4}
  {\partial \over \partial y}
  \left(
  {1 \over \sigma^2}
  {\dd \omegay \over \dd y}
  \right)
  \ .
\end{align}
\end{subequations}
In the spacetimes presented in this paper and its companions,
the spin-$1$ component~(\ref{weylspin1}) always vanishes.

\section{Electrovac solutions}
\label{electrovacappendix}

The standard electrovac solutions can be derived from the assumptions of
strict stationarity, axisymmetry, and strict separability as follows.
As shown by
\cite{Carter:1968c},
the line-element takes the form~(\ref{lineelement})
with a separable conformal factor $\rhosep$, equation~(\ref{rhosep}).

Given strict stationary ($\vel = 0$) and separability,
the Einstein component $G_{xy}$, which has zero electrovac source, is
\begin{equation}
\label{Gxy}
  \rhosep^2
  G_{xy}
  =
  - \frac{3}{2}
  \sqrt{- \Deltax \Deltay}
  {\ppartial \ln ( \rhosep^2 / \sigma^2 ) \over \partial x \partial y}
  \ .
\end{equation}
Homogeneous solution of this equation implies the form~(\ref{rhoseps})
of the conformal factor $\rhosep$.
At this point the constants $g_0$ and $g_1$
can be adjusted arbitrarily without affecting either $\rhox$ or $\rhoy$:
the overall normalization of $g_0$ and $g_1$
is cancelled by the normalizing factor
of $1 / \sqrt{f_0 g_1 {+} f_1 g_0}$ in $\rhox$ and $\rhoy$,
and the relative sizes of $g_0$ and $g_1$ can be changed
by adjusting an arbitrary constant in the split
between $\rhox^2$ and $\rhoy^2$.

The Einstein component $G_{t\phi}$, which also has zero electrovac source, is
\begin{equation}
\label{Gtphi}
  \rhosep^2
  G_{t\phi}
  =
  -
  {\sqrt{- \Deltax \Deltay} \over 2 \rhosep^2}
  \left[
  {\partial \over \partial x}
  \left(
  {\rhosep^2 \over \sigma^2}
  {\dd \omegax \over \dd x}
  \right)
  -
  {\partial \over \partial y}
  \left(
  {\rhosep^2 \over \sigma^2}
  {\dd \omegay \over \dd y}
  \right)
  \right]
  \ ,
\end{equation}
which,
given the expression~(\ref{rhoseps}) for the conformal factor $\rhosep$,
reduces to
\begin{equation}
\label{Gtphisep}
  \rhosep^2
  G_{t\phi}
  =
  \frac{1}{2}
  {\sqrt{- \Deltax \Deltay} \over \sigma^2}
  \left[
  {\dd \omegax \over \dd x}
  {\dd \ln ( f_0 + f_1 \omegax ) \over \dd x}
  -
  {\dd \omegay \over \dd y}
  {\dd \ln ( f_1 + f_0 \omegax ) \over \dd y}
  \right]
  \ .
\end{equation}
Homogeneous solution of this equation can be accomplished by
separation of variables,
setting each of the two terms inside square brackets,
the first of which is a function only of $x$,
while the second is a function only of $y$,
to the same separation constant $2 f_2$.
The result is
\begin{subequations}
\label{domegadvarpif2}
\begin{align}
\label{domegag2}
  {\dd \omegax \over \dd x}
  &=
  2
  \sqrt{
  \left( f_0 + f_1 \omegax \right)
  \left[ g_0 + {1 \over f_0} ( f_1 g_0 + f_2 ) \omegax \right]
  }
  \ ,
\\
\label{dvarpig2}
  {\dd \omegay \over \dd y}
  &=
  2
  \sqrt{
  \left( f_1 + f_0 \omegay \right)
  \left[ g_1 + {1 \over f_1} ( f_0 g_1 + f_2 ) \omegay \right]
  }
  \ ,
\end{align}
\end{subequations}
for some constants $g_0$ and $g_1$,
which can be taken without loss of generality to equal those
in the conformal factor~(\ref{rhoseps}).

The Einstein components
$G_{xx} + G_{tt}$
and
$G_{yy} - G_{\phi\phi}$,
which too have zero electrovac source, are
\begin{subequations}
\label{Gxxttyyphiphi}
\begin{align}
\label{Gxxtt}
  \rhosep^2
  \left(
  G_{xx} + G_{tt}
  \right)
  &=
  - \,
  {2 \Deltax \over \sigma^2}
  \left[
  \rhosep
  {\partial \over \partial x}
  \left(
  \sigma^2
  {\partial ( 1 / \rhosep ) \over \partial x}
  \right)
  +
  {1 \over 4 \sigma^2}
  \left(
  {\dd \omegay \over \dd y}
  \right)^2
  \right]
  \ ,
\\
\label{Gyyphiphi}
  \rhosep^2
  \left(
  G_{yy} - G_{\phi\phi}
  \right)
  &=
  {2 \Deltay \over \sigma^2}
  \left[
  \rhosep
  {\partial \over \partial y}
  \left(
  \sigma^2
  {\partial ( 1 / \rhosep ) \over \partial y}
  \right)
  +
  {1 \over 4 \sigma^2}
  \left(
  {\dd \omegax \over \dd x}
  \right)^2
  \right]
  \ ,
\end{align}
\end{subequations}
which with the conformal factor $\rhosep$ given by equation~(\ref{rhoseps})
and $\dd \omegax / \dd x$ and $\dd \omegay / \dd y$
given by equations~(\ref{domegadvarpif2}) reduce to
\begin{subequations}
\label{Gxxttyyphiphisepf2}
\begin{align}
\label{Gxxttsepf2}
  \rhosep^2
  \left(
  G_{xx} + G_{tt}
  \right)
  &=
  - \,
  2 \Deltax
  {( f_2 + f_0 g_1 + f_1 g_0 )
  ( f_0 + f_1 \omegax )^2
  \over
  f_0 f_1
  \sigma^4}
  \ ,
\\
\label{Gyyphiphisepf2}
  \rhosep^2
  \left(
  G_{yy} - G_{\phi\phi}
  \right)
  &=
  2 \Deltay
  {( f_2 + f_0 g_1 + f_1 g_0 )
  ( f_1 + f_0 \omegay )^2
  \over
  f_0 f_1
  \sigma^4}
  \ .
\end{align}
\end{subequations}
These vanish provided that the constant $f_2$ satisfies
\begin{equation}
\label{f2}
  f_2
  =
  -
  ( f_0 g_1 + f_1 g_0 )
  \ .
\end{equation}
Inserting this value into equations~(\ref{domegadvarpif2})
yields equations~(\ref{domegadvarpi}).

The four off-diagonal Einstein components
$G_{xt}$,
$G_{yt}$,
$G_{x\phi}$,
and $G_{y\phi}$
vanish identically.

Solution of the Einstein equations for the remaining two Einstein components
$G_{xx} - G_{tt}$ and $G_{yy} + G_{\phi\phi}$
was already discussed in \S\ref{nonlinearinflationeqs}.
Inserting the homogeneous solutions~(\ref{Yxhomog}) and (\ref{Yyhomog})
into the defining equations~(\ref{Yx}) and (\ref{Yy}) for $\Yx$ and $\Yy$,
and setting $\Ux = \Uy = 0$ (which eliminates inflation),
yields differential equations for the radial and angular horizon functions
$\Deltax$ and $\Deltay$,
solution of which, with appropriate boundary conditions,
recovers the Kerr solution.

Solutions including the energy-momentum
of a static electromagnetic field
fall out with little extra work.
With appropriate boundary conditions,
this is the Kerr-Newman solution.
Electrovac solutions
have $G_{tt} = - G_{xx}$ and $G_{\phi\phi} = G_{yy}$.
Electrovac solutions with
$G_{yy} = G_{xx}$,
as is true for a static radial electromagnetic field,
are found by taking the difference of
equations~(\ref{GUv})
and separating that difference in the pattern of equation~(\ref{sepG}).
The solution is a sum of a homogeneous solution
and a particular solution
\begin{equation}
\label{Yxye}
  \Yx
  =
  {2 
  ( \Qelecbh^2 + \Qmagbh^2 ) ( f_0 + f_1 \omegax )^2 \over \dd \omegax / \dd x}
  \ , \quad
  \Yy
  =
  0
  \ .
\end{equation}
Inserting equations~(\ref{Yxye})
into the Einstein expressions~(\ref{GUv})
yields Einstein components that have precisely the form
$G_{mn} = \left[ ( \Qelecbh^2 + \Qmagbh^2 ) / \rhosep^4 \right] \diag ( 1 , -1 , 1 , 1 )$
of the tetrad-frame energy-momentum tensor of a static
radial electromagnetic field.

Similarly,
solutions including vacuum energy,
which has
$G_{yy} = - G_{xx}$,
can be found by separating the sum of equations~(\ref{GUv})
in the pattern of equation~(\ref{sepG}).
A particular solution is
\begin{equation}
\label{Yxyvac}
  \Yx
  =
  {2 \Lambda \over f_1^2 \dd \omegax / \dd x}
  \ , \quad
  Y_\phi
  =
  {2 \Lambda \omegay^2 \over f_1^2 \dd \omegax / \dd x}
  \ .
\end{equation}
Inserting equations~(\ref{Yxyvac}) into the Einstein expressions~(\ref{GUv})
yields Einstein components that have precisely
the form of a cosmological constant,
$G_{mn} = - \Lambda \eta_{mn}$.

Solving equations~(\ref{Yx}) and (\ref{Yy})
with $\Yx$ and $\Yy$ given by
a sum of the homogeneous,
equations~(\ref{Yxhomog}) and (\ref{Yyhomog}),
electromagnetic,
equation~(\ref{Yxye}),
and vacuum,
equation~(\ref{Yxyvac}),
contributions,
yields the general electrovac solution
for the radial and angular horizon functions $\Deltax$ and $\Deltay$,
\begin{subequations}
\label{Deltaxy}
\begin{align}
\label{Deltax}
  \Deltax
  &=
  ( f_0 + f_1 \omegax )
  \left[
  ( k_0 + k_1 \omegax )
  -
  {2
  \Mbh
  \sqrt{( f_0 + f_1 \omegax ) ( g_0 - g_1 \omegax )}
  \over ( f_0 g_1 + f_1 g_0 )^{3/2}}
  +
  {( \Qelecbh^2 + \Qmagbh^2 )
  ( f_0 + f_1 \omegax ) \over f_0 g_1 + f_1 g_0}
  \right]
  -
  {\Lambda ( g_0 - g_1 \omegax ) \over 3 f_1 ( f_0 g_1 + f_1 g_0 )^2}
  \ ,
  \qquad
\\
\label{Deltay}
  \Deltay
  &=
  ( f_1 + f_0 \omegay )
  \left[
  ( k_1 + k_0 \omegay )
  -
  {2
  \NUTbh
  \sqrt{( f_1 + f_0 \omegay ) ( g_1 - g_0 \omegay )}
  \over ( f_0 g_1 + f_1 g_0 )^{3/2}}
  \right]
  +
  {\Lambda \omegay ( g_1 - g_0 \omegay ) \over 3 f_1 ( f_0 g_1 + f_1 g_0 )^2}
  \ ,
\end{align}
\end{subequations}
where
$k_0$ and $k_1$ are arbitrarily adjustable constants
arising from the freedom of choice in the constants $h_0$ and $h_1$
of the homogeneous solution.
The constant $\Mbh$
in the expression~(\ref{Deltax}) for $\Deltax$ is the black hole's mass.
The constant
$\NUTbh$
in the expression~(\ref{Deltay}) for $\Deltay$
is the NUT parameter
\cite{Taub:1951,Newman:1963,Stephani:2003,Kagramanova:2010bk},
which is to the mass $\Mbh$
as magnetic charge $\Qmagbh$ is to electric charge $\Qelecbh$.

The electrovac Weyl tensor~(\ref{CWeyl}) has only a spin-$0$ component,
and is
\begin{equation}
\label{CWeylev}
  C
  =
  - {1 \over ( \rhox - \im \rhoy )^3}
  \left(
  \Mbh + \im \NUTbh
  +
  {\Qelecbh^2 + \Qmagbh^2 \over \rhox + \im \rhoy}
  \right)
  \ ,
\end{equation}
with $\rhox$ and $\rhoy$ given by equations~(\ref{rhoxy}).

\section{Integrals along the path of a particle}
\label{estimateintegrals}

This Appendix derives conditions under which integrals
along the path of a particle can be deemed small,
in the conformally stationary limit of small accretion rate.
The results confirm that
the tetrad-frame momentum $p_k$
and the density $N$
predicted by the Hamilton-Jacobi equations for a massive particle
are accurate,
by demonstrating that the integrals of
equations~(\ref{dpkdx}) and (\ref{Dknk})
are adequately small.

Consider an integral of the form
\begin{equation}
\label{I}
  I
  =
  \int | \Deltax |^\alpha (\rho/\rhosep)^\beta \, \dd \lambda
  \ .
\end{equation}
Along the path of a particle,
$\dd \lambda = ( \rho^2 / | \Px | ) \, \dd x$,
and
$\rho/\rhosep = \ee^{\vel t - \xi} = \left[ ( \uel \mp \vel ) / ( \Ux \mp \vel) \right]^{1/2}$
from equations~(\ref{rhoinf}) and (\ref{evtparticle}).
The integration interval
$\dd x$
may be converted to either
$\dd \Deltax$
or
$\dd \Ux$
using either of equations~(\ref{DDUx}).
The integral~(\ref{I}) thus becomes
\begin{equation}
\label{IDU}
  I
  =
  \int 
  {| \Deltax |^{1+\alpha} \over | \Px |}
  \left( {\uel \mp \vel \over \Ux \mp \vel} \right)^{1 + \beta/2}
  {\dd \ln | \Deltax | \over
  \Deltax^\prime - 3 \Ux}
  =
  \int 
  {| \Deltax |^{1+\alpha} \over | \Px |}
  \left( {\uel \mp \vel \over \Ux \mp \vel} \right)^{1 + \beta/2}
  {\dd \Ux \over
  2 ( \Ux^2 - \vel^2 )}
  \ .
\end{equation}
The horizon function $\Deltax$ is given as a function of $\Ux$
by equation~(\ref{Dinf}).

Inflation ignites when $| \Deltax | \sim \vel$,
and the integral~(\ref{I}) may be counted from this point.
At the beginning of inflation,
before $| \Deltax |$ has been driven to an exponentially small value,
$\Ux$ changes little from its initial value of $\uel$,
and then the middle expression of equations~(\ref{IDU}) yields
\begin{equation}
\label{Iearly}
  I
  \approx
  {| \Deltax |^{1+\alpha} \over ( 1 + \alpha ) | \Px | \Deltax^\prime}
  \sim
  \vel^{1 + \alpha}
  \ .
\end{equation}
This integral is small in the conformally stationary limit $\vel \rightarrow 0$
provided that $\alpha > -1$,
but as to whether an integral in an actual situation can be judged small
will depend on additional factors of $\vel$ that the integral may be
multiplied by.

By the time that $\Ux$ has reached of order unity times its initial value $\uel$,
the horizon function
$| \Deltax |$ has become exponentially small.
In this regime further contribution to the integral is exponentially small
provided that
\begin{equation}
\label{a}
  \alpha > -1
  \ .
\end{equation}

During collapse, both $| \Deltax |$ and $\Ux$ increase exponentially
($| \Deltax |$ from an exponentially small starting point).
As long as $| \Deltax | \lesssim 1$,
further contribution to the integral remains exponentially small
provided that
\begin{equation}
\label{b}
  \beta > -4
  \ .
\end{equation}

At the end of collapse,
there is a regime where
$| \Deltax | \gtrsim 1$ and $| \Px | \sim \sqrt{- \Deltax}$
but $| \Deltax | \ll \Ux$,
while $\Ux$ is exponentially huge.
Here further contribution to the integral remains exponentially small provided that
\begin{equation}
\label{ablate}
  \mbox{either} \quad
  \beta \geq -3 + 2 \alpha
  \quad \mbox{or} \quad
  \beta > -4
  \ .
\end{equation}

In the case of equation~(\ref{dpkdx}) for $\dd p_k / \dd \lambda$
there are two terms.
The first has
$\alpha = 0$ and $\beta = -3$.
The second has
$\alpha = 0$ and $\beta = -1$,
except that the $x$ component has
$\alpha = -1/2$ and $\beta = -1$
when $| \Deltax | \lesssim 1$.
Both terms have pre-factors proportional to $\vel$
during early inflation
(including the term proportional to
$\partial ( \rho^2 - \rhoy^2 ) \ \partial y$).
For both terms,
the main contribution to the integral 
is from early inflation,
equation~(\ref{Iearly}),
which, multiplied by the $\vel$ pre-factor
yields $\sim \vel^2$ for the first term,
and $\sim \vel^{3/2}$ for the second.
The ratio of the integral of $\dd p_k / \dd \lambda$ to $p_k$ itself is,
since $p_x \propto 1/\sqrt{| \Deltax |}$
when $| \Deltax | \lesssim 1$,
\begin{equation}
\label{Dlnpk}
  {\Delta p_k \over p_k}
  \equiv
  {\int {\dd p_k / \dd \lambda} \, \dd \lambda
  \over p_k}
  \sim \vel^2
  \ .
\end{equation}
Both terms satisfy the conditions~(\ref{a}), (\ref{b}), and (\ref{ablate})
so the contributions to the integrals after early inflation
are exponentially small.

In the case of equation~(\ref{Dknk}) for $D_k n^k$,
the exponents are
$\alpha = 0$ and $\beta = -2$,
and the pre-factor is proportional to $\vel$ during early inflation.
Again, the main contribution to the integral is during early inflation,
giving
\begin{equation}
\label{DlnN}
  \Delta \ln N
  \equiv
  {\int {\dd p_k / \dd \lambda} \, \dd \lambda
  \over p_k}
  \sim \vel^2
  \ .
\end{equation}
Again the conditions~(\ref{a}), (\ref{b}), and (\ref{ablate})
are satisfied,
so the contribution to the integral after early inflation
is exponentially small.

\end{document}